\documentclass[11pt]{article}
\usepackage[authoryear]{natbib}
\usepackage{graphicx}
\usepackage[latin1]{inputenc}
\usepackage{amssymb,amsthm,amsmath,epsfig}
\usepackage{mathabx}
\usepackage{latexsym}
\usepackage{times}
\usepackage{bm,color}
\usepackage{epstopdf}
\usepackage{slashbox} 
 \usepackage{pict2e} 
\usepackage{booktabs,multirow}
\usepackage{rotating}
\usepackage{array,tabularx}
\usepackage{scalefnt}
\usepackage{subfigure}
\usepackage{cancel}
\usepackage{footnote}
\usepackage{subfig}
\usepackage[dvipsnames]{xcolor}
\usepackage{url}
\usepackage{caption, makecell}

\usepackage[flushleft]{threeparttable}

\def\R{{\mathbb R}}
\def\Z{{\mathbb Z}}
\def\P{{\mathbb P}}
\def\E{{\mathbb E}}
\def\N{{\mathbb N}}

\newcommand{\Beta}{{\rm Beta}}
\newcommand{\e}{{\rm e}}

\newcommand{\diag}{{\rm diag}}
\newcommand{\mat}[1]{\mbox{\boldmath\(#1\)\unboldmath}}

\newcommand{\longurl}[1]{%
	{\expandafter\def\expandafter\UrlBreaks\expandafter{\UrlBreaks\UrlOrds%
			\do\/\do\a\do\b\do\c\do\d\do\e\do\f%
			\do\g\do\h\do\i\do\j\do\k\do\l\do\m%
			\do\n\do\o\do\p\do\q\do\r\do\s\do\t%
			\do\u\do\v\do\w\do\x\do\y\do\z%
			\do\A\do\B\do\C\do\D\do\E\do\F\do\G%
			\do\H\do\I\do\J\do\K\do\L\do\M\do\N%
			\do\O\do\P\do\Q\do\R\do\S\do\T\do\U%
			\do\V\do\W\do\X\do\Y\do\Z}%
		\url{#1}}%
}

\title{Robust estimation in beta regression via maximum L$_q$-likelihood }

\author{Terezinha K. A.~Ribeiro\\
{\small {\em Department of Statistics, University of S\~ao Paulo, Brazil}}\\
Silvia L.P.~Ferrari\footnote{Corresponding author: {\tt silviaferrari@usp.br }}\\
{\small {\em Department of Statistics, University of S\~ao Paulo, Brazil}}
}
\date{}

\begin{document}
\maketitle

\begin{abstract}

Beta regression models are widely used for modeling continuous data limited to the unit interval, such as proportions, fractions, and rates. The inference for the parameters of beta regression models is commonly based on maximum likelihood estimation. However, it is known to be sensitive to discrepant observations. In some cases, one atypical data point can lead to severe bias and erroneous conclusions about the features of interest. In this work, we develop a robust estimation procedure for beta regression models based on the maximization of a reparameterized L$_q$-likelihood. The new estimator offers a trade-off between robustness and efficiency through a tuning constant. To select the optimal value of the tuning constant, we propose a data-driven method which ensures full efficiency in the absence of outliers.  
We also improve on an alternative robust estimator by applying our data-driven method to select its optimum tuning constant. Monte Carlo simulations suggest marked robustness of the two robust estimators with little loss of efficiency. 
Applications to three datasets are presented and discussed. As a by-product of the proposed methodology, 
residual diagnostic plots based on robust fits highlight outliers that would be masked under maximum likelihood estimation.\\


\end{abstract}

\section{Introduction}\label{Int}

Random phenomena that produce data in the unit interval $(0, 1)$ are common in many research areas, including medicine, finance, economics, environment, hydrology, and psychology. Data of this type are usually rates, proportions, percentages, or fractions. A few examples are unemployment rates, the proportion of exports in total sales, the fraction of surface covered by vegetation, Gini's index, the loss given default, health-related quality of life scores, and neonatal mortality rates. Beta regression models \citep{ferrari2004beta,smithson2006better} are widely employed for modeling continuous data bounded within the unit interval. The response variable is assumed to follow a beta distribution indexed by the mean and a precision parameter, which in turn are modeled through a variety of regression structures.

 Let the response variable $y_i$ have a beta distribution with  mean  $0<\mu_i<1$ and precision parameter $\phi_i>0$. The probability density function  of $y_i$ is
\begin{eqnarray}
 f(y_i;\mu_i,\phi_i)=\frac{1}{B(\mu_i\phi_i,(1-\mu_i)\phi_i)}y_i^{\mu_i\phi_i-1}(1-y_i)^{(1-\mu_i)\phi_i-1}, \mbox{ }0<y_i<1,
 \label{fdpbeta}
\end{eqnarray}
where $B(.,.)$ is the beta function. We write $y_i\sim\Beta(\mu_i,\phi_i)$, $i=1,\ldots,n$, and assume that $y_1,\ldots,y_n$ are independent. Since ${\rm Var}(y_i)=\mu_i(1-\mu_i)/(1+\phi_i)$, the parameter $\phi_i$ is taken as a precision parameter because it is inversely related to the dispersion of $y_i$, for fixed mean $\mu_i$. The beta density (\ref{fdpbeta}) assumes different forms depending on the parameters. It has a single mode in $(0,1)$ if $\mu_i\phi_i>1$ and $(1-\mu_i)\phi_i>1$, and its possible forms include J, inverted-J, tilde, inverted tilde, bathtub and uniform shapes. When $\mu_i\phi_i<1$ or $(1-\mu_i)\phi_i<1$ the beta density is unbounded at one or both of the boundaries. We will assume that the regression structures for $\mu_i$ and $\phi_i$ are, respectively,
\begin{eqnarray}
 g_\mu(\mu_i) = \mat{X}_i^\top{\boldsymbol \beta}, \quad {\rm (mean \ submodel)}
 \label{ligmu}
\end{eqnarray}
\begin{eqnarray}
 g_\phi(\phi_i)= \mat{Z}_i^\top{\boldsymbol \gamma}, \quad {\rm (precision \ submodel)}
 \label{ligphi}
\end{eqnarray}
where ${\boldsymbol \beta}= (\beta_1,\ldots,\beta_{p_1})^\top\in\mathbb{R}^{p_1}$ and $\mat{\gamma} = (\gamma_1,\ldots,\gamma_{p_2})^\top\in\mathbb{R}^{p_2}$ are vectors of unknown regression parameters, $\mat{X}_i=(x_{i1},\ldots,x_{ip_1})^\top$ and $\mat{Z}_i=(z_{i1},\ldots,z_{ip_2})^\top$  are vectors of covariates of lengths $p_1$ and $p_2$ ($p_1+p_2 = p <n$), respectively, which are assumed fixed and known. In addition, $g_\mu(\cdot): (0,1)\rightarrow \mathbb{R}$ and $g_\phi(\cdot): (0,\infty)\rightarrow \mathbb{R}$  are strictly monotonic and twice differentiable link functions. The model introduced by \citet{ferrari2004beta} is a constant precision (or, equivalently, constant dispersion) beta regression model in the sense that $\phi_i=\phi$, for $i=1,\ldots,n$.

Inference for the beta regression model (\ref{fdpbeta})-(\ref{ligphi}) is commonly based on the maximum likelihood estimator (MLE). However, it may be highly influenced by the presence of outlying observations in the data.

\begin{figure}[h]
	\centering
	\includegraphics[width=6cm,height=6cm]{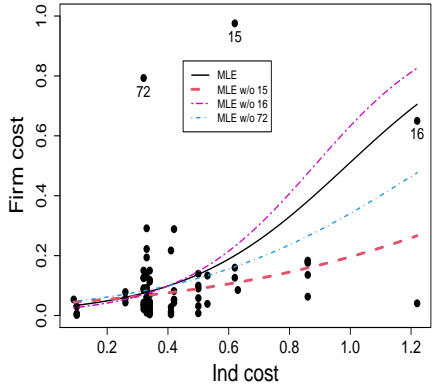}
	\caption{Scatter plot of Firm cost versus Ind cost along with the fitted lines based on the MLE for the full data and the data without outliers.}
\label{scatter_Indcost_introd}
\end{figure}

As an illustration of the sensitivity of the MLE to outliers, we consider a dataset on risk management practices of 73 firms introduced and analyzed by \cite{schmit1990cost}; see also \cite{gomez2014log}. The response variable is $\mbox{Firm cost}$, defined as premiums plus uninsured losses as a percentage of the total assets. Smaller values of $\mbox{Firm cost}$ are attributed to firms that have a good risk management performance. We take only two covariates, namely $\mbox{Ind cost}$, a measure of the firm's industry risk, and $\mbox{Size log}$, the logarithm of its total asset. Figure \ref{scatter_Indcost_introd} displays the scatter plot of $\mbox{Firm cost}$ versus $\mbox{Ind cost}$ along with the fitted lines based on the MLE for a beta regression with varying precision for the full data and the data without the most eye-catching outliers, labeled in the plots as observations 15, 16 and 72.\footnote{The scatter plots were produced by setting the value of the remaining covariate at its sample median value.} These observations have  disproportionate effects on the fitted regression curves.

To deal with atypical observations, models based on distributions that are more flexible than the beta distribution may be employed. For instance, \cite{bayes2012new}, \cite{migliorati2018new}, and \cite{dibrisco2019} proposed different kinds of mixtures of beta distributions. However, more flexibility comes at the cost of model complexity and additional parameters, making model specification, inference, and interpretation more difficult. Our approach keeps the model simple, i.e. the beta distribution remains as the postulated model, but the inference procedure is replaced by a robust
method.

\cite{ghosh2019robust} proposed a robust estimator in beta regression models based on the minimization of an empirical density power divergence introduced by \cite{basu1998}. The estimator, called the minimum density power divergence estimator (MDPDE),  is similar to the approach we will propose here.  The main difference between the two estimation procedures is that they are based on two different families of divergences that have the Kullback-Leibler divergence as a special case. In other words, both have the MLE as a special case. \cite{ghosh2019robust} presented some simulations of the MDPDE, but only for constant precision beta regression models with fixed values for the tuning constant. Here, we propose a data-driven method to select the tuning constant of the MDPDE which ensures full efficiency in the absence of contamination. Additionally, we evaluate the performance of the MDPDE under constant and varying precision using the proposed selection algorithm.

The chief goal of this paper is to propose a new robust estimator under the beta regression model (\ref{fdpbeta})-(\ref{ligphi}) with a scheme determined by the data to choose its optimum tuning constant. In Section \ref{Lq}, we introduce a robust estimator based on the maximization of a reparameterized L$_q$-likelihood, and derive some properties. In Section \ref{qchoice}, we introduce a data-driven method to the selection of the tuning constant. In Section \ref{MC}, we provide Monte Carlo evidence on the proposed robust inference approach.   Section \ref{app} contains applications of the robust estimation for three datasets.  Finally,  Section \ref{concl} closes the paper with some concluding remarks. Some technical details are left for an appendix. The on-line Supplementary Material contains further technical information and additional numerical results. The codes for the simulations and applications and the datasets are available at \url{https://github.com/terezinharibeiro/RobustBetaRegression}.

\section{Robust inference}\label{Lq}

Let $y_1,\ldots,y_n$ be $n$ independent observations, where $y_i$ comes from a postulated model indexed by a common unknown parameter $\mat{\theta}\in \Theta \subset \mathbb{R}_p$. For the beta regression model, the postulated density for $y_i$ is given by (\ref{fdpbeta})-(\ref{ligphi}), and we will denote it, from now on, by $f_{\boldsymbol \theta}(\cdot;\mu_i,\phi_i)$. Let $\ell_q(\mat{\theta})$ denote the L$_q$-likelihood \citep{ferrari2010maximum},
\begin{eqnarray}
\ell_{q}({\boldsymbol \theta})=\displaystyle\sum_{i=1}^{n}L_q(f_{\boldsymbol \theta}(y_i;\mu_i,\phi_i)),
\label{vero_Lq}
\end{eqnarray}
where $0<q\le1$ is a tuning constant, and
\begin{align*}
L_q(u)=\begin{cases}
  (u^{1-q}-1)/(1-q), \mbox{ if } q\neq 1,\\
  \log(u) , \mbox{ if } q=1,
 \end{cases}
\end{align*}
is the Box-Cox transformation \citep{box1964analysis}, also called distorted logarithm. The population version of $-\ell_q(\mat{\theta})/n$ with respect to the density $g(\cdot)$ is
\begin{eqnarray*}
H_q(f_{\boldsymbol \theta},g)=-\displaystyle\int L_q\left(f_{\boldsymbol \theta}({y};{\mu},{\phi})\right)g({y})d{y},
\end{eqnarray*}
which is the $q$-entropy of $f_{\boldsymbol \theta}$ with respect to $g$  \citep{tsallis1988possible} and reduces to the Shannon entropy when $q=1$.  For $q=1$, $\ell_q(\mat{\theta})$ is the usual log-likelihood function and maximizing $\ell_q(\mat{\theta})$ leads to the maximum likelihood estimator (MLE).

The estimator obtained by maximizing (\ref{vero_Lq}), will be denoted by  $\widetilde{{\boldsymbol\theta}}_{q}$. It comes from the estimating equation
\begin{eqnarray}
 \sum_{i=1}^{n}U(y_i,{\boldsymbol \theta})f_{\boldsymbol \theta}(y_i;\mu_i,\phi_i)^{1-q} = \mat{0},
\label{esteq}
\end{eqnarray}
where $U(y_i,{\boldsymbol \theta})=\nabla_{\boldsymbol \theta}\log[f_{\boldsymbol \theta}(y_i;\mu_i,\phi_i)]$, with $\nabla_{{\boldsymbol\theta}}$ denoting the gradient with respect to $\mat{\theta}$. The score vector for $\mat{\theta}$ corresponding to the $i$-th observation is given by
\begin{eqnarray}
U(y_i;\mat{\theta})=\left(
\begin{array}{ccc}
\phi_i   \frac{(y_i^{\star}-\mu_i^{\star})}{g_\mu'(\mu_i)}\mat{X}_i^\top, & \frac{\mu_i(y_i^{\star}-\mu_i^{\star})+y_i^{\dagger}-
\mu_i^{\dagger}}{g_\phi'(\phi_i)}\mat{Z}_i^\top  \\
\end{array}
\right)^\top,
\label{escore1}\end{eqnarray}
where $ y_i^{\star}= \log(y_i/(1-y_i))$,   $y_i^{\dagger}=\log(1-y_i)$,  $\mu_i^{\star}=\mbox{E}(y_i^{\star})=\psi(\mu_i\phi_i)-\psi((1-\mu_i)\phi_i)$, and $\mu_i^{\dagger}=\mbox{E}(y_i^{\dagger})=\psi((1-\mu_i)\phi_i)-\psi(\phi_i)$, with $\psi(\cdot)$ denoting the digamma function.

The score vector of the  $i$-th observation is weighted by $f_{\boldsymbol \theta}(y_i;\mu_i,\phi_i)^{1-q}$, which depends on the postulated model and the tuning constant $q$. A choice of $q$ not too close to 1 produces a robust estimation  procedure, because observations that are inconsistent with the assumed model receive smaller weights. Furthermore, (\ref{esteq}) defines an M-estimation procedure, first introduced  by \cite{huber1964robust}. The class of M-estimators is widely used for obtaining robust estimators, and has interesting properties such as the asymptotic normality \citep{huber1981j}.

The estimating function $U(\mat{y}; \mat{\theta})=\sum_{i=1}^{n}U(y_i,{\boldsymbol \theta})f_{\boldsymbol \theta}(y_i;\mu_i,\phi_i)^{1-q}$ is biased unless $q=1$. Hence, the estimator $\widetilde{{\boldsymbol\theta}}_{q}$ is not Fisher-consistent.  \cite{Ferrari2012} showed that a suitable calibration function may be applied to rescale the parameter estimates and to achieve Fisher-consistency if the family of postulated distributions is closed under the power transformation. Given a density $h$ and a constant $\alpha\in (0, \infty)$, the power transformation is defined as
\begin{eqnarray}
h^{(\alpha)}(y) = \frac{h(y)^\alpha}{\int h(y)^\alpha dy} \ \propto \ h(y)^\alpha, \ \forall y \mbox{ in the support},
\label{proph}
\end{eqnarray}
provided that $\int h(y)^\alpha dy <\infty$, for $0<\alpha<\infty$. For a family of postulated densities $\{h_{\boldsymbol\theta}(\cdot), {\boldsymbol \theta}\in \Theta\}$ that is closed under (\ref{proph}) for all $0<\alpha<\infty$, let $\tau_{\alpha}(\mat{\theta}): \Theta \mapsto \Theta$ be an invertible continuous function satisfying $h_{\tau_\alpha(\boldsymbol \theta)}(y)=h_{\boldsymbol \theta}^{(\alpha)}(y)$ for all $y$ in the support, which is assumed not to depend on $\mat{\theta}$. Note that the function  $\tau_\alpha(\cdot)$  maps each density $h_{\boldsymbol \theta}(\cdot)$ with a unique power-transformed density $h_{\tau_\alpha(\boldsymbol \theta)}(\cdot)$.

 The $q$-entropy $H_q(f_{\boldsymbol \theta}, g)$ is related with the family of divergences of  $f_{\boldsymbol \theta}$ with respect to $g$
 $$D_q(f_{\boldsymbol \theta}, g)=-q^{-1}\displaystyle\int L_q\left(\frac{f_{\boldsymbol \theta}({y};{\mu},{\phi})}{g(y)}\right)g({y})d{y}.$$
 \citet[Lemma 1]{Ferrari2012} showed that $D_q(f_{\boldsymbol \theta}, g^{(1/q)}) \ \propto \ H_q(f_{\boldsymbol \theta}, g) - H_q(g^{(1/q)}, g)$. Consequently, the minimizer $\widetilde{{\boldsymbol\theta}}_{q}$ of the empirical $q$-entropy with respect to $\mat{\theta}$ equals the minimizer of the empirical version of $D_q(f_{\boldsymbol \theta},g^{(1/q)})$. In the minimization of $D_q(f_{\boldsymbol \theta},g^{(1/q)})$, the target density is the transformed density $g^{(1/q)}$ instead of $g$. \citet[Proposition 1]{Ferrari2012} show that $\widetilde{{\boldsymbol\theta}}_{q}$ is not Fisher-consistent for $\mat{\theta}$ but it is Fisher-consistent for $\tau_{1/q}(\mat{\theta})$. Hence, $\widehat{{\boldsymbol\theta}}_{q}=\tau_q(\widetilde{{\boldsymbol\theta}}_{q})$ is Fisher-consistent for $\mat{\theta}$. Alternatively, $\widehat{{\boldsymbol\theta}}_{q}$ is obtained by maximizing the L$_q$-likelihood in the parameterization $\tau_{1/q}(\boldsymbol \theta)$ \citep{la2015robust}. Another proof of the Fisher-consistency of $\widehat{{\boldsymbol\theta}}_{q}$ is given in the Supplementary Material. 

For the beta regression model (\ref{fdpbeta})-(\ref{ligphi}), $\widehat{{\boldsymbol\theta}}_{q}$ is the maximizer of
\begin{eqnarray*}
\ell^{*}_{q}({\boldsymbol \theta})=\displaystyle\sum_{i=1}^{n}L_q(f_{\tau_{1/q}(\boldsymbol \theta)}(y_i;\mu_i,\phi_i))=
\displaystyle\sum_{i=1}^{n}L_q\left(f_{\boldsymbol \theta}^{(1/q)}(y_i;\mu_i,\phi_i)\right),
\label{likelihoodLq}
\end{eqnarray*}
where $f^{(1/q)}_{\boldsymbol \theta}(y_i;\mu_i,\phi_i)= f_{\boldsymbol \theta}(y_i;\mu_{i,q^{-1}},\phi_{i,q^{-1}})$, for $0<q<1$, with
\begin{eqnarray}\label{muphi}
\mu_{i,q} = {\phi_{i,q}}^{-1}[q(\mu_{i}\phi_{i}-1)+1], \qquad \phi_{i,q}= q(\phi_{i}-2)+2,
\end{eqnarray}
provided that $0<\mu_{i,q^{-1}}<1$ and $\phi_{i,q^{-1}}>0$, or equivalently, $\mu_i\phi_i>1-q$ and $(1-\mu_i)\phi_i>1-q$. Then, $f_{\boldsymbol \theta}(y_i;\mu_i,\phi_i)$ satisfies (\ref{proph}) for all $\alpha = 1/q>0$,  if $\mu_i\phi_i\ge1$ and $(1-\mu_i)\phi_i\ge 1$, i.e. if $f_{\boldsymbol \theta}(y_i;\mu_{i},\phi_{i})$ is bounded. From (\ref{ligmu}) and (\ref{ligphi}), $f^{(1/q)}_{\boldsymbol \theta}(y_i;\mu_i,\phi_i)$ corresponds to a modified beta regression model defined by the beta density (\ref{fdpbeta}) with mean and precision submodels given respectively by
\begin{eqnarray*}
g^{*}_{ \boldsymbol\mu}(\mu_i) =g_{ \boldsymbol\mu}(\mu_{i,q}), \quad  g^{*}_{ \boldsymbol\phi}(\phi_i)= g_{ \boldsymbol\phi}(\phi_{i,q}),
\label{lig2}
\end{eqnarray*}
and we denote it by $f_{\boldsymbol \theta}^*(y_i;\mu_i,\phi_i)$.

Therefore, we propose the estimator that  is obtained through the maximization of
\begin{eqnarray}
\ell^{*}_{q}({\boldsymbol \theta})=\displaystyle\sum_{i=1}^{n}L_q(f^{*}_{\boldsymbol \theta}(y_i;\mu_i,\phi_i)),
\label{vero_pseudo_aditiva2}
\end{eqnarray}
which leads to the following estimating equation
\begin{eqnarray}
\sum_{i=1}^{n}U^{*}(y_i,{\boldsymbol \theta})f^{*}_{\boldsymbol \theta}(y_i;\mu_i,\phi_i)^{1-q} = \mat{0},
\label{eqest22}\end{eqnarray}
where $U^{*}(y_i,{\boldsymbol \theta})=\nabla_{\boldsymbol \theta}\log[f^{*}_{\boldsymbol \theta}(y_i;\mu_i,\phi_i)]$ being a modified score vector for $\mat{\theta}$ corresponding to the $i$-th observation given by
\begin{eqnarray}
U^{*}(y_i;\mat{\theta})=\left(
\begin{array}{ccc}
q^{-1}\phi_{i,q}   \frac{(y_i^{\star}-\mu_i^{\star})}{g_\mu'(\mu_{i,q})}\mat{X}_i^\top, &q^{-1}  \frac{ \mu_{i,q}(y_i^{\star}-\mu_i^{\star}) + y_i^{\dagger}-\mu_i^{\dagger}}{g_\phi'(\phi_{i,q})}\mat{Z}_i^\top
\end{array}
\right)^\top.
\label{Modescore}\end{eqnarray}
The final estimator, $\widehat{{\boldsymbol\theta}}_{q}$, will be simply called the surrogate maximum likelihood estimator (SMLE).

From the theory of M-estimation, $\widehat{{\boldsymbol\theta}}_{q}\mbox{ }{\overset{a}{\sim}}\mbox{ }\mbox{N}(\mat{\theta},V_q(\mat{\theta})),$ where $\overset{a}{\sim}$ denotes asymptotic distribution and 
\begin{equation}\label{cov}
V_q(\mat{\theta}) = J_q(\mat{\theta})^{-1}K_q(\mat{\theta}){J_q(\mat{\theta})^{-1}}^{\top},
\end{equation}
with
\begin{eqnarray*}
J_q(\mat{\theta}) = \sum_{i=1}^{n} \mbox{E}\left\{\nabla_{{\boldsymbol \theta}^\top}\left[ U^{*}(y_i,{\boldsymbol \theta})f^{*}_{\boldsymbol \theta}(y_i;\mu_i,\phi_i)^{1-q}\right]\right\}, \ \ 
K_q(\mat{\theta}) = \sum_{i=1}^{n} \mbox{E}\left\{ U^{*}(y_i,{\boldsymbol \theta})U^{*}(y_i,{\boldsymbol \theta})^\top f^{*}_{\boldsymbol \theta}(y_i;\mu_i,\phi_i)^{2(1-q)}\right\};
\end{eqnarray*}
see \citet[Section 3.2]{Ferrari2012}. For the beta regression model (\ref{fdpbeta})-(\ref{ligphi}), $J_q(\mat{\theta})$ and $K_q(\mat{\theta})$ are given in the Appendix. The matrix $V_q(\mat{\theta})$ is well-defined provided that $0<\mu_{i,2-q}<1$ and $\phi_{i,2-q}>0$, or equivalently  $\mu_{i}\phi_{i} > 2(1-q)/(2-q)$ and $(1-\mu_{i})\phi_{i} > 2(1-q)/(2-q)$. These conditions hold for all $0<q\le 1$, if $\mu_{i}\phi_{i} \ge 1$ and $(1-\mu_{i})\phi_{i} \ge 1$. 

Likelihood-based tests are usually constructed from the three well-known test statistics, namely the likelihood ratio, Wald, and score statistics, and they rely on maximum likelihood estimation. Following \cite{heritier1994robust} (see also \citet[Section 2.5.3]{heritier2009robust}), robust versions of these statistics may be derived by replacing the MLE, the log-likelihood function, the score vector, and the asymptotic covariance matrix respectively by the SMLE, the reparameterized $L_q$-likelihood in (\ref{vero_pseudo_aditiva2}), the estimating function in (\ref{eqest22}), and the matrix in (\ref{cov}). In this paper, we focus on the robust Wald statistic, referred here as the Wald-type statistic, of the null hypothesis $\beta_k=\beta_k^{(0)}$ against a two-sided alternative defined as $\{(\widehat \beta_{kq}-\beta_k^{(0)})/{\rm se}(\widehat\beta_{kq})\}^2$, which has a $\chi^2_1$ distribution under the null hypothesis. Here, $\widehat\beta_{kq}$ is the SMLE of $\beta_k$ and ${\rm se}(\widehat\beta_{kq})$ is its asymptotic standard error that is computed from (\ref{cov}).

\begin{figure}[!htb]
\centering
\subfigure{\includegraphics[width=5.0cm,height=5.0cm]{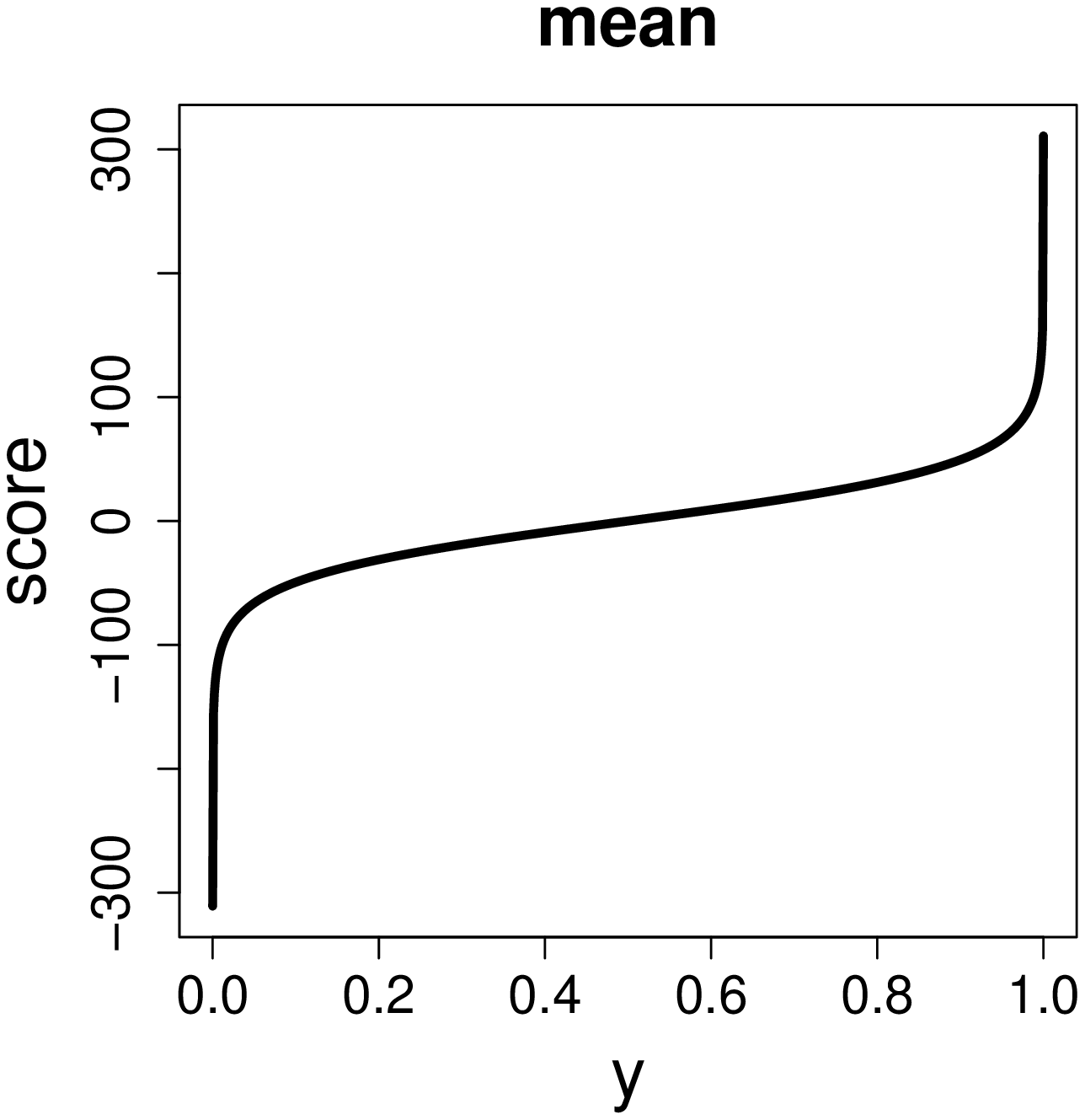}}
\qquad
\subfigure{\includegraphics[width=5.0cm,height=5.0cm]{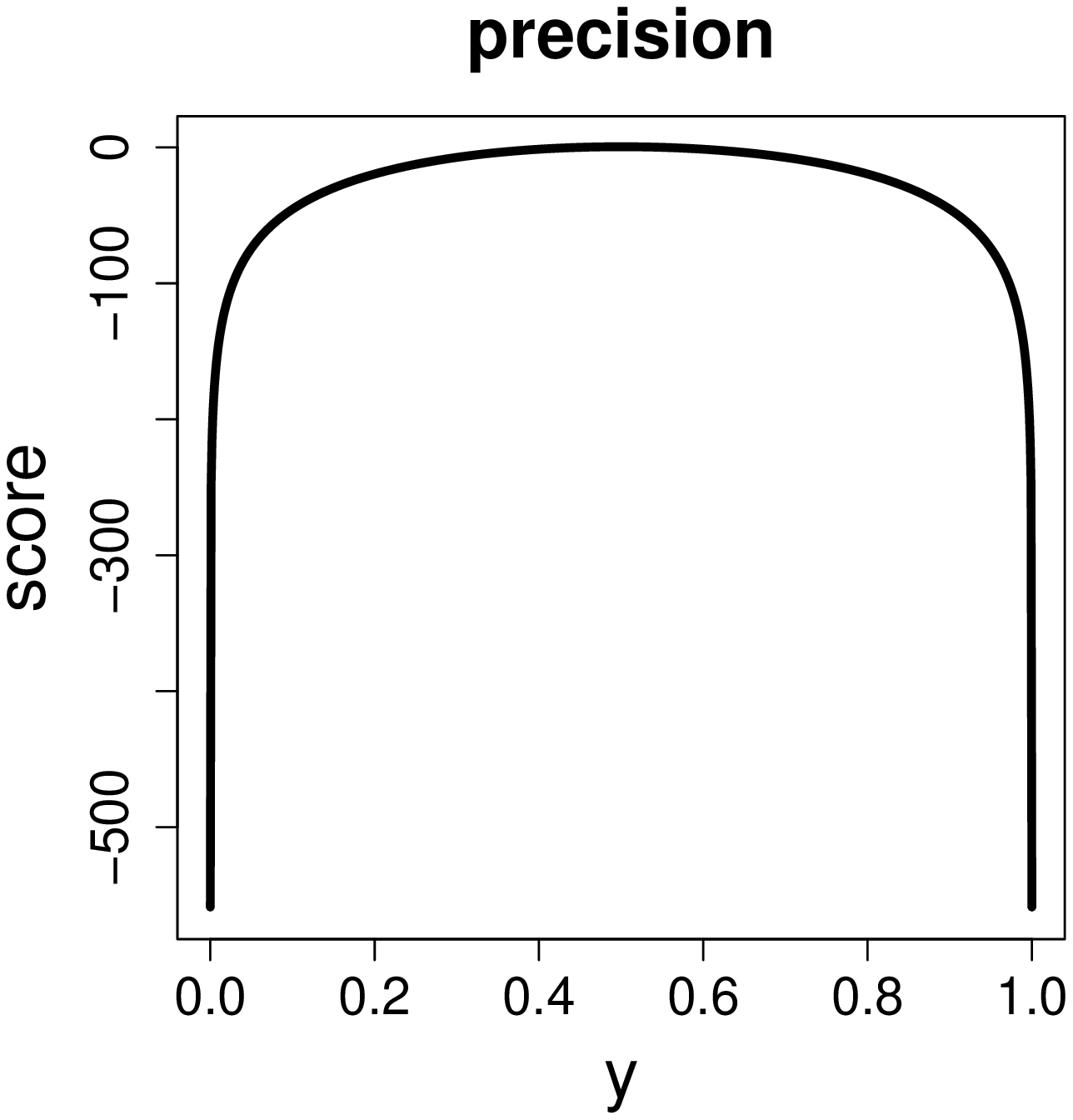}}\\
\subfigure{\includegraphics[width=5.0cm,height=5.0cm]{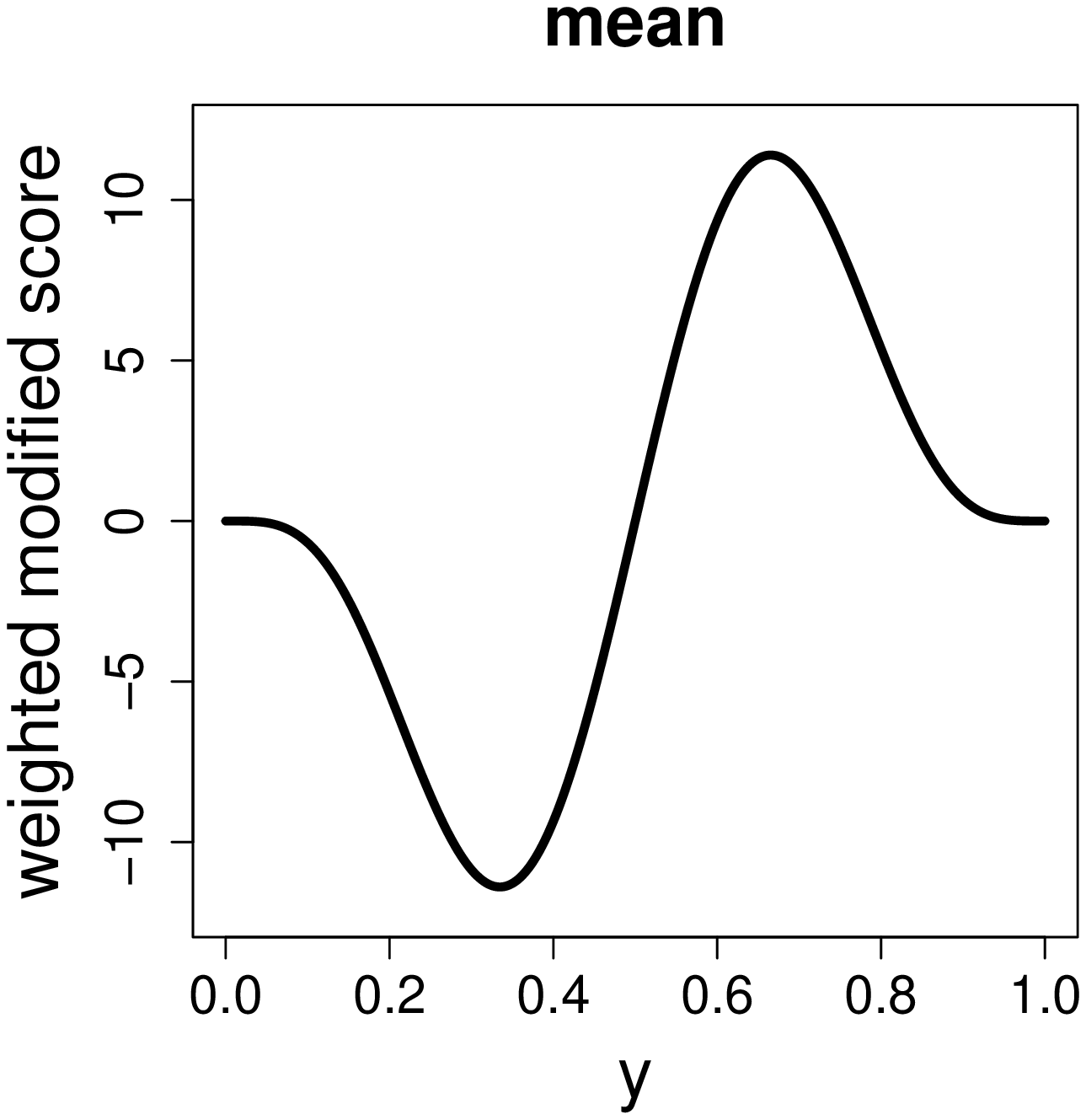}}
\qquad
\subfigure{\includegraphics[width=5.0cm,height=5.0cm]{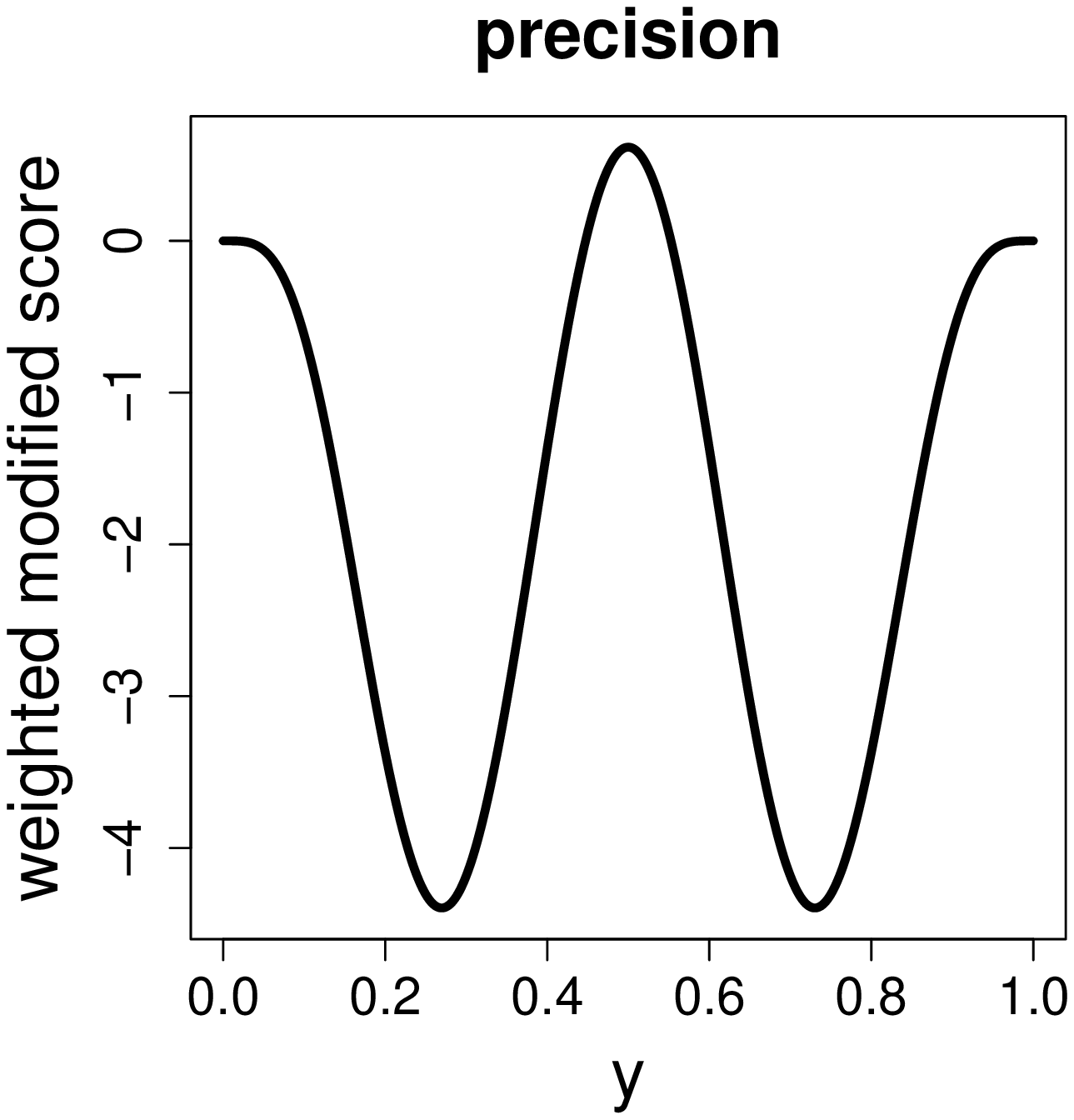}}\\
\caption{Components of the score vector and of the weighted modified score vector versus $y$; $\mu=0.5$, $\phi= 90$, and $q=0.9$.}
\label{Score_functions}
\end{figure}

Since the MLE ($\widehat{\mat{\theta}}$) and the SMLE are M-estimators, their influence functions for the beta regression model (\ref{fdpbeta})-(\ref{ligphi}) are $\mbox{IF}(y; \widehat{\mat{\theta}})=-K_1({\mat{\theta}})^{-1}U(y;{\mat{\theta}})$ and $\mbox{IF}(y;\widehat{{\boldsymbol\theta}}_{q})=-J_q(\mat{\theta})^{-1}U^{*}(y,{\boldsymbol \theta})f^{*}_{\boldsymbol \theta}(y;\mu,\phi)^{1-q}$, where
$U(y;{\mat{\theta}})$ and $U^{*}(y,{\boldsymbol \theta})$ are given in (\ref{escore1}) and (\ref{Modescore}), respectively, $K_1({\mat{\theta}})$ is the Fisher information matrix (see the Appendix), and $J_q(\mat{\theta})$ is given in (\ref{cov}). The influence function measures the asymptotic bias of the estimator caused by an infinitesimal contamination introduced in the data point $y$
\citep[Section 2.1.b]{hampel2011robust}. The individual score vector $U(y;{\mat{\theta}})$ is unbounded when $y\rightarrow 0$ or $y\rightarrow 1$, i.e., for $y$ tending to one of the boundaries of the support set. Hence, the influence function of the MLE is also unbounded. On the other hand, for bounded beta densities the weighted modified score vector $U^{*}(y,{\boldsymbol \theta})f^{*}_{\boldsymbol \theta}(y;\mu,\phi)^{1-q}$ and the influence function of the SMLE are bounded, which implies that the SMLE is  B-robust \citep[Section 2.1]{hampel2011robust}. Figure \ref{Score_functions} illustrates the behavior of the score vector and of the weighted modified score vector for the i.i.d. case. Note that the components of the score vector grow or decrease unboundedly when $y$ approaches zero or one, while the components of the weighted modified score vector has a redescending shape, approaching zero at the boundaries. This means that observations close to zero or one may be influential in the MLE fit but not in the SMLE fit.

We also studied the behavior of the change-of-variance function, that measures the bias on the covariance matrix caused by an infinitesimal contamination in the data point $y$. For the MLE, the change-of-variance function is unbounded. However, as presented in \citet[Proposition 2]{Ferrari2012}, the change-of-variance function of SMLE is bounded if the weighted modified score vector and its first derivative are bounded. We prove the boundedness of theses quantities for bounded beta densities, which implies that the SMLE is V-robust \citep[Section 2.5]{hampel2011robust}; see the Supplementary Material for details. 

\cite{ghosh2019robust} proposed a robust estimator in beta regression models that minimizes an empirical version of the density power divergence introduced by \cite{basu1998}. The minimum density power divergence estimator (MDPDE) for the beta regression model (\ref{fdpbeta})-(\ref{ligphi}) comes from the estimating equations
\begin{eqnarray}
 \sum_{i=1}^{n}\left\{U(y_i,{\boldsymbol \theta})f_{\boldsymbol \theta}(y_i;\mu_i,\phi_i)^{1-q} - E_{i,q}(\boldsymbol \theta)\right\}= \mat{0},
\label{estG}
\end{eqnarray}
where $E_{i,q}(\boldsymbol \theta)= {\rm E}(U(y_i,{\boldsymbol \theta})f_{\boldsymbol \theta}(y_i;\mu_i,\phi_i)^{1-q})$. The components of $E_{i,q}(\boldsymbol \theta)$ may be found in \cite{ghosh2019robust}, and the asymptotic covariance matrix of the MDPDE is given in the Supplementary Material. The expectation $E_{i,q}(\boldsymbol \theta)$ requires that $\mu_i\phi_i>(1-q)/(2-q)$ and $(1-\mu_i)\phi_i>(1-q)/(2-q)$. Besides, the asymptotic covariance matrix of the MDPDE requires that $\mu_i\phi_i> 2(1-q)/(3-2q)$ and $(1-\mu_i)\phi_i> 2(1-q)/(3-2q)$. Hence, these quantities are well-defined, for all $0<q\le 1$, only for bounded beta densities.

Note that the estimating functions in (\ref{estG}) are unbiased. They agree with those derived from the $L_q$-likelihood in (\ref{esteq}) except for the term $E_{i,q}(\boldsymbol \theta)$ that plays the role of correcting the score bias. Recall that the score bias correction in the $L_q$-likelihood approach adopted here is achieved by rescaling the parameter estimates or, equivalently by reparameterizing the $L_q$-likelihood.

\cite{ghosh2019robust} presented simulations of the MDPDE only for constant precision beta regression models with fixed values for the tuning constant. However, in practical applications the choice of such a constant becomes a crucial issue. In Section \ref{qchoice}, we will propose a data-driven method to select the tuning constant of the SMLE and the MDPDE which ensures full efficiency in the absence of contamination.

\section{Selection of the tuning constant $q$}\label{qchoice}

An important aspect in real data applications is the choice of the optimal value of the tuning constant $q$. Smaller values of $q$ increase robustness at the expense of reduced efficiency. \cite{la2015robust} suggest to select the tuning constant closest to 1 (i.e., closest to MLE) such that the estimates of the parameters are sufficiently stable, ensuring full efficiency in the absence of contamination. This is the approach we follow here, but we standardize the estimates by `$\sqrt{n}\times\mbox{std. error}$' to simultaneously remove the effect of the sample size and of the magnitude of estimates of different parameters. We then propose a data-driven algorithm that starts with $q_0=1$ (MLE) and follows the steps below.

\begin{enumerate}

\item Define an equally spaced grid for $q$: $q_0>q_1>q_2>\cdots>q_m,$ with $q_m\ge q_{\rm min}$;

\item compute the estimates $\widehat{\theta}^{1}_{q_k}, \ldots, \widehat{\theta}^{p}_{q_k}$ with the corresponding asymptotic standard errors, $\mbox{se}(\widehat{\theta}^{1}_{q_k}), \ldots,$ $\mbox{se}(\widehat{\theta}^{p}_{q_k})$; 

\item compute the vectors of standardized estimates,
$$\mat{z}_{q_k}=\left(\frac{\widehat{\theta}^{1}_{q_k}}{\sqrt{n}\ \mbox{se}(\widehat{\theta}^{1}_{q_k})},\cdots, \frac{\widehat{\theta}^{p}_{q_k}}{\sqrt{n}\ \mbox{se}(\widehat{\theta}^{p}_{q_k})}\right)^\top,$$
and the standardized quadratic variations (SQV) defined by $$\mbox{SQV}_{q_k} = p^{-1}|| \mat{z}_{q_k} - \mat{z}_{q_{k+1}}||;$$

\item for each $q_k$, check whether the stability condition defined by $\mbox{SQV}_{q_k} < L$ is satisfied, where $L>0$ is a pre-defined threshold;

\item if the stability condition is satisfied for all $q_k$, set the optimal value of $q$ at $q^{*}= \max q_k$; 

\item  if the stability condition is not satisfied for some $q_k$, set a new grid starting from the smallest $q_k$ for which the condition does not hold;

\item repeat steps 1-6 until achieving stability of the estimates for the current grid or reaching the minimum value $q_{\rm min}$; if stability is achieved, set the optimal value of $q$ at $q^{*}= \max q_k$, otherwise at $q^{*}= 1$.

\end{enumerate}

This procedure chooses the optimum value for the tuning constant that is closest to 1 (MLE) and still guarantees stability of the estimates for $m$ consecutive equally spaced values of $q$. It may occur that such an optimal $q$ is not reached. In this case, the algorithm sets the optimal $q$ at 1 i.e. the MLE is chosen. This choice may seem unreasonable but recall that the SMLE is not expected to be robust for unbounded beta densities, which might lead to unstable estimates even for reasonably small $q$. In such a situation, it is not advisable to replace the MLE by the SMLE. Our experience with simulated samples suggests to set the grid spacing at 0.02, the size of the grids at $m=3$, the minimum value of $q$ at $q_{\rm min} = 0.5$ and the threshold at $L= 0.02$. These values were used in the simulations and applications reported below and worked very well.

\section{Monte Carlo simulation results}\label{MC}

To evaluate the performance and compare the  MLE, SMLE and MDPDE, Monte Carlo simulations in the presence and absence of contamination in the data were carried out. The sample sizes considered are $n=40, 80, 160, 320$. The covariate values were set for the sample size $n=40$ and replicated twice, four times and eight times to obtain the covariate values corresponding to the sizes $n= 80,$ 160, 320, respectively. This scheme guarantees that the heteroskedasticity intensity, $\mbox{max}(\phi_i)/\mbox{min}(\phi_i)$, be constant for all the sample sizes \citep{espinheira2017nonlinear, espinheira2019model}. For all the scenarios we employ logit and logarithmic links in the mean and precision submodels,  respectively (\ref{ligmu}) and (\ref{ligphi}). The models include intercepts, i.e. $x_{1i} = z_{1i}=1$, for all $i=1,2,\ldots,n$, and the other covariates for the mean submodel are taken as random draws from a standard uniform distribution and kept constant throughout the simulated samples. For non-constant precision scenarios, the same covariates are used in the mean and precision submodels. The percentage of contamination in the sample for all the scenarios was set at 5\%. The selection of the tuning constant $q$ for SMLE and MDPDE was conducted following the proposed algorithm described in Section \ref{qchoice}. All the results are based on 1,000 Monte Carlo replications and were carried out using the {\tt R} software environment \citep{Rreference}.

Different configurations of parameter values and patterns of data contamination were considered. We now describe the results for two scenarios; results for other scenarios are collected in the Supplementary Material.

\vspace{-0.2cm}
\paragraph{\bf Scenario 1: Constant precision, one covariate in the mean submodel, mean response values close to zero.}  The parameter values were set at $\beta_1=-1.8,$ $\beta_2=-2.0$, and   $\gamma_1 = 4.5$,  which yield $\mu \in (0.02,  0.14)$ with $\mbox{median}(\mu)= 0.06$ and $\phi = \exp(4.5)=90$. The contaminated sample replaces the observations generated with the 5\% smallest  mean responses by observations generated from a beta regression model with mean $\mu_i'=(1+\mu_i)/2$ (and precision $\phi$). For instance, if $\mu_i\approx 0.1$, then $\mu_i' \approx 0.55$.

\vspace{-0.2cm}
\paragraph{\bf Scenario 2: Varying precision, two covariates  in the mean and precision submodels, mean response values around 0.4.} The parameter values were set at
$\beta_1=0.8,$ $\beta_2=-1.2$, $\beta_3=-1.2$, $\gamma_1=3.8,$ $\gamma_2=0.7$, and $\gamma_3=0.7$,  which result in $\mu \in (0.22, 0.64)$ with $\mbox{median}(\mu)= 0.43$ and $\phi \in (51, 148)$ with $\mbox{median}(\phi)= 85.$ The contaminated sample replaces 5\% of the sample as follows: the observations generated with the 2.5\% largest values of $\mu$ and those generated with the 2.5\% smallest values of $\mu$ are replaced by independent draws of a beta regression model with mean $\mu^{(1)}_i= a_1c_i/(1+a_1c_i)$ and $\mu^{(2)}_i= a_2c_i/(1+a_2c_i)$, respectively, where $c_i = \mu_i/(1-\mu_i)$, $a_1 = 0.01$ and $a_2=6.0$. For instance, if $\mu\approx 0.6$, then $\mu^{(1)} \approx 0.01$, and if $\mu\approx 0.2$, then  $\mu^{(2)} \approx 0.6$.

\begin{figure}[htbp]
	\centering
	\subfigure{\includegraphics[width=5.0cm,height=5.0cm]{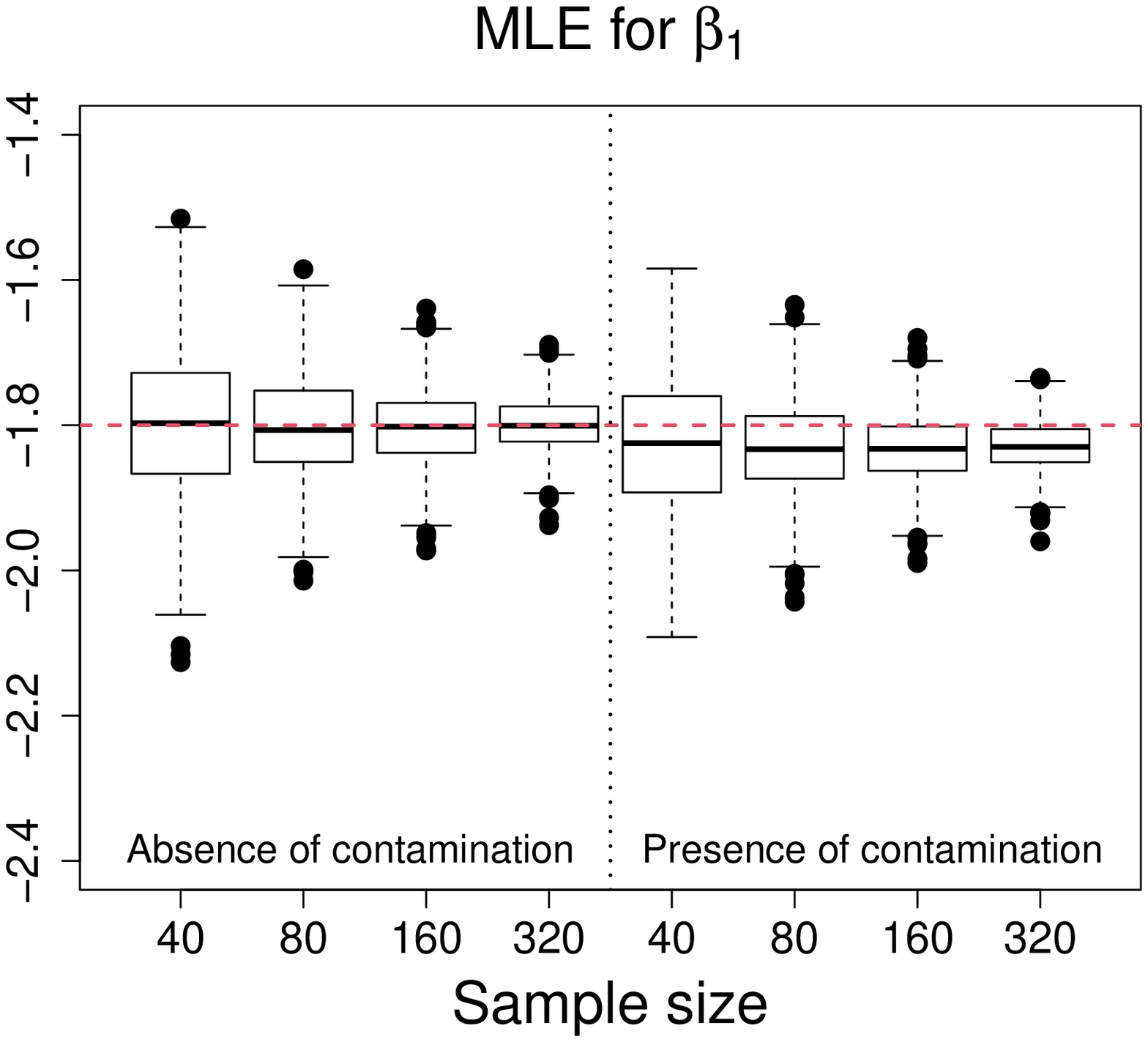}}
	\qquad
	\subfigure{\includegraphics[width=5.0cm,height=5.0cm]{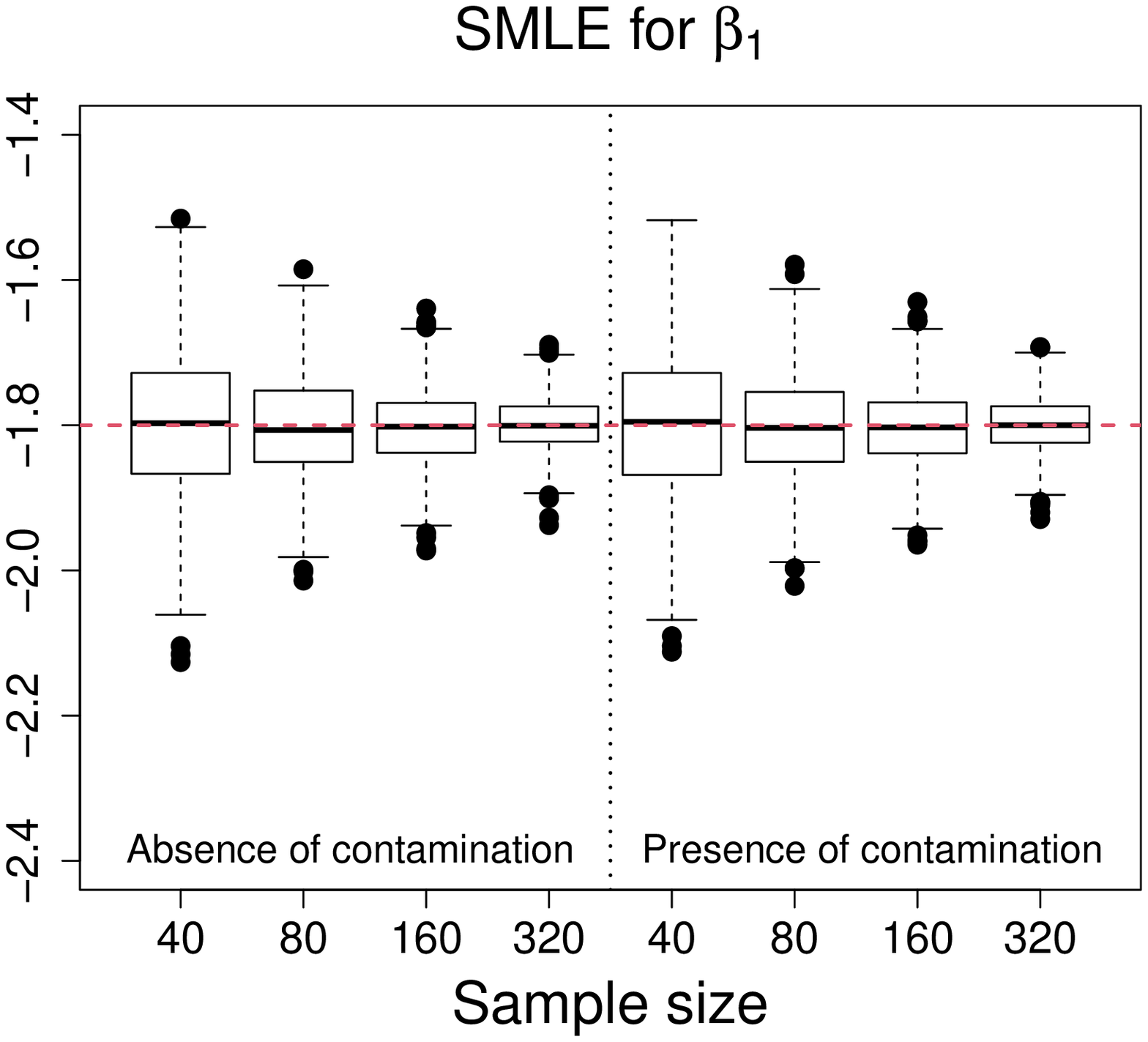}}
	\qquad
	\subfigure{\includegraphics[width=5.0cm,height=5.0cm]{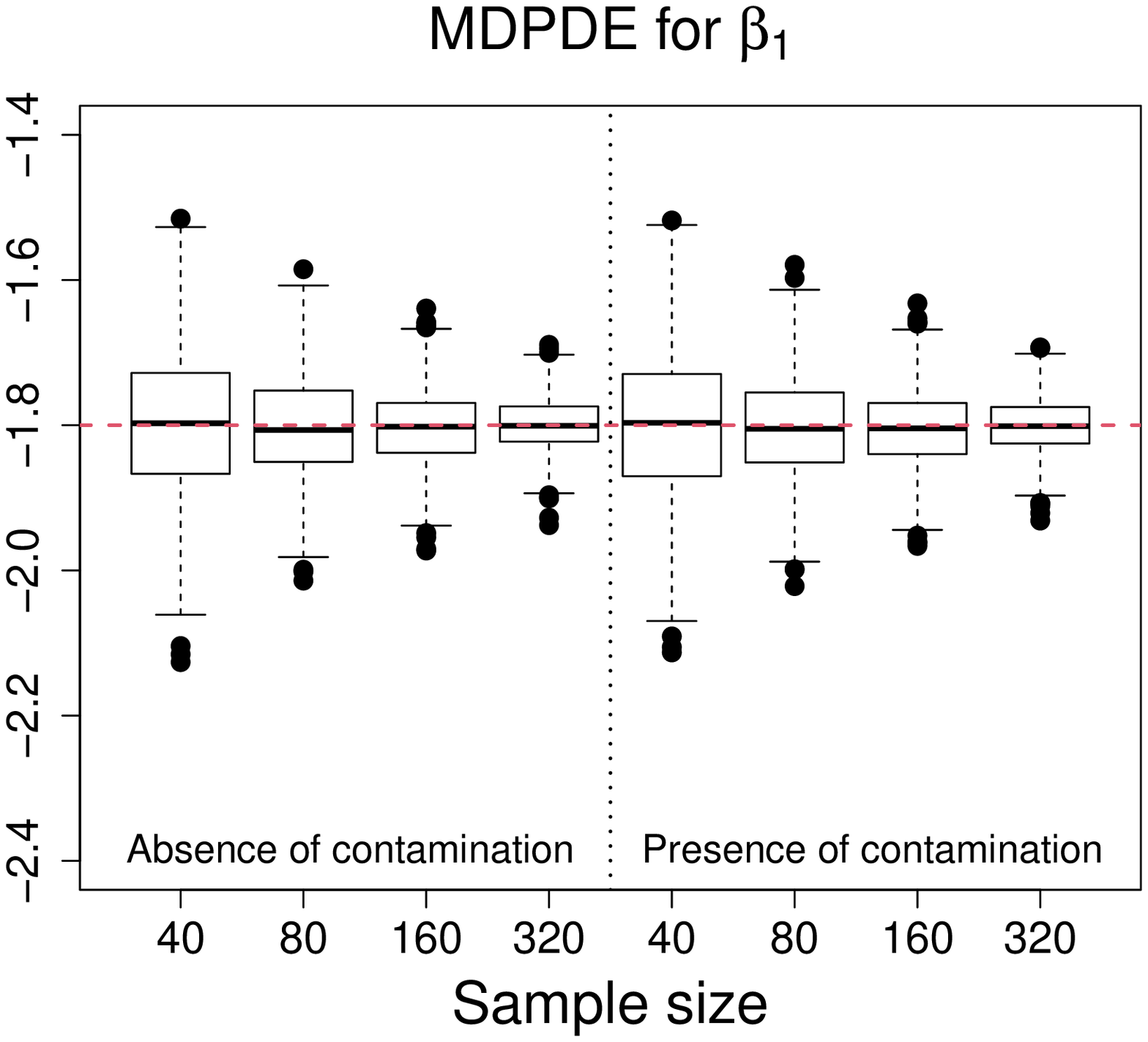}}
	\qquad
	\subfigure{\includegraphics[width=5.0cm,height=5.0cm]{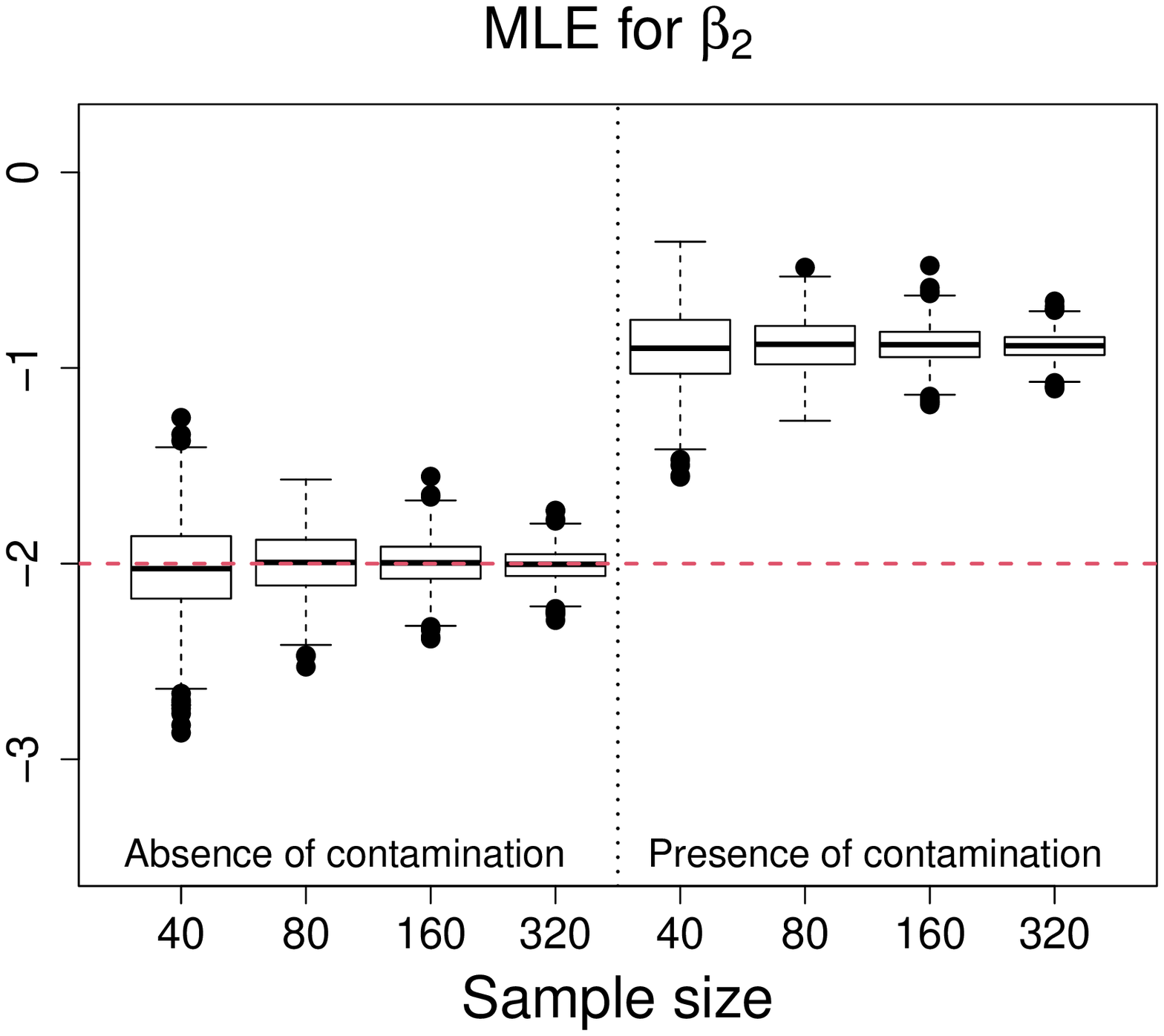}}
	\qquad
	\subfigure{\includegraphics[width=5.0cm,height=5.0cm]{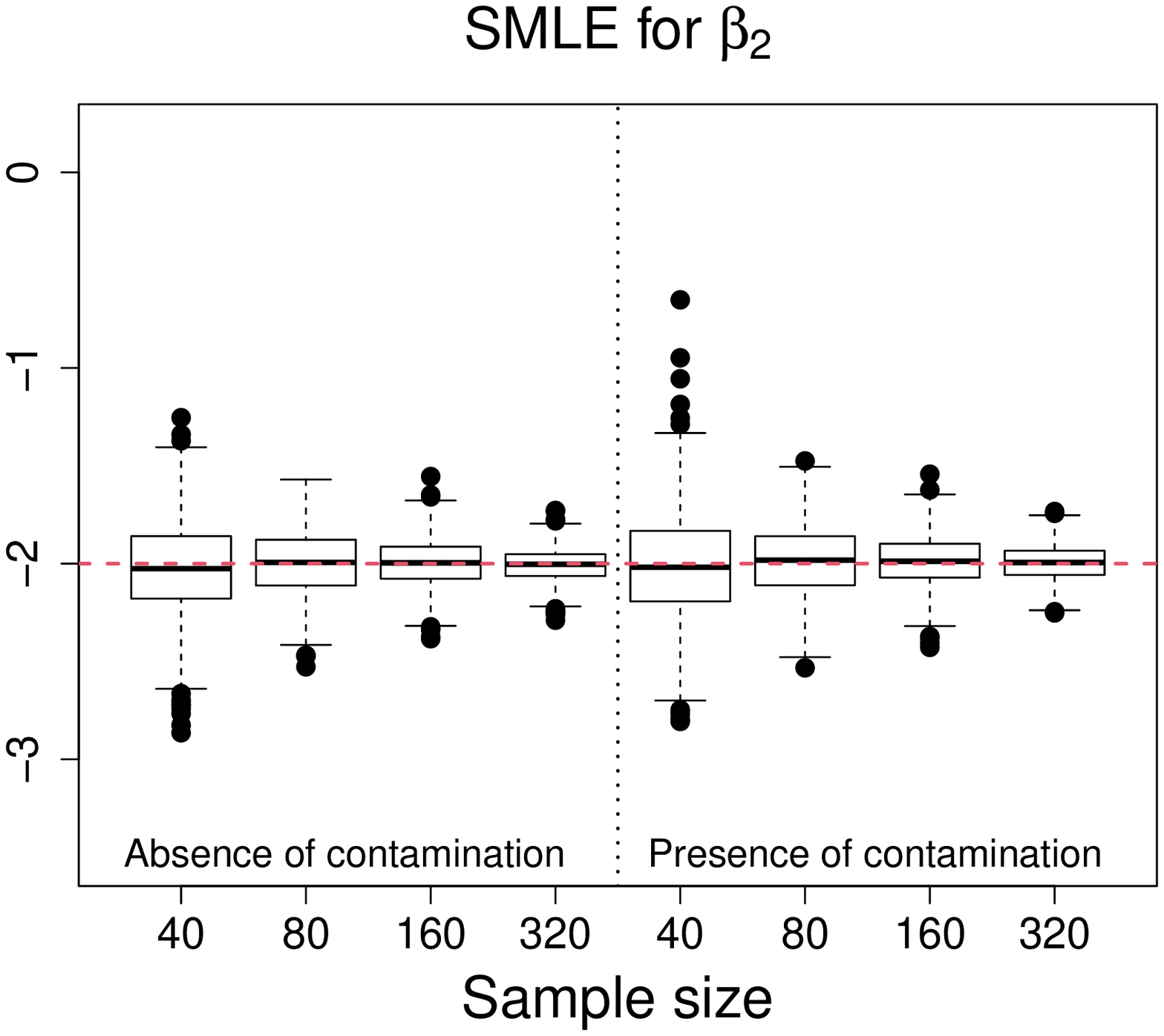}}
	\qquad
	\subfigure{\includegraphics[width=5.0cm,height=5.0cm]{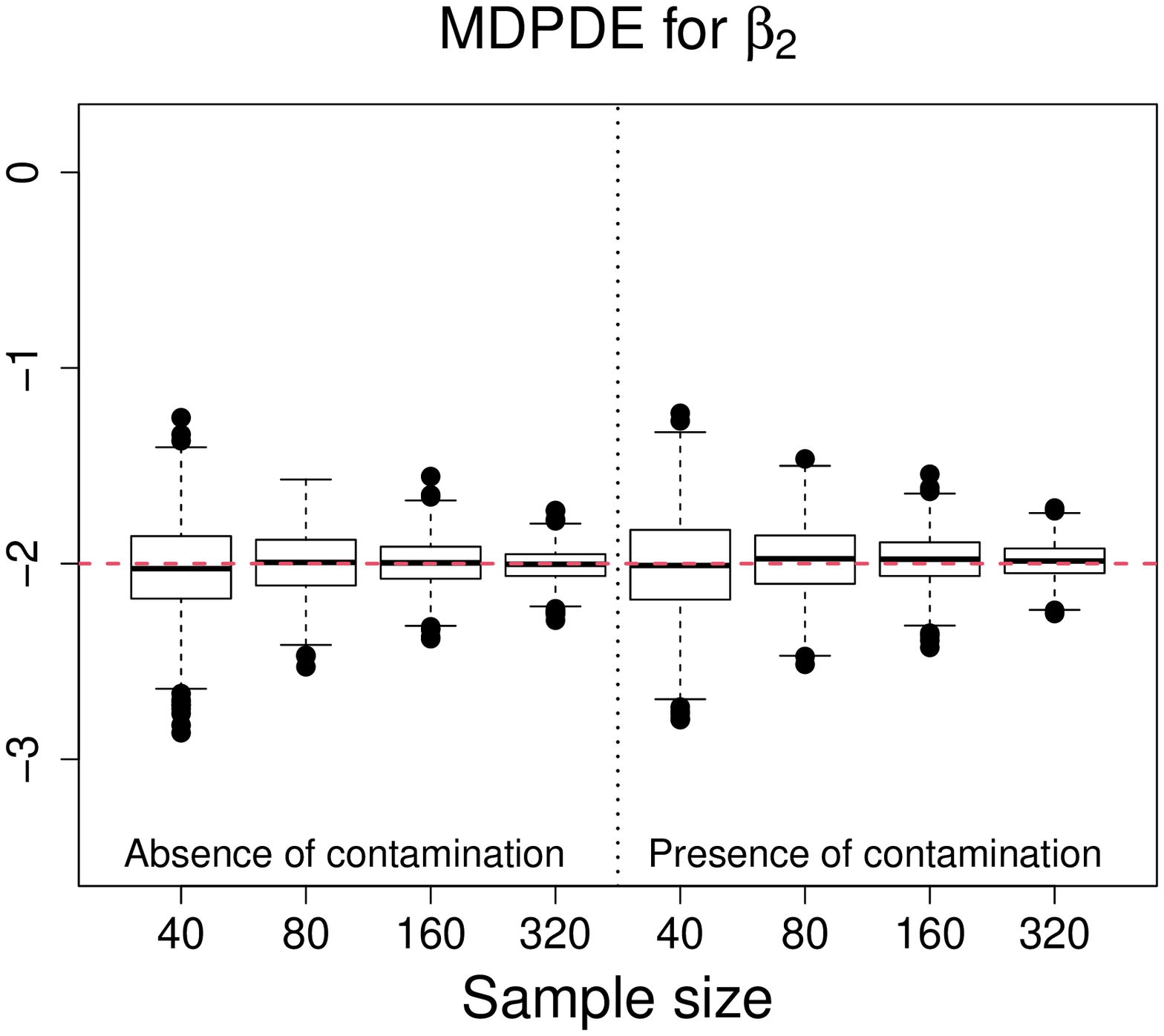}}
	\qquad
	\subfigure{\includegraphics[width=5.0cm,height=5.0cm]{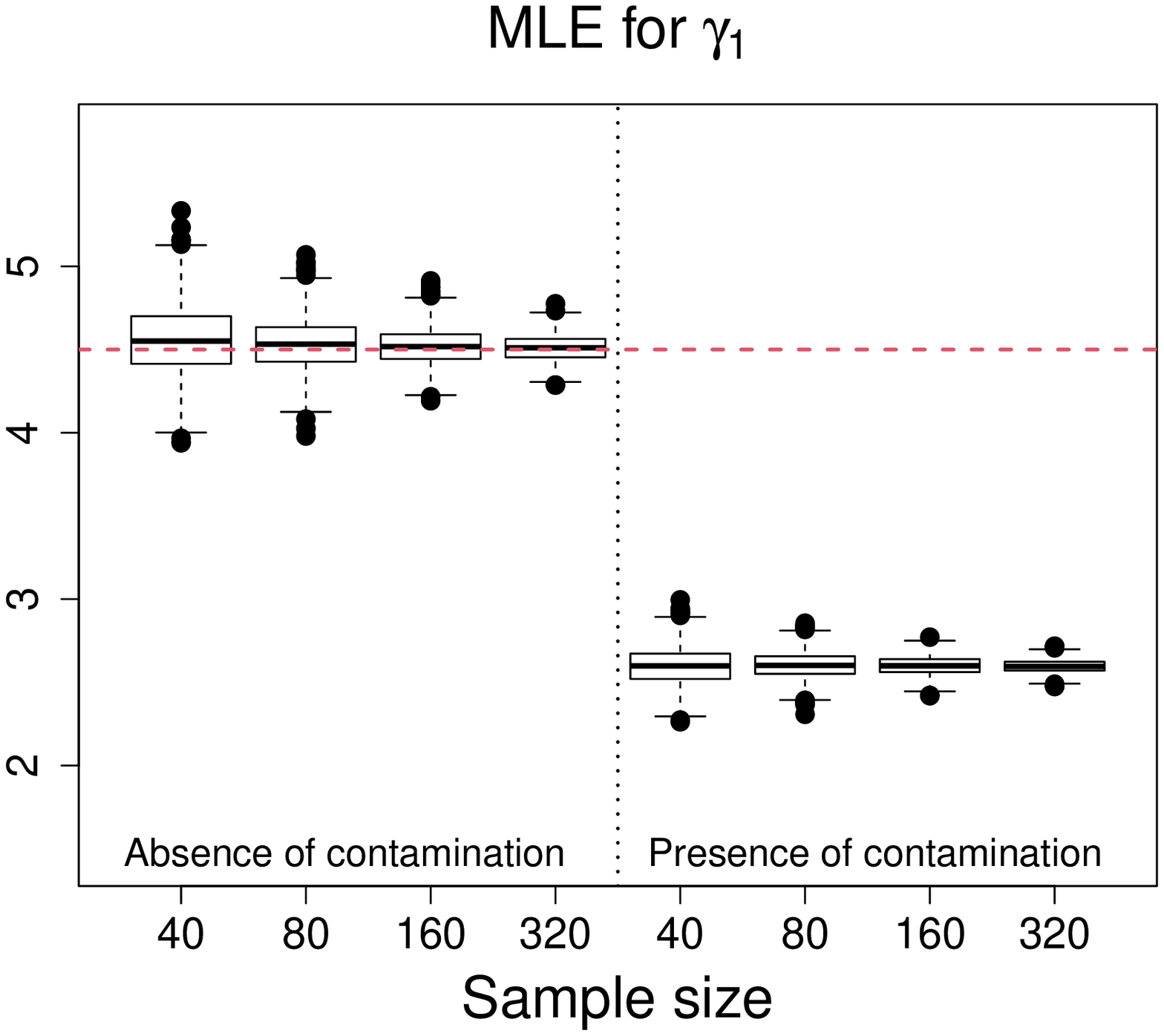}}
	\qquad
	\subfigure{\includegraphics[width=5.0cm,height=5.0cm]{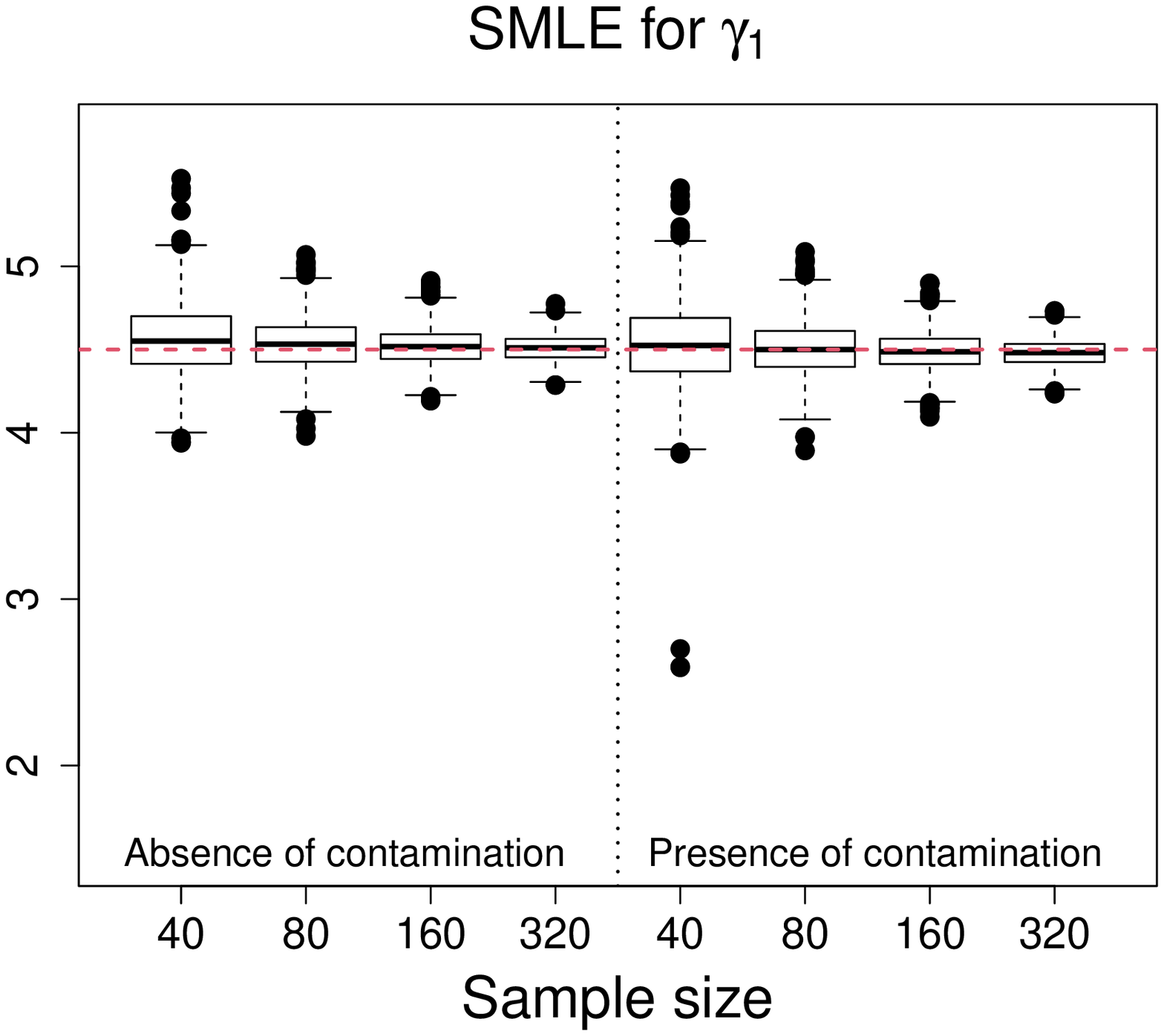}}
	\qquad
	\subfigure{\includegraphics[width=5.0cm,height=5.0cm]{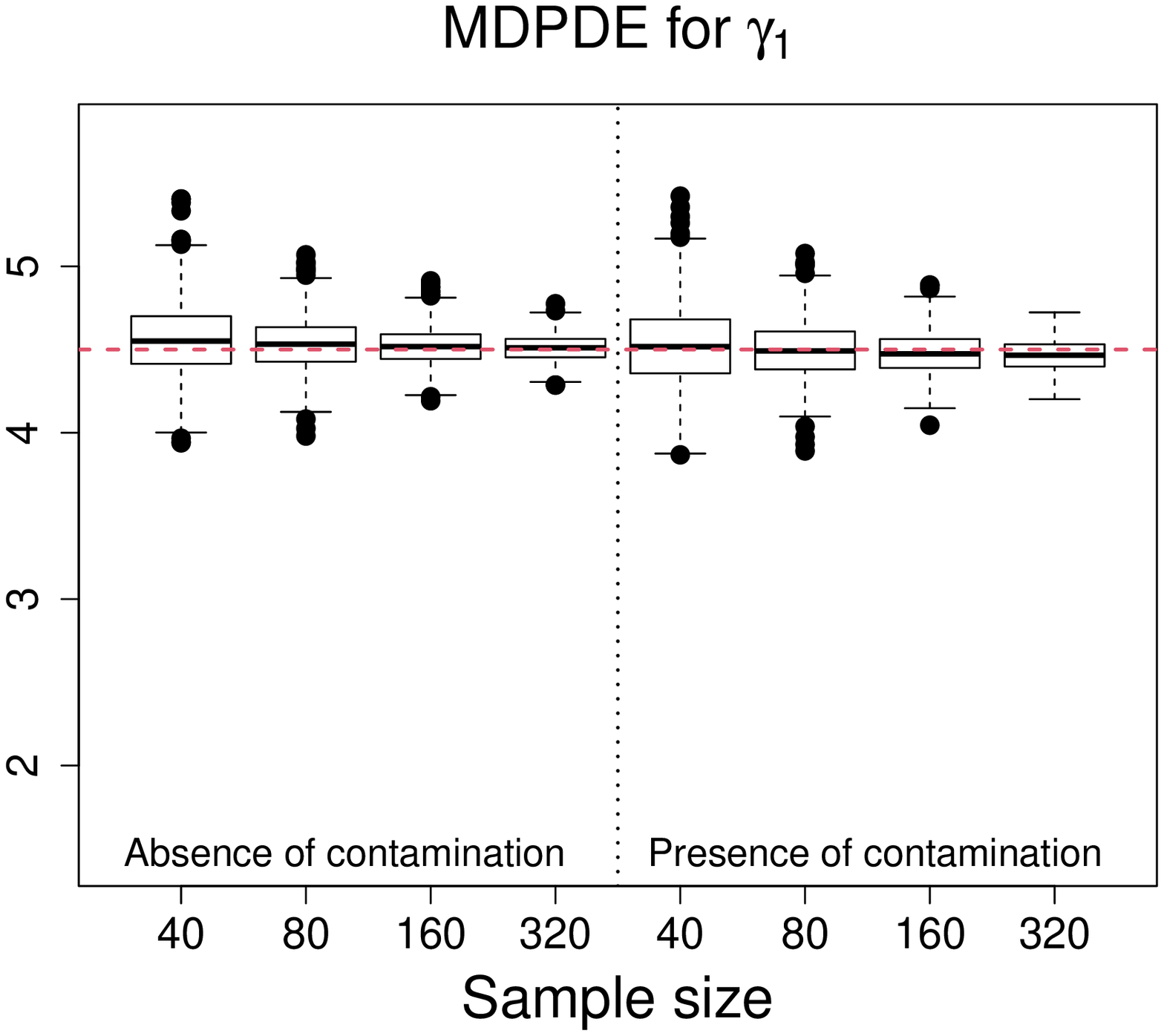}}
	\caption{Boxplots of parameter estimates under Scenario 1: MLE (left), SMLE (center), and MDPDE (right).
		The red dashed line represents the true parameter value.}
	\label{boxplotsa01}
\end{figure}

\begin{figure}[htbp]
	\centering
	\subfigure{\includegraphics[width=5.0cm,height=5.0cm]{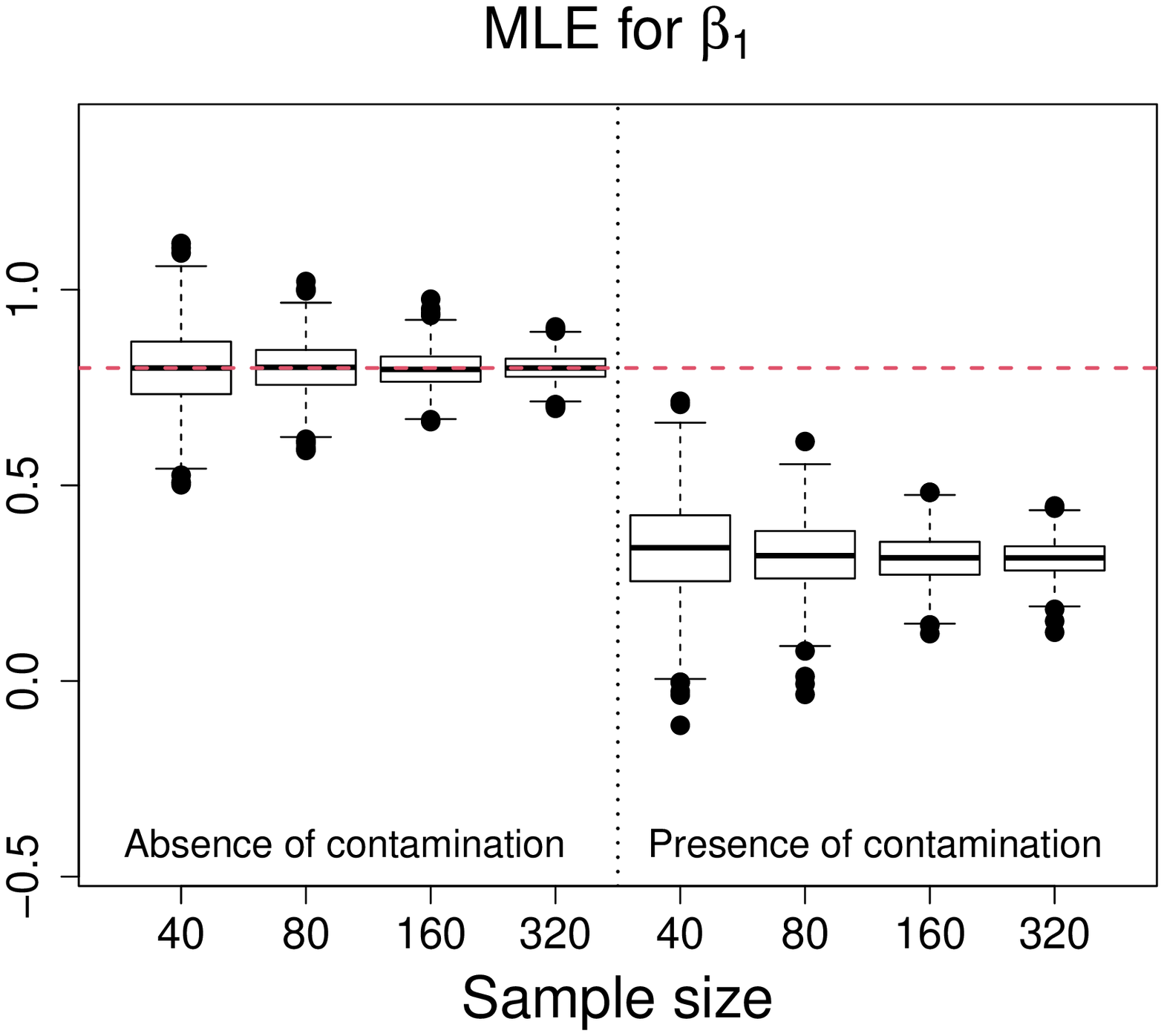}}
	\qquad
	\subfigure{\includegraphics[width=5.0cm,height=5.0cm]{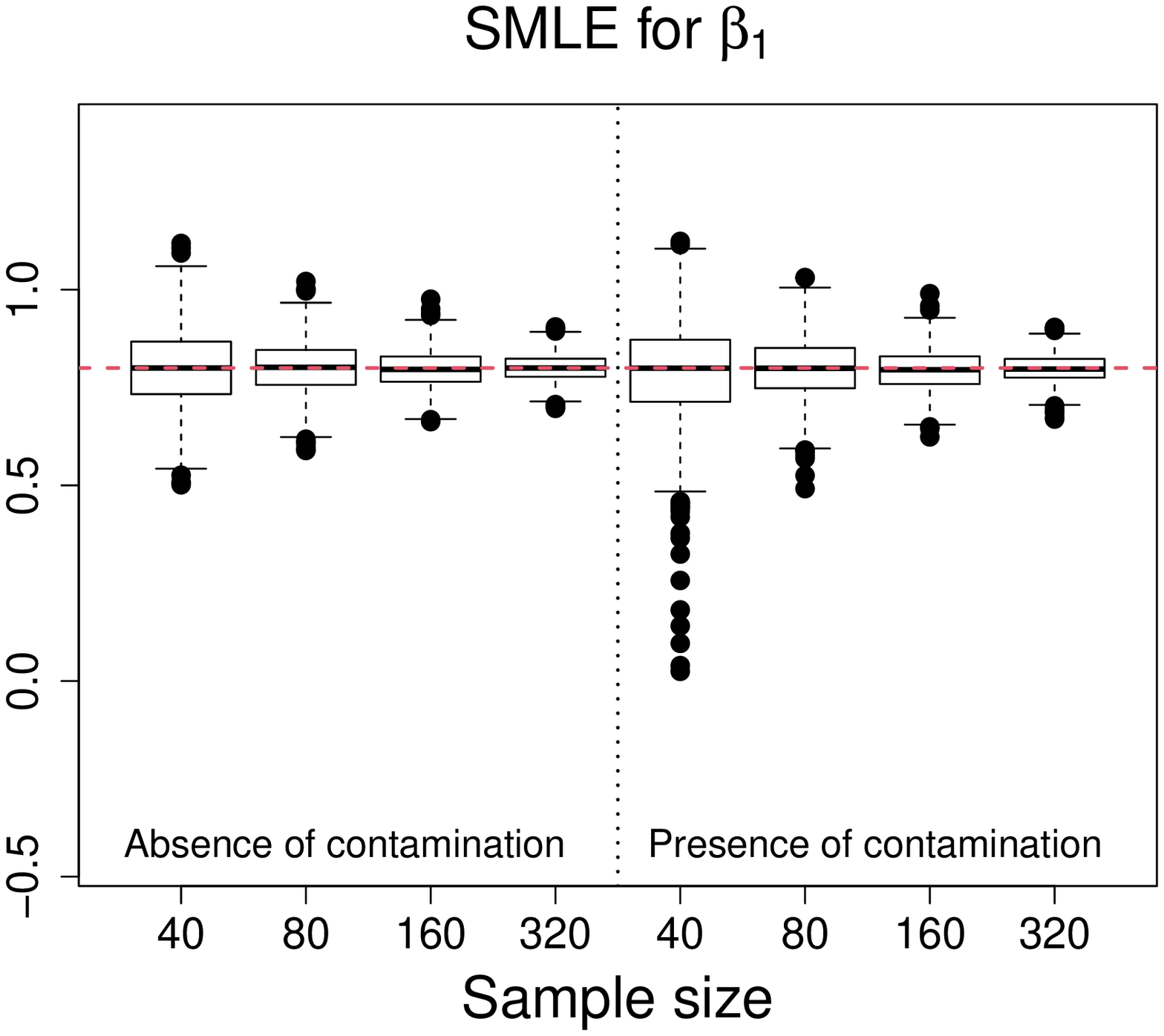}}
	\qquad
	\subfigure{\includegraphics[width=5.0cm,height=5.0cm]{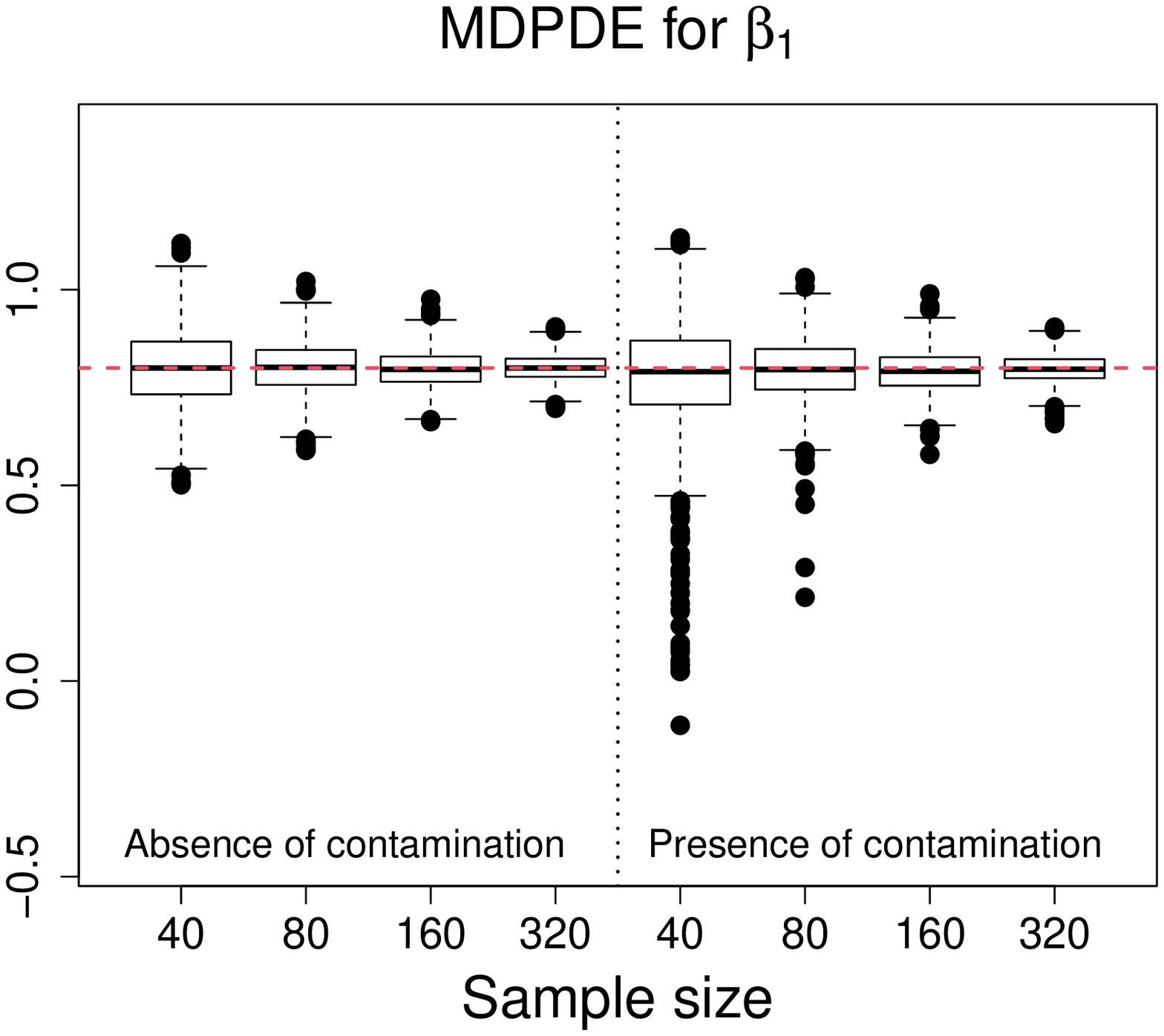}}
	\qquad
	\subfigure{\includegraphics[width=5.0cm,height=5.0cm]{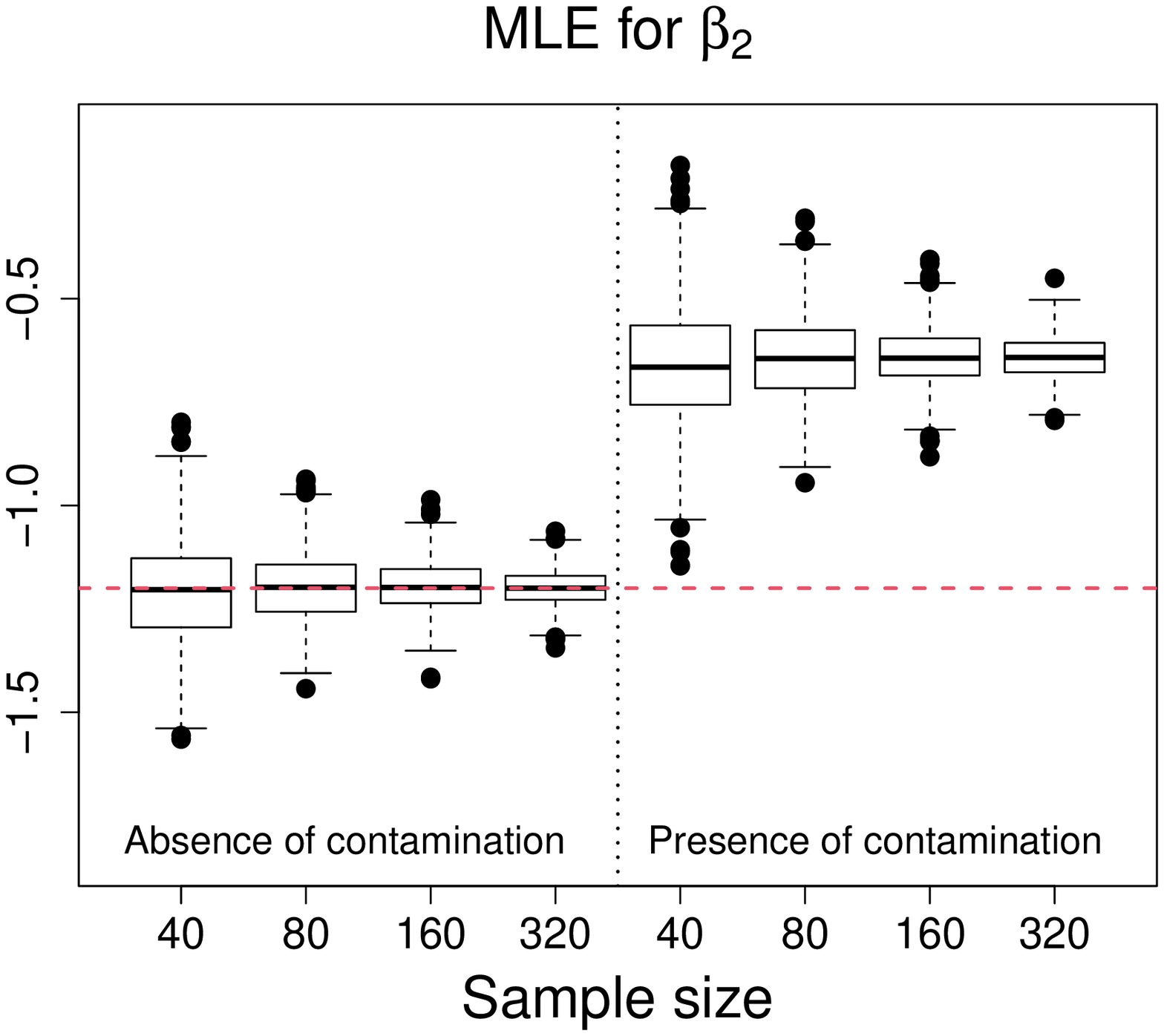}}
	\qquad
	\subfigure{\includegraphics[width=5.0cm,height=5.0cm]{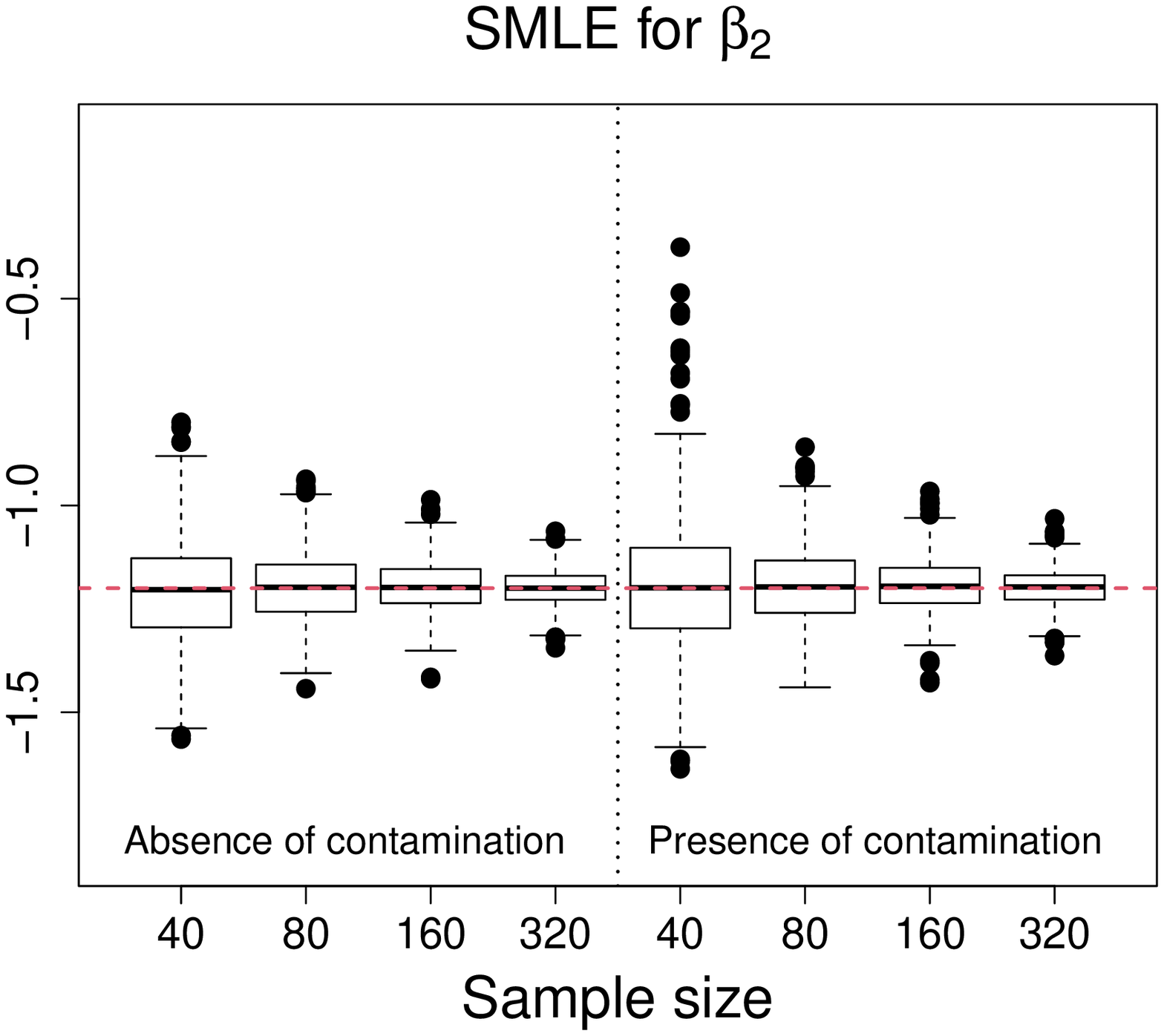}}
	\qquad
	\subfigure{\includegraphics[width=5.0cm,height=5.0cm]{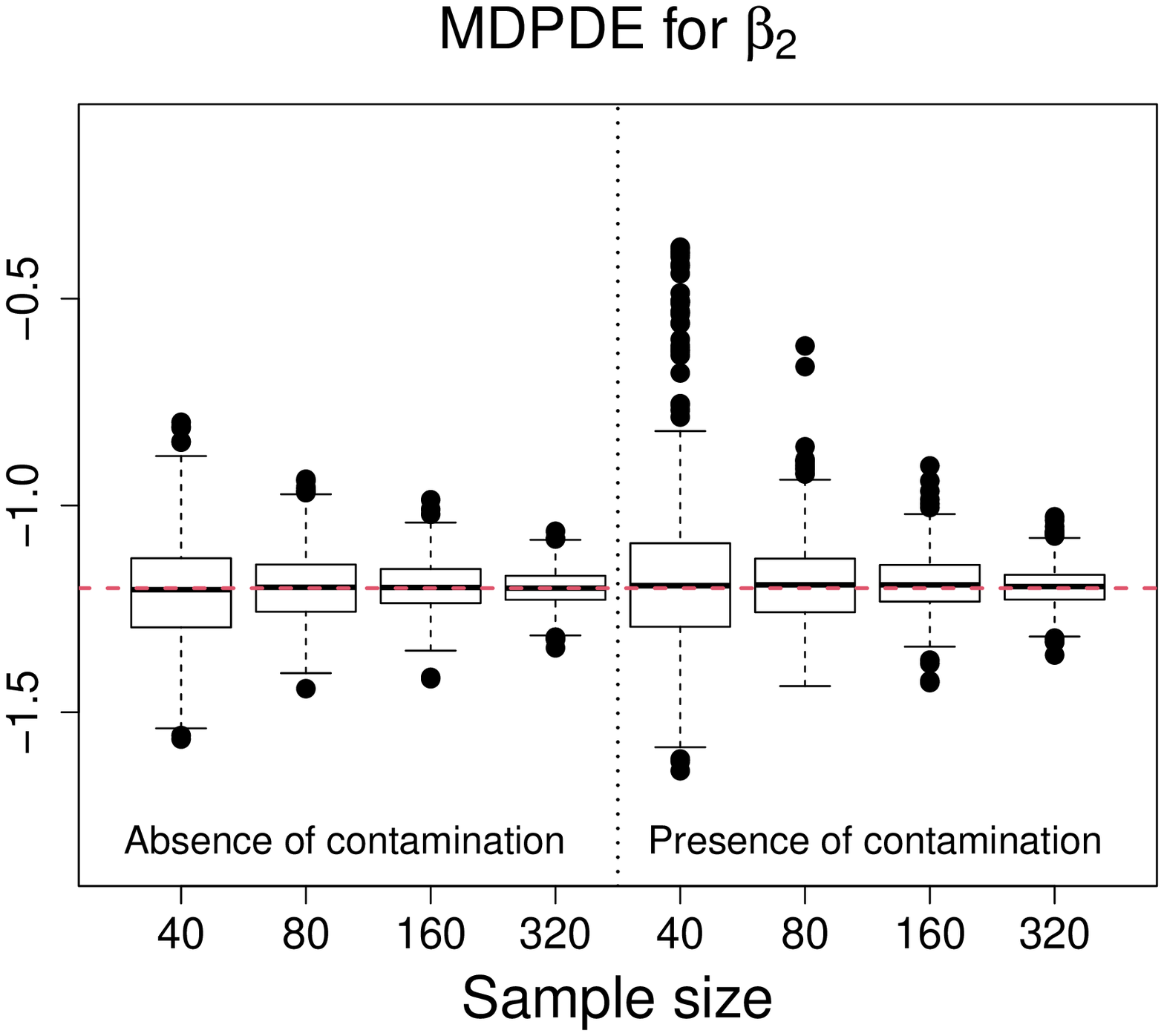}}
	\qquad
	\subfigure{\includegraphics[width=5.0cm,height=5.0cm]{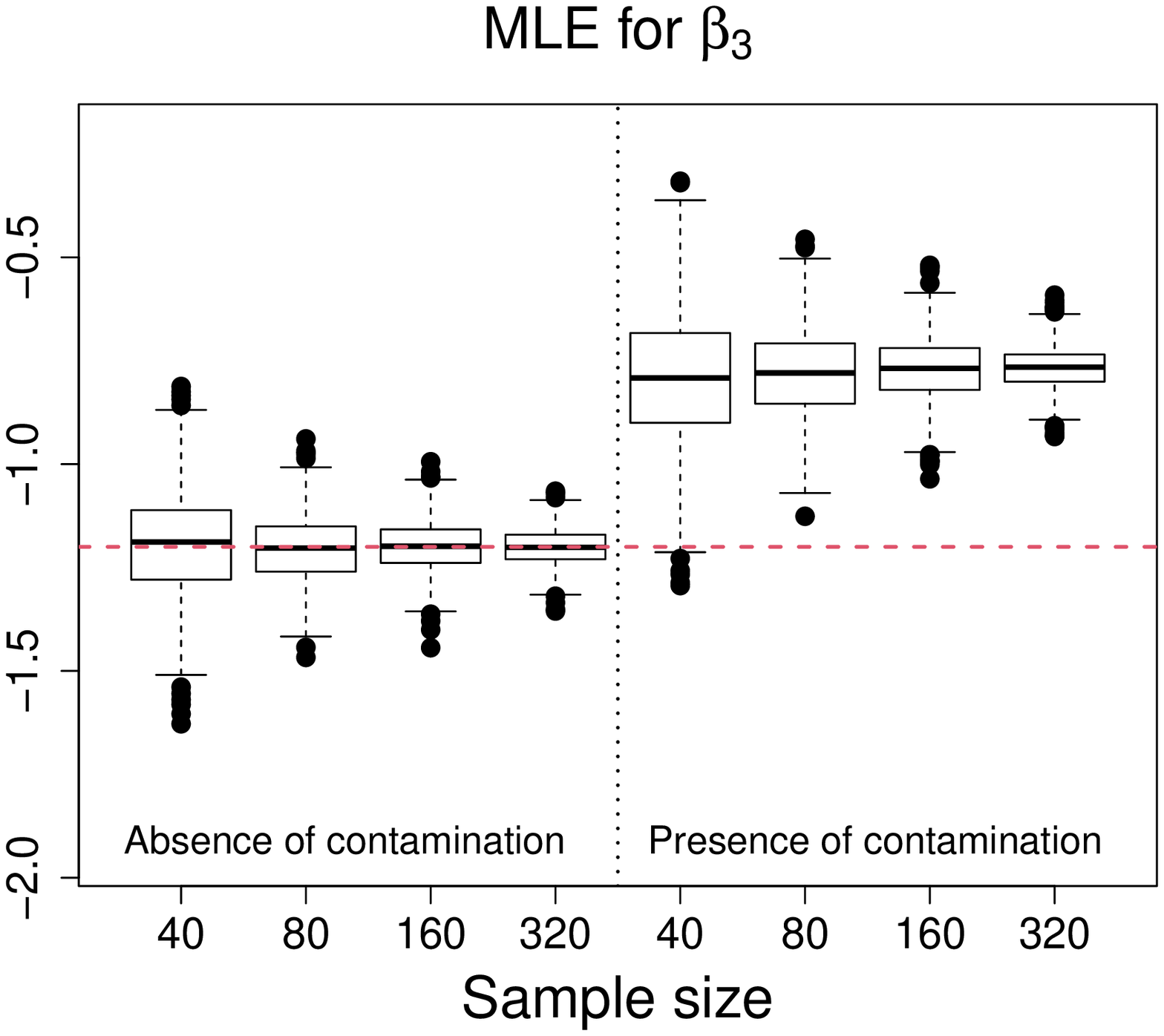}}
	\qquad
	\subfigure{\includegraphics[width=5.0cm,height=5.0cm]{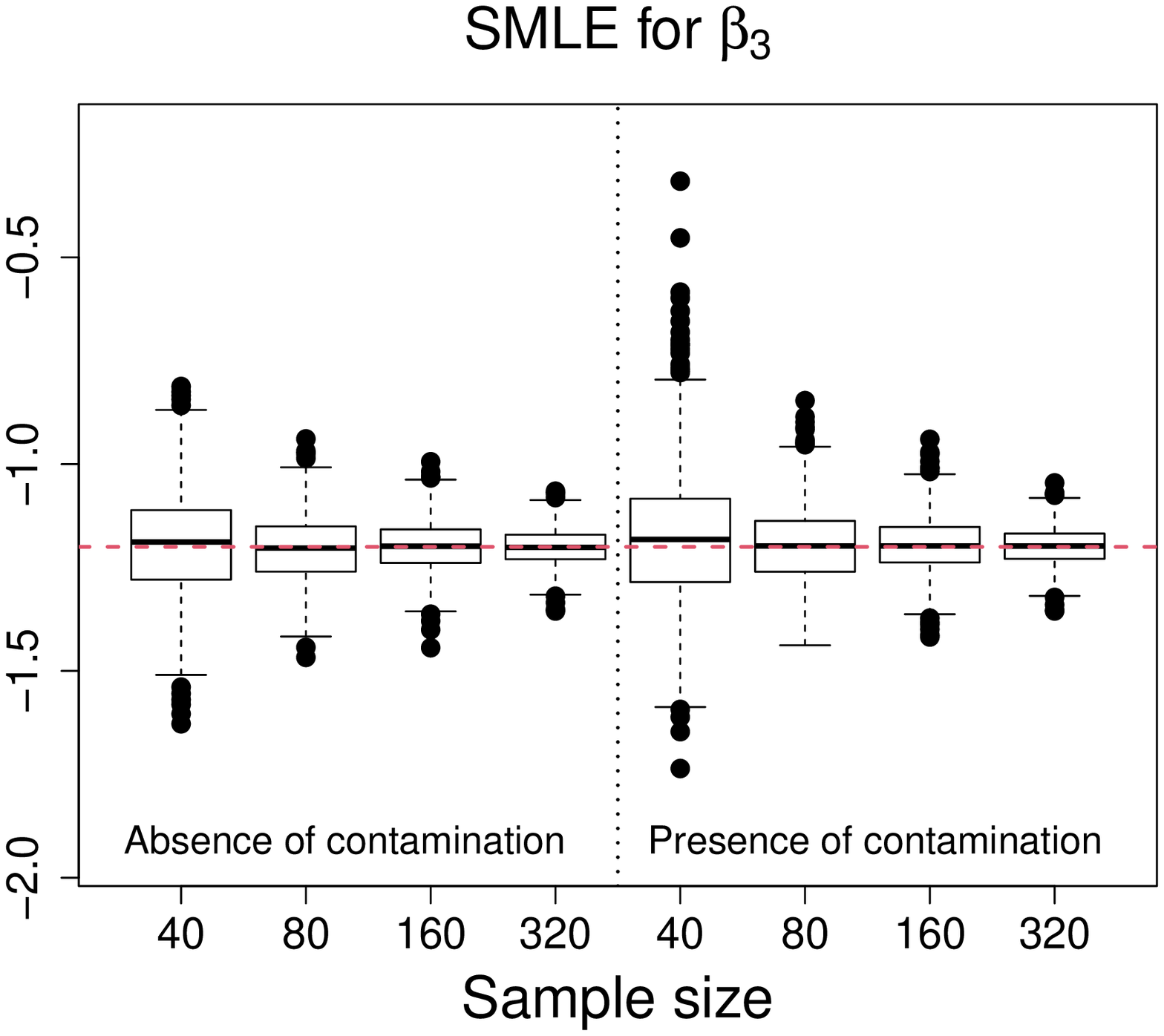}}
	\qquad
	\subfigure{\includegraphics[width=5.0cm,height=5.0cm]{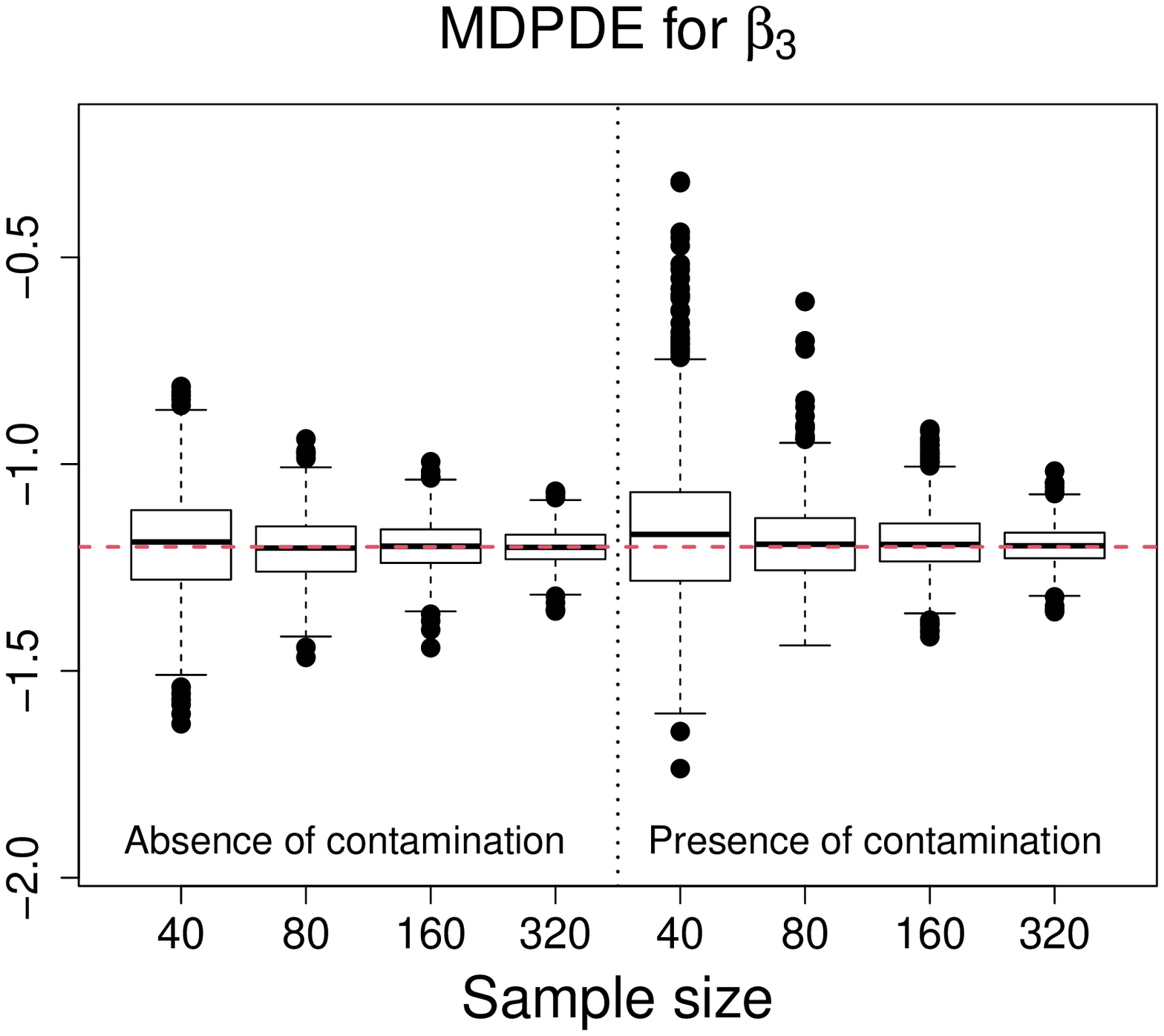}}
	\caption{Boxplots of estimates of  $\beta_1$, $\beta_2$, and $\beta_3$ under Scenario 2: MLE (left), SMLE (center), and MDPDE (right).
		The red dashed line represents the true parameter value.}
	\label{boxplotsa02_a}
\end{figure}

\begin{figure}[htbp]
	\centering
	\subfigure{\includegraphics[width=5.0cm,height=5.0cm]{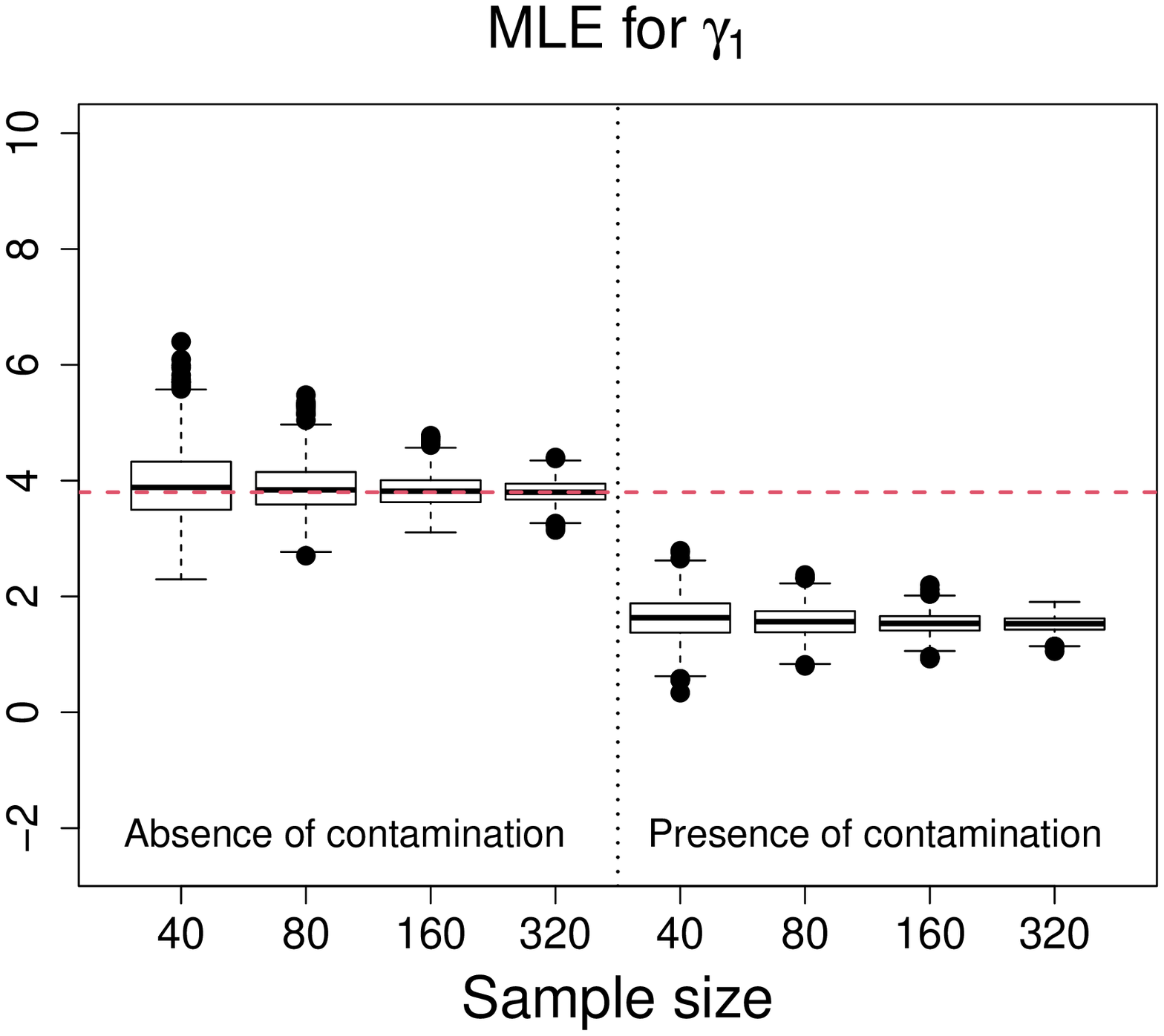}}
	\qquad
	\subfigure{\includegraphics[width=5.0cm,height=5.0cm]{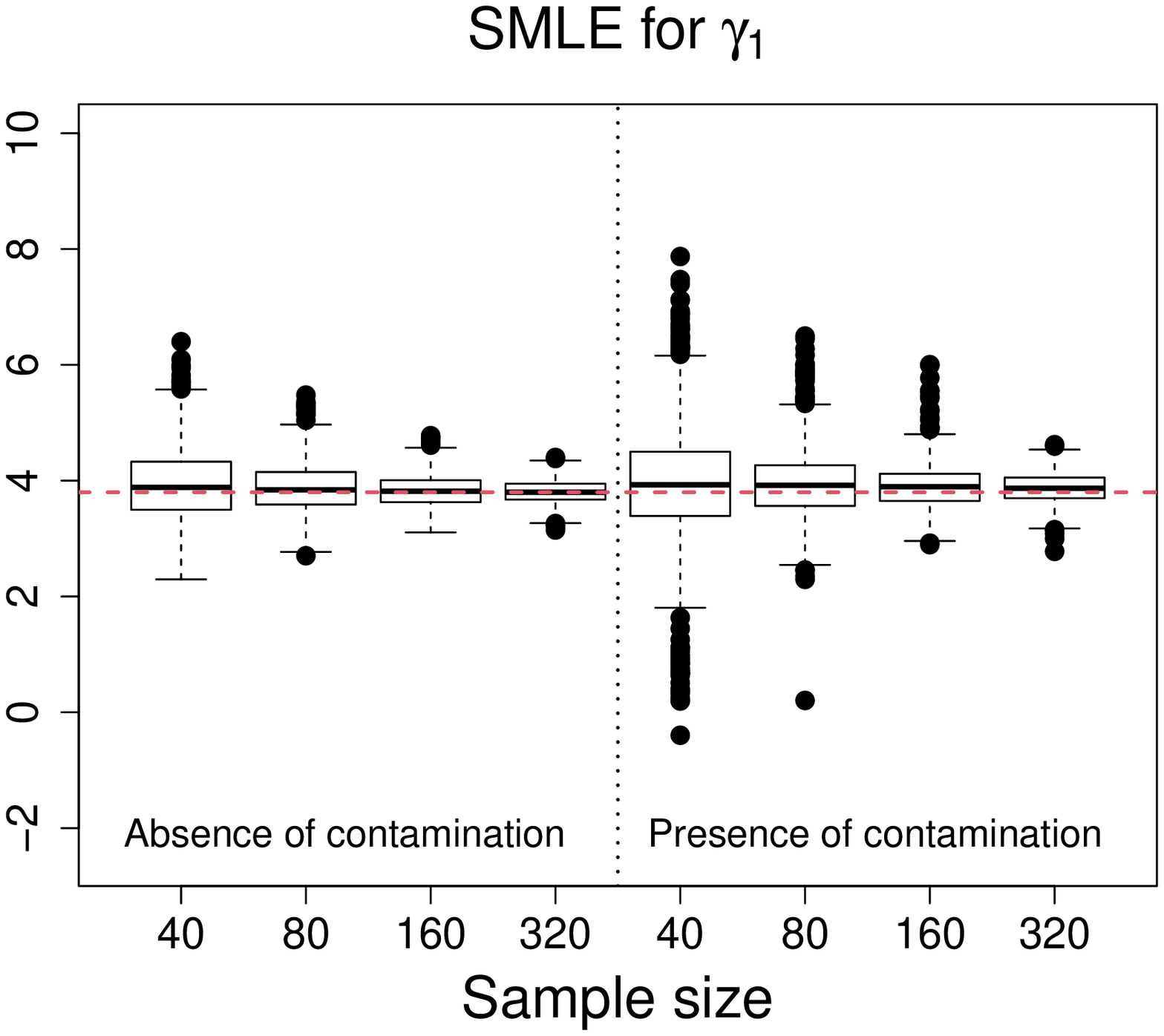}}
	\qquad
	\subfigure{\includegraphics[width=5.0cm,height=5.0cm]{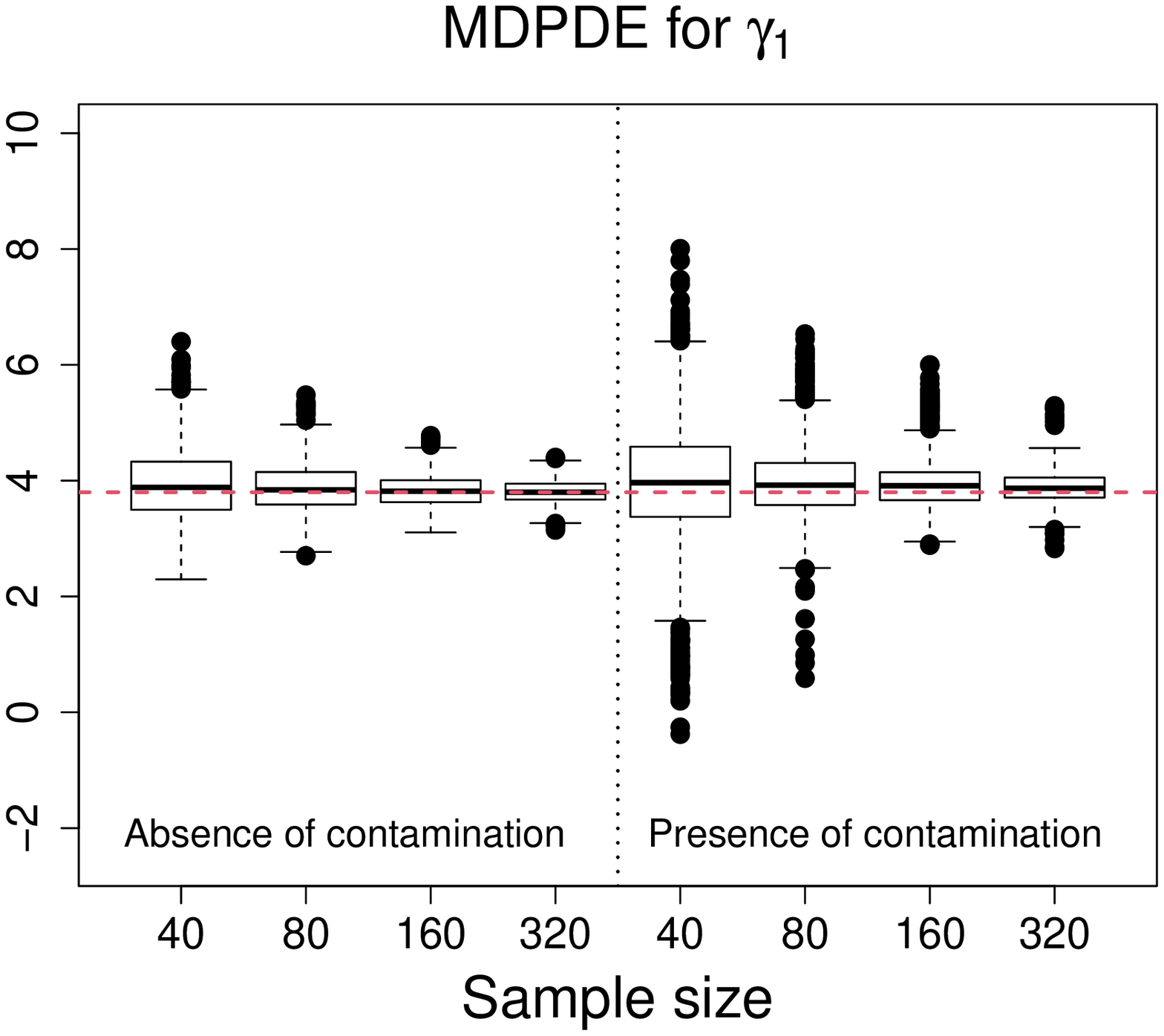}}
	\qquad
	\subfigure{\includegraphics[width=5.0cm,height=5.0cm]{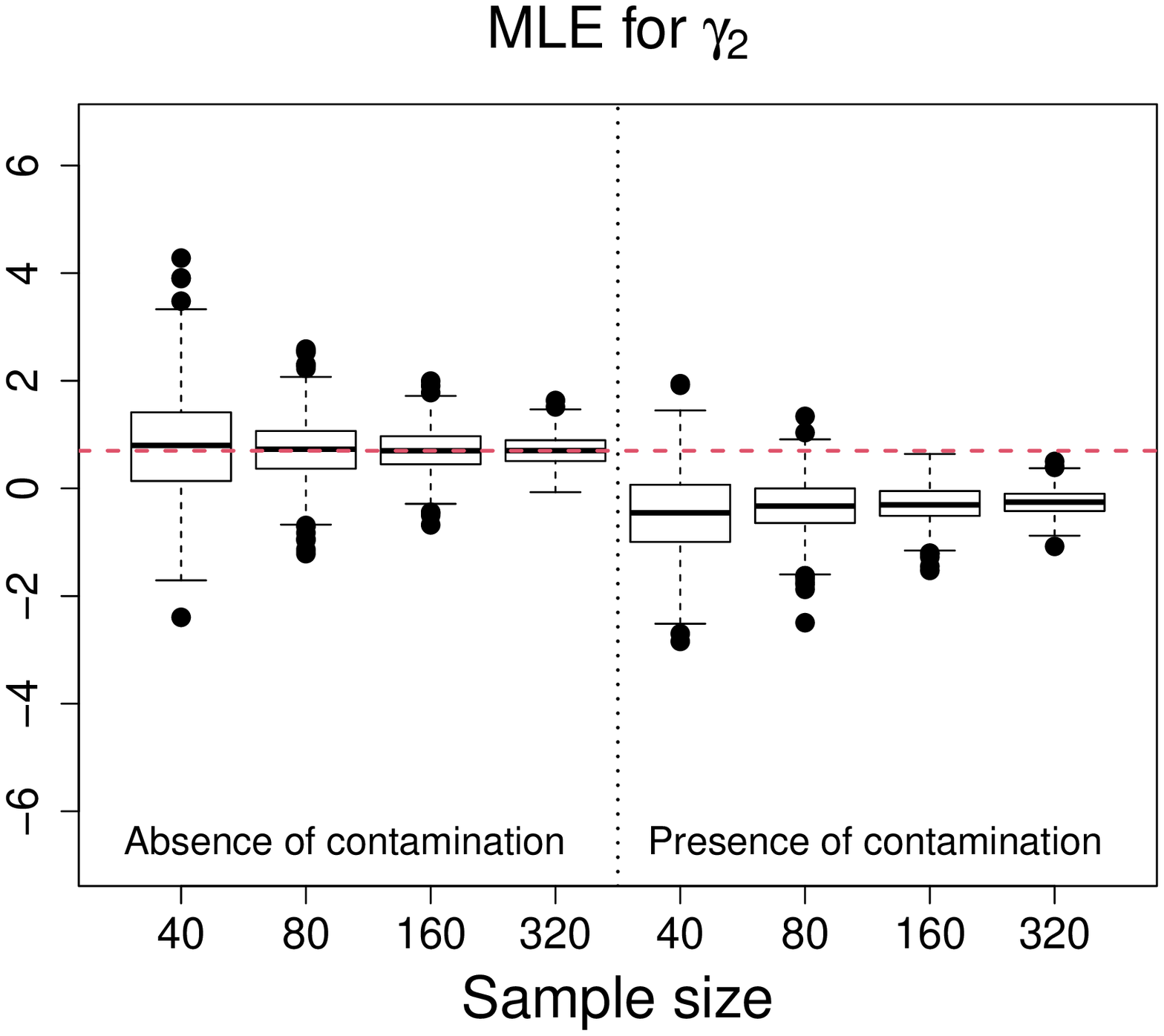}}
	\qquad
	\subfigure{\includegraphics[width=5.0cm,height=5.0cm]{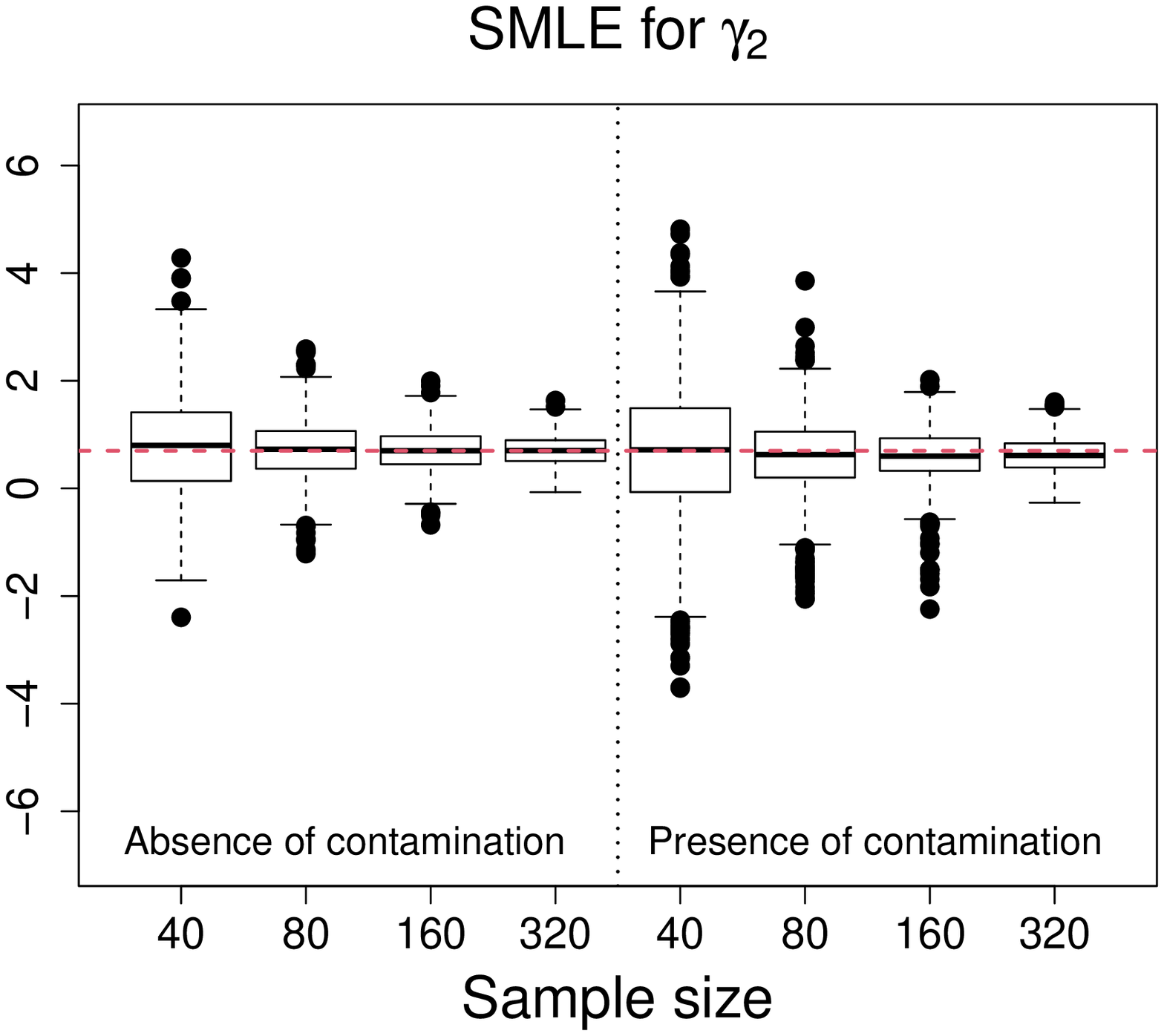}}
	\qquad
	\subfigure{\includegraphics[width=5.0cm,height=5.0cm]{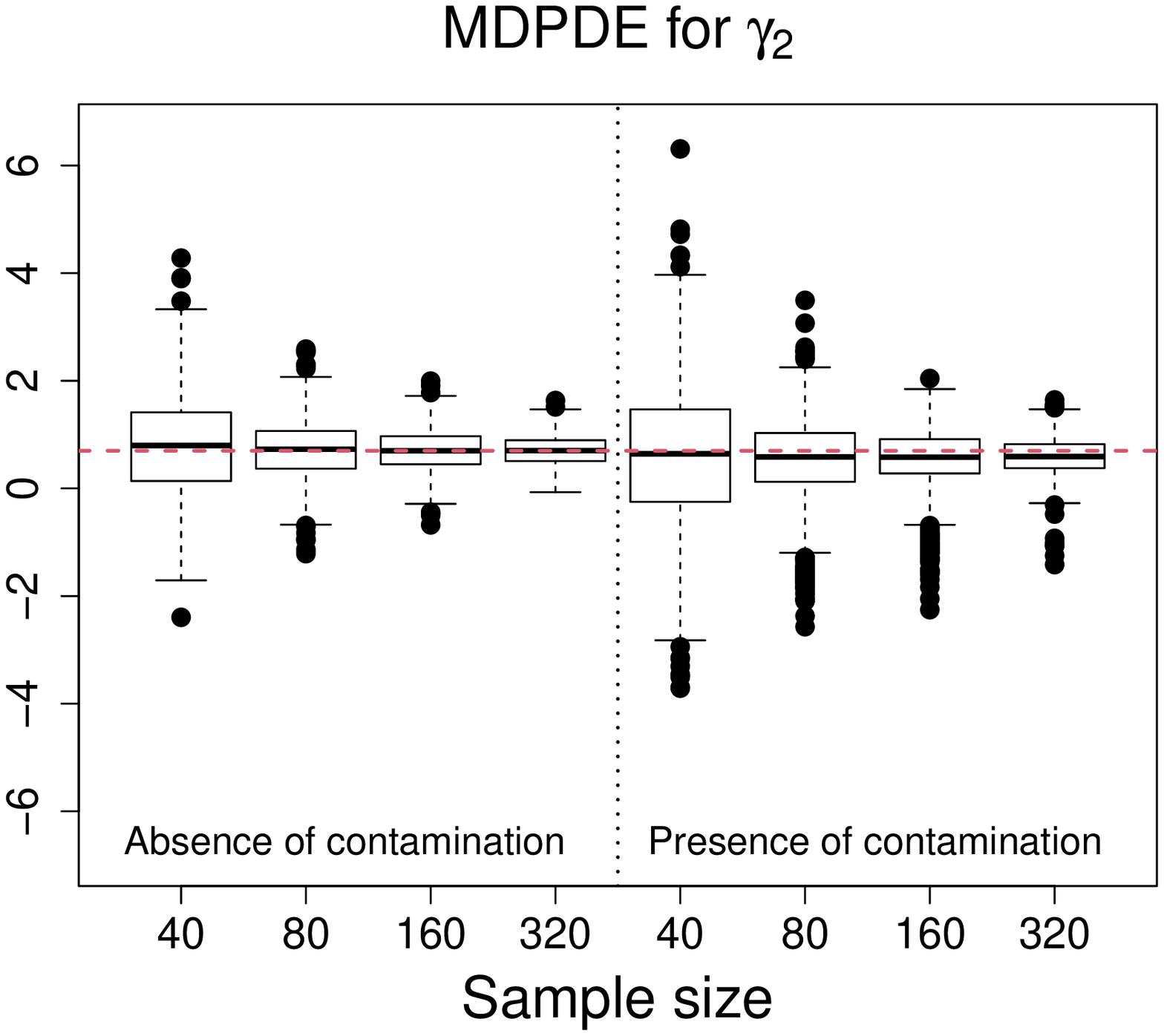}}
	\qquad
	\subfigure{\includegraphics[width=5.0cm,height=5.0cm]{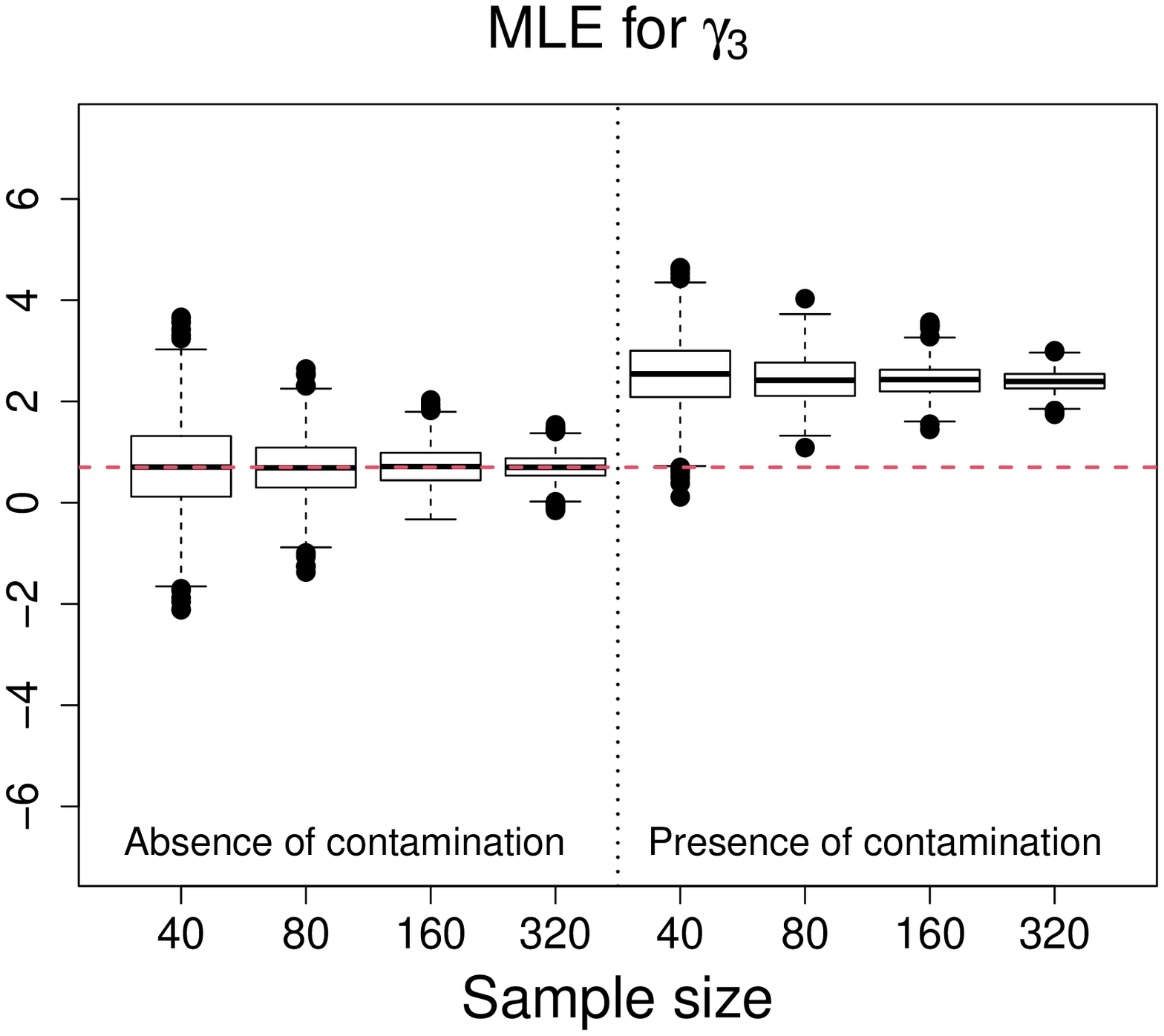}}
	\qquad
	\subfigure{\includegraphics[width=5.0cm,height=5.0cm]{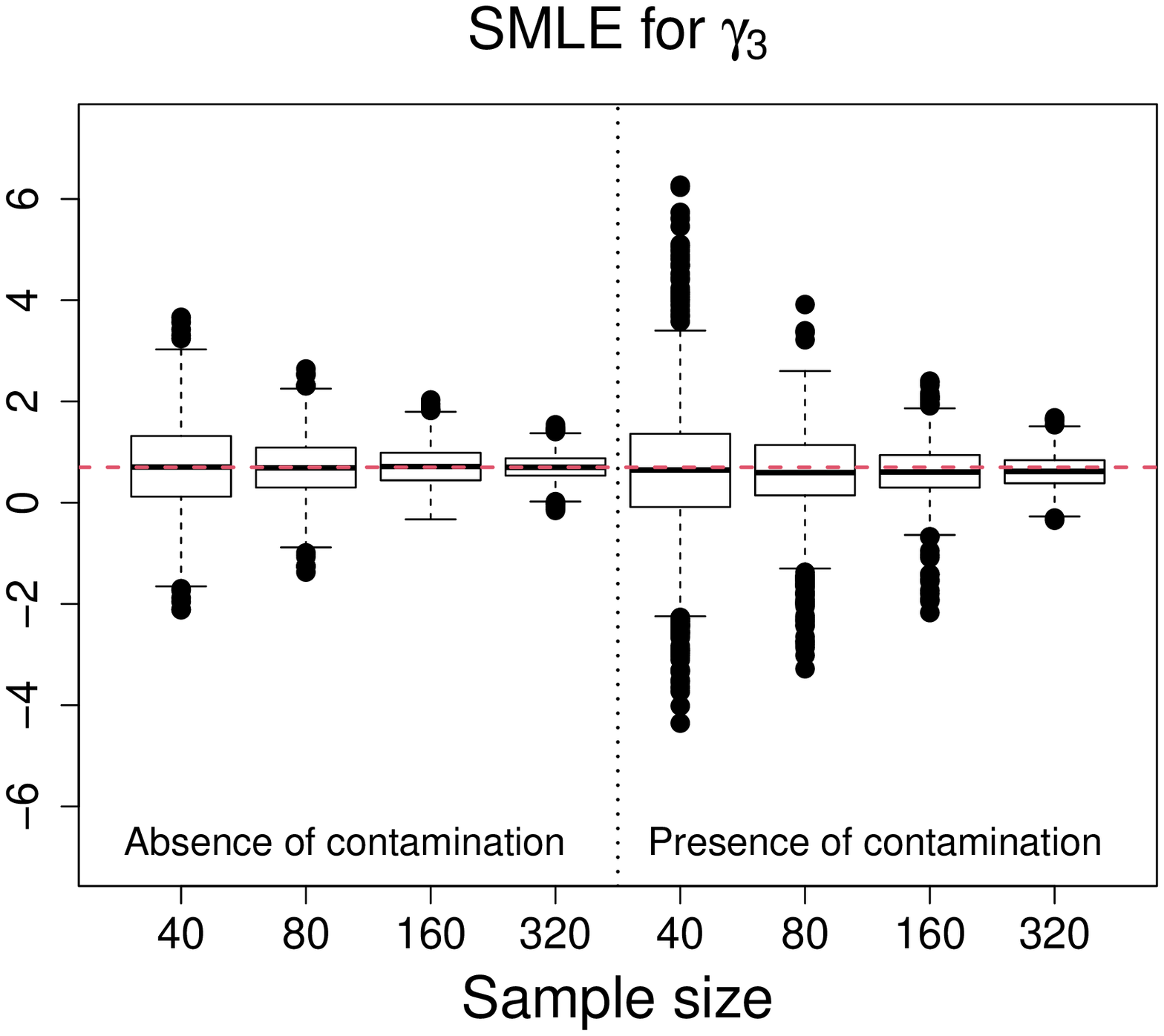}}
	\qquad
	\subfigure{\includegraphics[width=5.0cm,height=5.0cm]{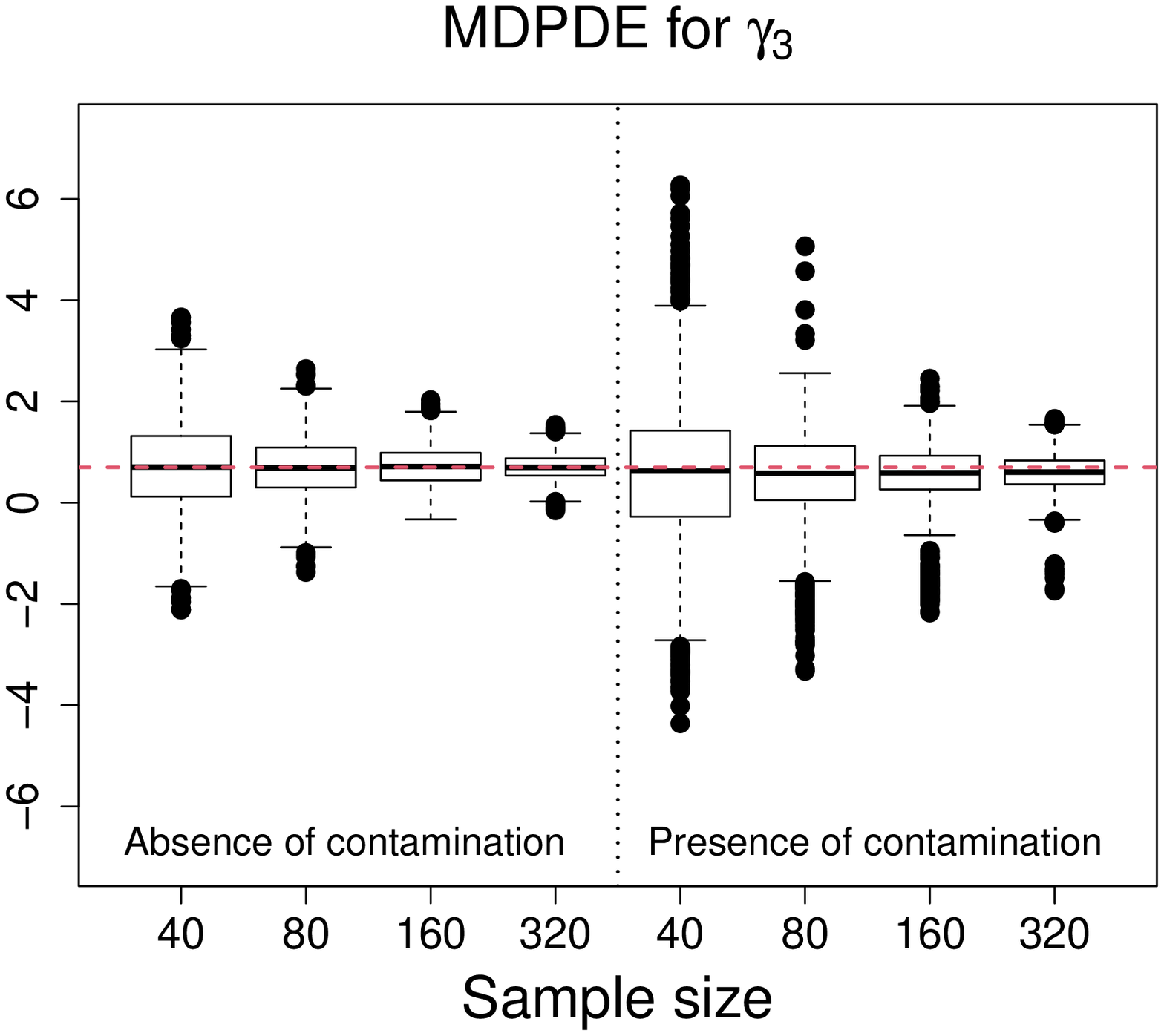}}
	\caption{Boxplots of estimates of  $\gamma_1$, $\gamma_2$, and $\gamma_3$ under Scenario 2: MLE (left), SMLE (center), and MDPDE (right).
		The red dashed line represents the true parameter value.}\label{boxplotsa02_b}
\end{figure}

\begin{figure}[!ht]
	\centering
	\subfigure{\includegraphics[width=5.0cm,height=5.0cm]{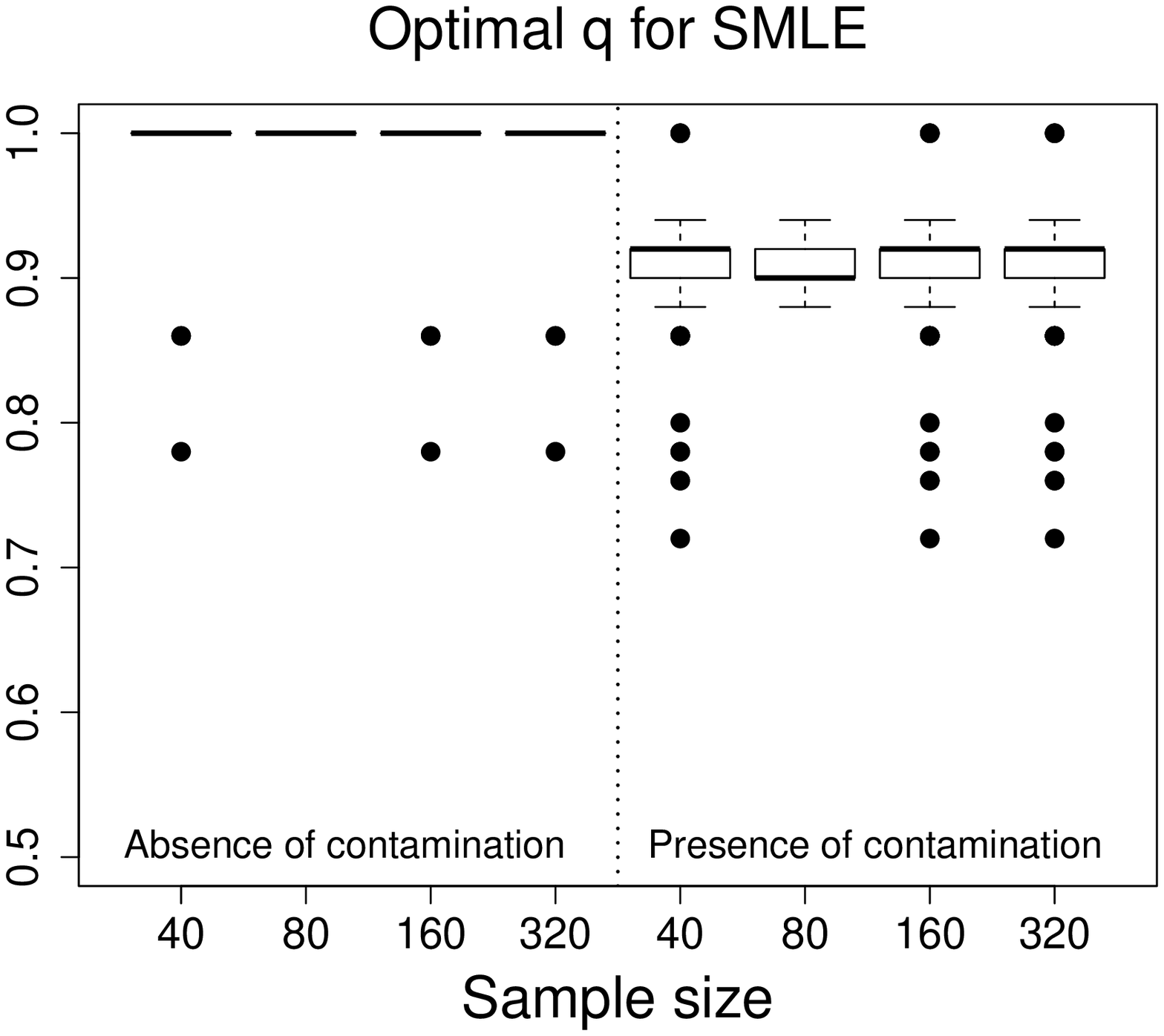}}
	\qquad
	\subfigure{\includegraphics[width=5.0cm,height=5.0cm]{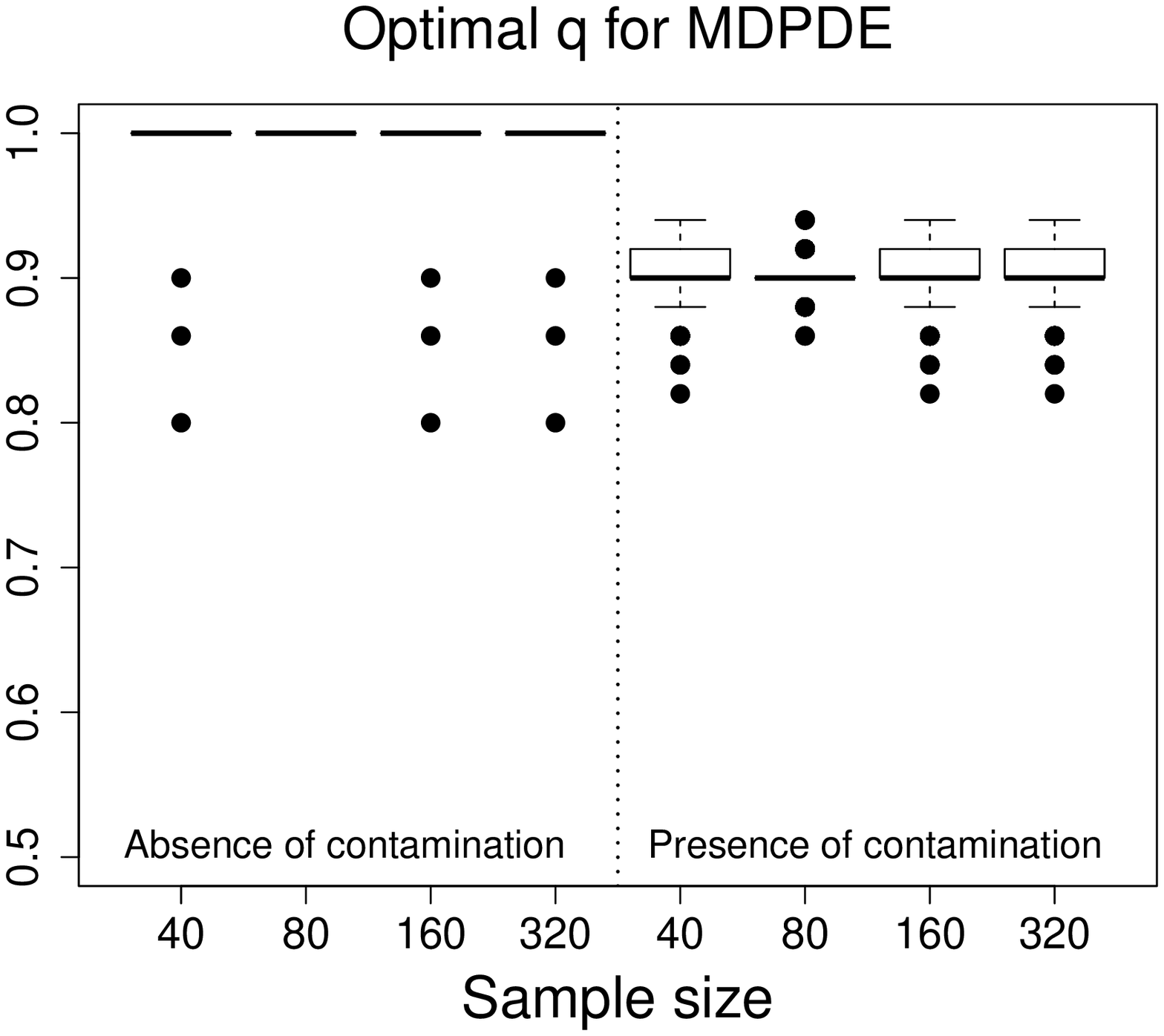}}\\
	\subfigure{\includegraphics[width=5.0cm,height=5.0cm]{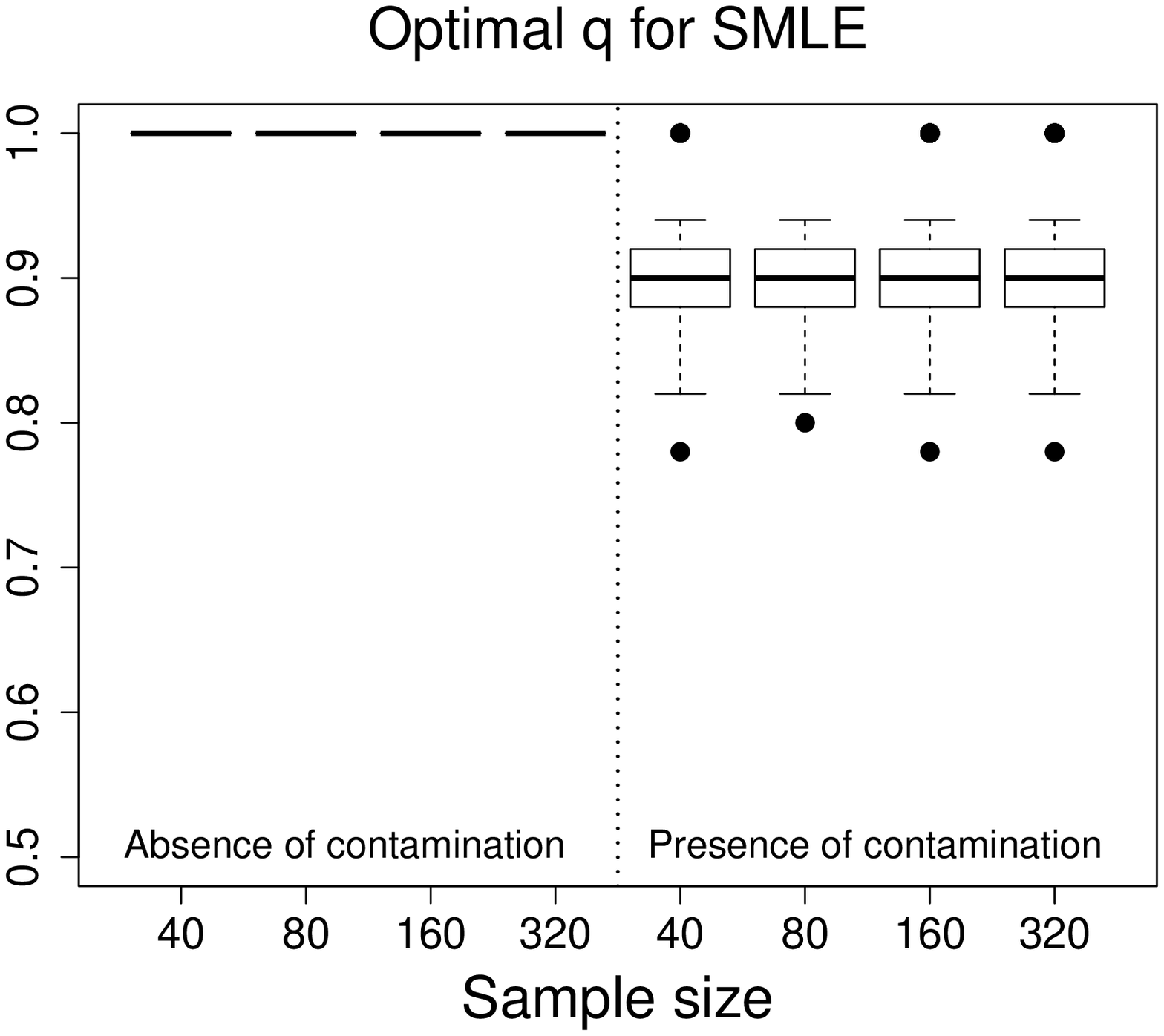}}
	\qquad
	\subfigure{\includegraphics[width=5.0cm,height=5.0cm]{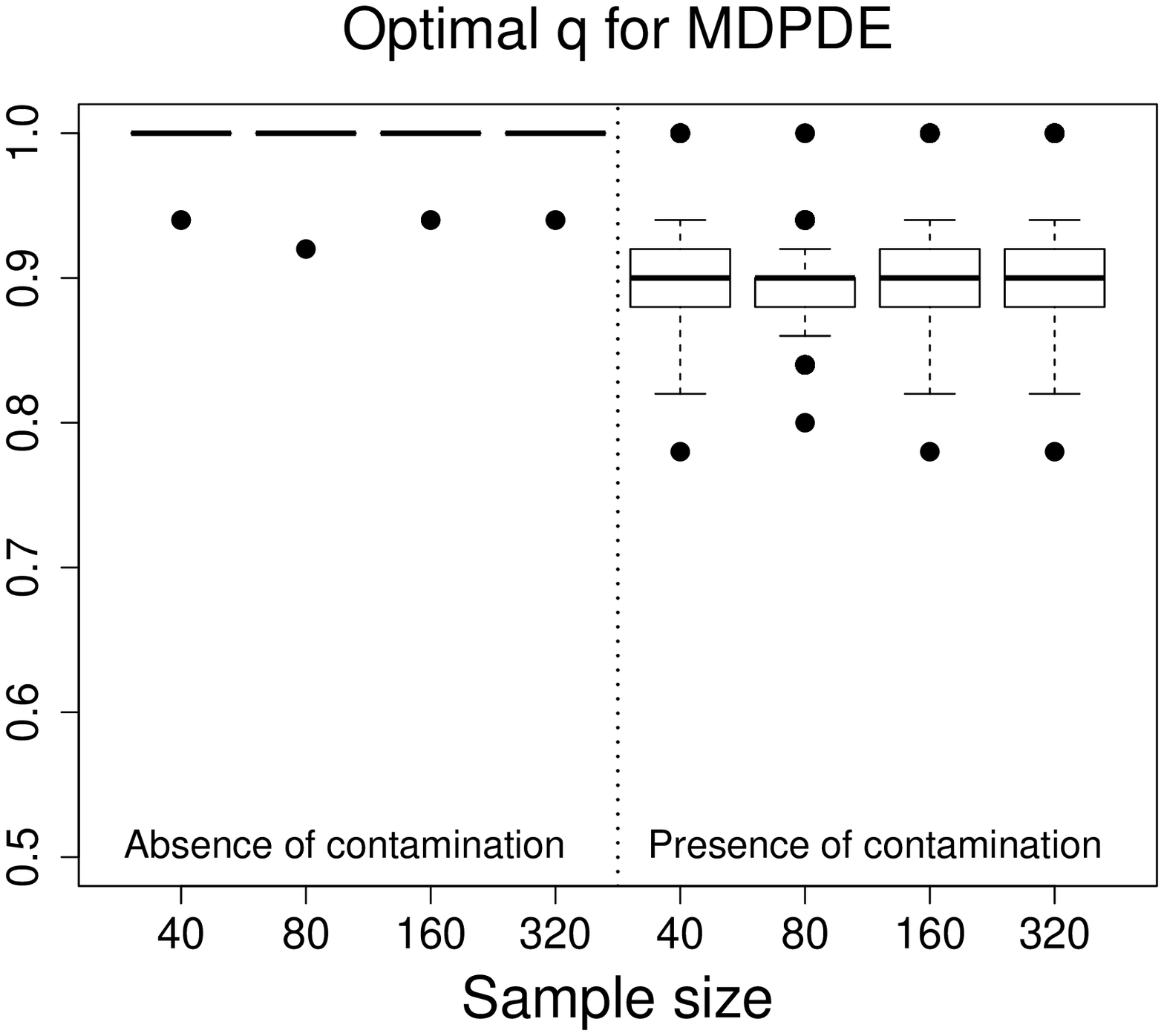}}
	\caption{Boxplots of the optimal values for the tuning constant $q$ for SMLE and MDPDE, under Scenarios 1 (first row) and 2 (second row).}
	\label{boxplots_tunings}
\end{figure}

\vspace{0.3cm}
Figure \ref{boxplotsa01} and Figures \ref{boxplotsa02_a}-\ref{boxplotsa02_b} display the boxplots of the parameter estimates using MLE, SMLE, and MDPDE under Scenarios 1 and 2, respectively, for the data without contamination and for the contaminated data. Some general tendencies become apparent from these figures.
First, the MLE of the parameters in the mean submodel and the precision submodel may be heavily affected by contamination in the data.
Second, the robust estimators behave similarly to the MLE under non contaminated data.
Third, the SMLE and MDPDE have similar behavior.
Fourth, the SMLE and MDPDE remain centered at the true parameter value when contamination in introduced in the data; the robustness comes at the cost of some extra variability in small samples.
Fifth, the data-driven selection of the optimum tuning constants worked well. In fact, as evidenced by Figure \ref{boxplots_tunings} the optimal values of $q$ for the SMLE and MDPDE are equal to 1 for the great majority of the non contaminated samples and around 0.9 for contaminated samples. Recall that $q=1$ corresponds to the MLE. This means that the proposed mechanism is able to identify whether the sample requires a robust procedure or not, and correctly selects lower values of $q$ for contaminated samples. As a consequence, the robust estimators are full efficient in the absence of contamination. This is confirmed from a study (not shown) of the asymptotic efficiency of the robust estimators relative to the MLE measured by the ratio of the traces of the respective asymptotic covariance matrices. For all the scenarios and all the sample sizes considered, the relative asymptotic efficiency under no contaminated data is equal to 1 in almost all the cases; the smallest observed value is 0.9999.

We now compare the efficiency of the estimators through their total mean squared errors (TMSE). Table \ref{MSEs} displays the ratios of TMSE of the MLE, SMLE and MDPDE under Scenarios 1 and 2. When the data are not contaminated, the efficiency of the three estimators is similar, i.e., the ratios of the TMSE are equal to or close to 1. However, if contamination is present, the TMSE of the robust estimators are much smaller than those of the MLE more so if the sample is large. For instance, for $n=160$ and Scenario 1, the TMSE of MLE is 141 times (134 times) larger than that of the SMLE (MDPDE). Under Scenario 2, the TMSE of the SMLE is smaller than that of the MDPDE for all the sample sizes.

\begin{table}
\centering
\caption{Ratio of total mean squared errors of the  estimators under Scenarios 1 and 2.}
\label{MSEs}
\scalefont{0.65}
\begin{tabular}{>{\raggedleft\arraybackslash}m{0.5cm}>{\raggedleft\arraybackslash}m{0.7cm}>{\raggedleft\arraybackslash}m{0.7cm}>{\raggedleft\arraybackslash}m{0.7cm}>{\raggedleft\arraybackslash}m{0.01cm}>{\raggedleft\arraybackslash}m{0.7cm}>{\raggedleft\arraybackslash}m{0.7cm}>{\raggedleft\arraybackslash}m{0.7cm}>{\raggedleft\arraybackslash}m{0.01cm}>{\raggedleft\arraybackslash}m{0.7cm}>{\raggedleft\arraybackslash}m{0.7cm}>{\raggedleft\arraybackslash}m{0.7cm}>{\raggedleft\arraybackslash}m{0.01cm}>{\raggedleft\arraybackslash}m{0.7cm}>{\raggedleft\arraybackslash}m{0.7cm}>{\raggedleft\arraybackslash}m{0.7cm}}\hline
	
            & \multicolumn{7}{c}{{\bf Scenario 1}}&& \multicolumn{7}{c}{{\bf Scenario 2}}\\\cline{2-8}\cline{10-16}
            & \multicolumn{3}{c}{{\bf Absence of contamination}}&& \multicolumn{3}{c}{{\bf Presence of contamination}}  && \multicolumn{3}{c}{{\bf Absence of contamination}}&& \multicolumn{3}{c}{{\bf Presence of contamination}}\\  \cline{2-4}\cline{6-8}\cline{10-12}\cline{14-16}
$n$         &$\frac{\rm MLE}{\rm SMLE}$&$\frac{\rm MLE}{\rm MDPDE}$&$\frac{\rm SMLE}{\rm MDPDE}$&
            &$\frac{\rm MLE}{\rm SMLE}$&$\frac{\rm MLE}{\rm MDPDE}$&$\frac{\rm SMLE}{\rm MDPDE}$&
						&$\frac{\rm MLE}{\rm SMLE}$&$\frac{\rm MLE}{\rm MDPDE}$&$\frac{\rm SMLE}{\rm MDPDE}$&
						&$\frac{\rm MLE}{\rm SMLE}$&$\frac{\rm MLE}{\rm MDPDE}$&$\frac{\rm SMLE}{\rm MDPDE}$\\  \cline{2-4}\cline{6-8}\cline{10-12}\cline{14-16}
$40$        &  $0.98$  &  $0.99$  &  $1.00$  &&  $31.1$  &  $34.4$  &  $1.11$  &&  $1.00$  &  $1.00$  &  $1.00$  &&  $2.37$  &  $1.85$  &  $0.78$  \\
$80$        &  $1.00$  &  $1.00$  &  $1.00$  &&  $73.5$  &  $72.8$  &  $0.99$  &&  $1.00$  &  $1.00$  &  $1.00$  &&  $5.90$  &  $4.16$  &  $0.70$ \\
$160$       &  $1.00$  &  $1.00$  &  $1.00$  && $141.1$  &  $133.7$ &  $0.95$  &&  $1.00$  &  $1.00$  &  $1.00$  &&  $14.0$  &  $8.40$  &  $0.60$\\
$320$       &  $1.00$  &  $1.00$  &  $1.00$  && $284.2$  &  $243.3$ &  $0.86$  &&  $1.00$  &  $1.00$  &  $1.00$  &&  $31.7$  &   $26.0$ &  $0.82$\\
\hline
\end{tabular}
\end{table}

Table \ref{Empirical_levels_I} shows the empirical levels of the Wald-type tests based on the MLE and the robust estimators. The Wald-type test of the null hypothesis ${\rm H}: \theta^j = \theta^j_0$ against a two-sided alternative rejects H at the nominal level $1-\alpha$ when $\{(\widehat{\theta}^{j} - {\theta}^{j}_0)/\mbox{se}(\widehat{\theta}^{j})\}^2$ exceeds the $1-\alpha$ quantile of the $\chi_1$ distribution. For non contaminated data, the tests behave similarly, the empirical levels being close 5\% except for the smallest sample size in Scenario 2 (the empirical levels were observed between 8\% and 11\%). Under contaminated data, the tests that employ the robust estimators suffer from some type I error probability inflation. On the other hand, the tests that use the MLE are erratic in most of the cases, their empirical levels being near 100\%. Hence, when the data contain outliers, the usual Wald-type tests should be avoided. The Wald-type tests computed from the robust estimators are reasonably reliable, although any decision based on a $p$-value close to the nominal level should be taken with care.
\begin{table}
	\centering
	\caption{Empirical null levels of Wald-type tests under Scenarios 1 and 2 at the 5\% nominal level.}
	\label{Empirical_levels_I}
	\scalefont{0.65}
	\begin{tabular}{>{\raggedleft\arraybackslash}m{1.1cm}>{\raggedleft\arraybackslash}m{0.3cm}>{\raggedleft\arraybackslash}m{0.3cm}>{\raggedleft\arraybackslash}m{0.3cm}>{\raggedleft\arraybackslash}m{0.01cm}>{\raggedleft\arraybackslash}m{0.3cm}>{\raggedleft\arraybackslash}m{0.3cm}>{\raggedleft\arraybackslash}m{0.3cm}>{\raggedleft\arraybackslash}m{0.01cm}>{\raggedleft\arraybackslash}m{0.3cm}>{\raggedleft\arraybackslash}m{0.3cm}>{\raggedleft\arraybackslash}m{0.3cm}>{\raggedleft\arraybackslash}m{0.3cm}>{\raggedleft\arraybackslash}m{0.3cm}>{\raggedleft\arraybackslash}m{0.3cm}>{\raggedleft\arraybackslash}m{0.01cm}>{\raggedleft\arraybackslash}m{0.3cm}>{\raggedleft\arraybackslash}m{0.3cm}>{\raggedleft\arraybackslash}m{0.3cm}>{\raggedleft\arraybackslash}m{0.3cm}>{\raggedleft\arraybackslash}m{0.3cm}>{\raggedleft\arraybackslash}m{0.3cm}}\hline
		     & \multicolumn{7}{c}{{\bf Scenario 1}}  && \multicolumn{13}{c}{{\bf Scenario 2}}\\\cline{2-8}\cline{10-22}
		$n=40$& \multicolumn{3}{c}{{\bf Abs. of cont.}}&& \multicolumn{3}{c}{{\bf Pres. of cont.}}  && \multicolumn{6}{c}{{\bf Abs. of cont.}}&& \multicolumn{6}{c}{{\bf Pres. of cont.}}\\  \cline{2-4}\cline{6-8}\cline{10-15}\cline{17-22}
		Estimator  & $\beta_1$&$\beta_2$&$\gamma_1$ && $\beta_1$&$\beta_2$&$\gamma_1$&& $\beta_1$&$\beta_2$&$\beta_3$ & $\gamma_1$&$\gamma_2$&$\gamma_3$&& $\beta_1$&$\beta_2$&$\beta_3$ & $\gamma_1$&$\gamma_2$&$\gamma_3$\\  \cline{2-4}\cline{6-8}\cline{10-15}\cline{17-22}
		MLE        &  $0.06$  &  $0.05$  &  $0.07$  &&  $0.00$  &  $0.94$  &  $1.00$  &&  $0.08$  &  $0.09$  &  $0.09$  &  $0.10$  &  $0.11$  &  $0.11$  &&  $0.61$  &  $0.41$  &  $0.20$  &  $1.00$  &  $0.39$  &  $0.72$\\
		SMLE       &  $0.06$  &  $0.05$  &  $0.07$  &&  $0.06$  &  $0.08$  &  $0.10$  &&  $0.08$  &  $0.09$  &  $0.09$  &  $0.10$  &  $0.11$  &  $0.11$  &&  $0.15$  &  $0.13$  &  $0.13$  &  $0.22$  &  $0.21$  &  $0.23$\\
		MDPDE      &  $0.06$  &  $0.05$  &  $0.07$  &&  $0.05$  &  $0.08$  &  $0.09$  &&  $0.08$  &  $0.09$  &  $0.09$  &  $0.10$  &  $0.11$  &  $0.11$  &&  $0.17$  &  $0.15$  &  $0.14$  &  $0.29$  &  $0.26$  &  $0.29$ \\ \cline{2-8}\cline{10-22} 
             & \multicolumn{7}{c}{{\bf Scenario 1}}  && \multicolumn{13}{c}{{\bf Scenario 2}}\\\cline{2-8}\cline{10-22}
		$n=80$& \multicolumn{3}{c}{{\bf Abs. of cont.}}&& \multicolumn{3}{c}{{\bf Pres. of cont.}}  && \multicolumn{6}{c}{{\bf Abs. of cont.}}&& \multicolumn{6}{c}{{\bf Pres. of cont.}}\\  \cline{2-4}\cline{6-8}\cline{10-15}\cline{17-22}
		Estimator  & $\beta_1$&$\beta_2$&$\gamma_1$ && $\beta_1$&$\beta_2$&$\gamma_1$&& $\beta_1$&$\beta_2$&$\beta_3$ & $\gamma_1$&$\gamma_2$&$\gamma_3$&& $\beta_1$&$\beta_2$&$\beta_3$ & $\gamma_1$&$\gamma_2$&$\gamma_3$\\  \cline{2-4}\cline{6-8}\cline{10-15}\cline{17-22}
		MLE        &  $0.05$  &  $0.05$  &  $0.05$  &&  $0.00$  &  $1.00$  &  $1.00$  &&  $0.08$  &  $0.07$  &  $0.05$  &  $0.07$  &  $0.06$  &  $0.08$  &&  $0.98$  &  $0.94$  &  $0.60$  &  $1.00$  &  $0.54$  &  $0.94$\\
	    SMLE       &  $0.05$  &  $0.05$  &  $0.05$  &&  $0.05$  &  $0.07$  &  $0.07$  &&  $0.08$  &  $0.07$  &  $0.05$  &  $0.07$  &  $0.06$  &  $0.08$  &&  $0.11$  &  $0.08$  &  $0.07$  &  $0.17$  &  $0.13$  &  $0.18$\\
	    MDPDE      &  $0.05$  &  $0.05$  &  $0.05$  &&  $0.05$  &  $0.07$  &  $0.07$  &&  $0.08$  &  $0.07$  &  $0.05$  &  $0.07$  &  $0.06$  &  $0.08$  &&  $0.12$  &  $0.09$  &  $0.08$  &  $0.22$  &  $0.19$  &  $0.24$ \\ \cline{2-8}\cline{10-22} 
		    & \multicolumn{7}{c}{{\bf Scenario 1}}&& \multicolumn{13}{c}{{\bf Scenario 2}}\\ \cline{2-8}\cline{10-22} 
		$n=160$& \multicolumn{3}{c}{{\bf Abs. of cont.}}&& \multicolumn{3}{c}{{\bf Pres. of cont.}}  && \multicolumn{6}{c}{{\bf Abs. of cont.}}&& \multicolumn{6}{c}{{\bf Pres. of cont.}}\\  \cline{2-4}\cline{6-8}\cline{10-15}\cline{17-22}
		Estimator  & $\beta_1$&$\beta_2$&$\gamma_1$ && $\beta_1$&$\beta_2$&$\gamma_1$&& $\beta_1$&$\beta_2$&$\beta_3$ & $\gamma_1$&$\gamma_2$&$\gamma_3$&& $\beta_1$&$\beta_2$&$\beta_3$ & $\gamma_1$&$\gamma_2$&$\gamma_3$\\  \cline{2-4}\cline{6-8}\cline{10-15}\cline{17-22}
	    MLE        &  $0.06$  &  $0.05$  &  $0.05$  &&  $0.00$  &  $1.00$  &  $1.00$  &&  $0.05$  &  $0.06$  &  $0.07$  &  $0.06$  &  $0.06$  &  $0.06$  &&  $1.00$  &  $1.00$  &  $0.97$  &  $1.00$  &  $0.79$  &  $1.00$\\
		SMLE       &  $0.06$  &  $0.05$  &  $0.05$  &&  $0.05$  &  $0.07$  &  $0.08$  &&  $0.05$  &  $0.06$  &  $0.07$  &  $0.06$  &  $0.06$  &  $0.06$  &&  $0.09$  &  $0.07$  &  $0.08$  &  $0.15$  &  $0.11$  &  $0.14$ \\
		MDPDE      &  $0.06$  &  $0.05$  &  $0.05$  &&  $0.05$  &  $0.07$  &  $0.08$  &&  $0.05$  &  $0.06$  &  $0.07$  &  $0.06$  &  $0.06$  &  $0.06$  &&  $0.11$  &  $0.08$  &  $0.09$  &  $0.19$  &  $0.16$  &  $0.18$\\  \cline{2-8}\cline{10-22} 
		    & \multicolumn{7}{c}{{\bf Scenario 1}}&& \multicolumn{13}{c}{{\bf Scenario 2}}\\ \cline{2-8}\cline{10-22} 
		$n=320$& \multicolumn{3}{c}{{\bf Abs. of cont.}}&& \multicolumn{3}{c}{{\bf Pres. of cont.}}  && \multicolumn{6}{c}{{\bf Abs. of cont.}}&& \multicolumn{6}{c}{{\bf Pres. of cont.}}\\  \cline{2-4}\cline{6-8}\cline{10-15}\cline{17-22}
		Estimator  & $\beta_1$&$\beta_2$&$\gamma_1$ && $\beta_1$&$\beta_2$&$\gamma_1$&& $\beta_1$&$\beta_2$&$\beta_3$ & $\gamma_1$&$\gamma_2$&$\gamma_3$&& $\beta_1$&$\beta_2$&$\beta_3$ & $\gamma_1$&$\gamma_2$&$\gamma_3$\\  \cline{2-4}\cline{6-8}\cline{10-15}\cline{17-22}
		MLE        &  $0.06$  &  $0.05$  &  $0.05$  &&  $0.00$  &  $1.00$  &  $1.00$  &&  $0.06$  &  $0.06$  &  $0.05$  &  $0.06$  &  $0.06$  &  $0.06$  &&  $1.00$  &  $1.00$  &  $1.00$  &  $1.00$  &  $0.98$  &  $1.00$\\
		SMLE       &  $0.06$  &  $0.05$  &  $0.05$  &&  $0.05$  &  $0.07$  &  $0.08$  &&  $0.06$  &  $0.06$  &  $0.05$  &  $0.06$  &  $0.06$  &  $0.06$  &&  $0.07$  &  $0.07$  &  $0.07$  &  $0.17$  &  $0.12$  &  $0.13$\\
		MDPDE      &  $0.06$  &  $0.05$  &  $0.05$  &&  $0.05$  &  $0.07$  &  $0.13$  &&  $0.06$  &  $0.06$  &  $0.05$  &  $0.06$  &  $0.06$  &  $0.06$  &&  $0.08$  &  $0.07$  &  $0.07$  &  $0.16$  &  $0.13$  &  $0.13$\\  \hline
	\end{tabular}
\end{table}

Overall, both the robust estimators performed similarly to the MLE when no contamination is present, and much better than the MLE under contamination in the
data. The algorithm for the data-driven choice of the tuning constant proposed in this paper proved to be very effective for both estimators. It tends to
select $q\approx 1$, i.e. the MLE, when the data are not contaminated, leading to full efficiency. When the data are contaminated, the selected $q$ tends
to be reasonably smaller than 1, providing robust estimation. Both robust estimators applied with this automated selection of the tuning constant produce
usable Wald-type tests, while the Wald-type tests that employ the MLE present unacceptable type I probabilities and should be avoided whenever there is
any evidence of the presence of outliers.
\section{Real data applications}
\label{app}
\subsection{The AIS Data}

\cite{bayes2012new}, \cite{ghosh2019robust}, \cite{migliorati2018new} and \cite{niekerk2018beta} illustrated the sensitivity of MLE to outliers in the beta
regression model with constant precision through the analysis of body fat percentages of 37 rowing athletes. The dataset, collected at the Australian
Institute of Sport (AIS), is available in the package {\tt sn} of software {\tt R}  (http://azzalini.stat.unipd.it/SN/index.html). The objective is to model the mean
of the body fat percentage (BFP) as a function of the lean body mass (LBM). The authors noted substantial changes in the parameter estimates when two
atypical observations are excluded from the data. \cite{ghosh2019robust} applied the MDPDE with different values of the tuning constant. Although the MDPDE
displayed a robust fit, the practitioner would be uncertain of which value of the tuning constant should be chosen. Here,  we will show that our proposed estimator with the optimal tuning constant selected as described in Section \ref{qchoice} provides a robust fit to the data.

\begin{figure}[h]
\centering
\subfigure{\includegraphics[width=6cm,height=6cm]{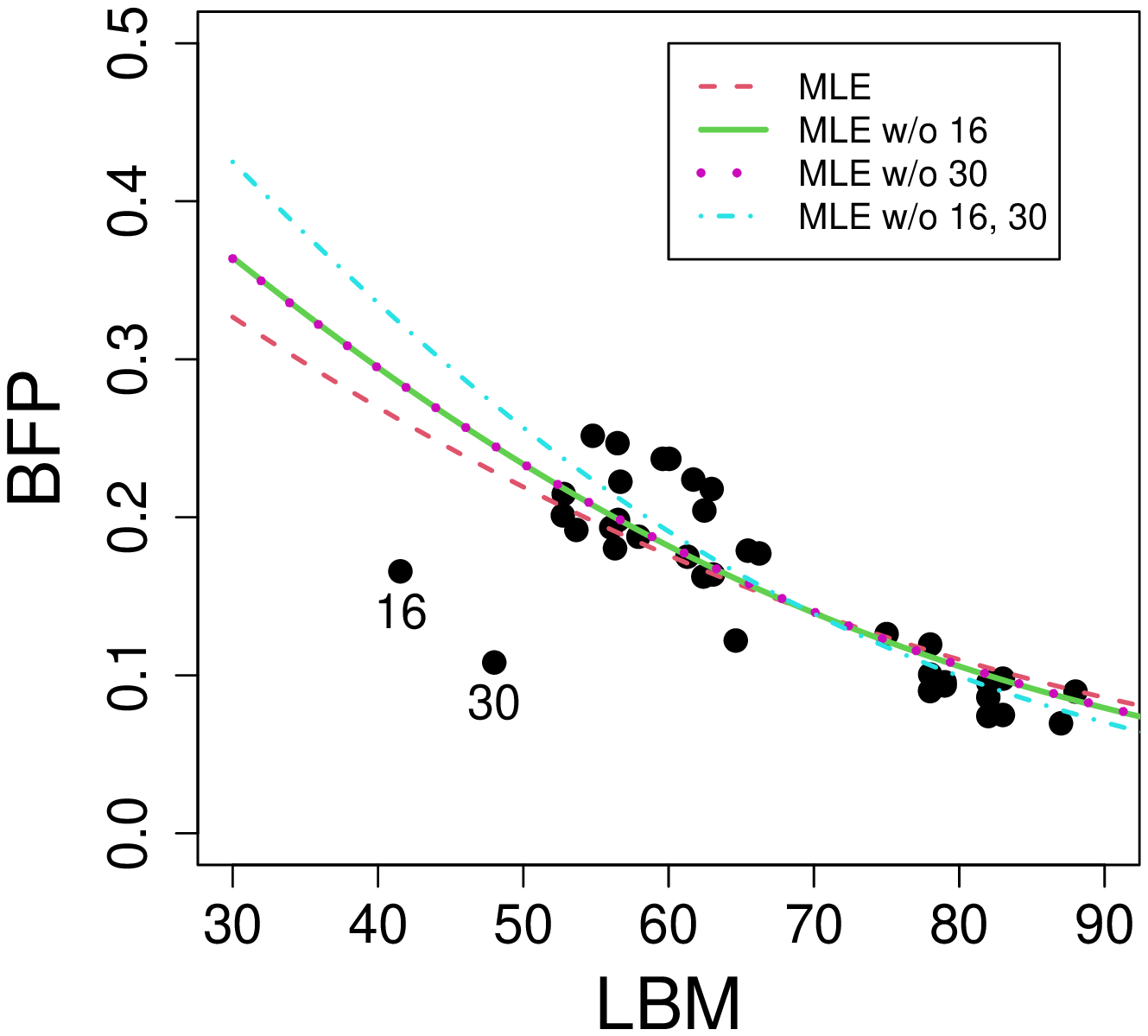}}
\qquad
\subfigure{\includegraphics[width=6cm,height=6cm]{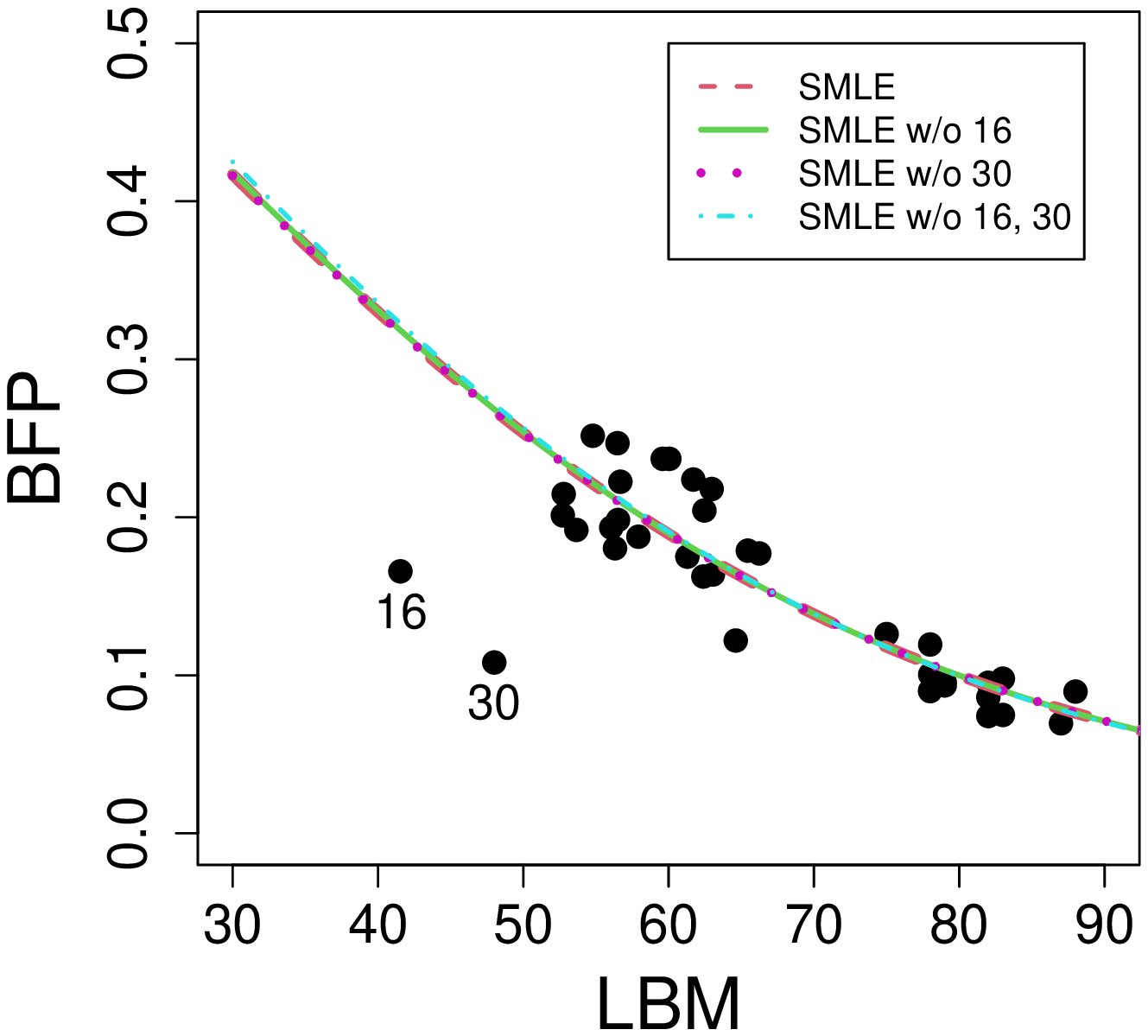}}
\caption{Scatter plots of BFP versus LBM along with the fitted lines based on the MLE and SMLE for the full data and the data without outliers; AIS data.}
\label{scatter_AISdata}
\end{figure}

Following the same specification of the model as in the papers mentioned above, we consider a beta regression model with constant precision given by
$\log({\mu_i}/{(1-\mu_i)}) =\beta_1+\beta_2\mbox{LBM}_i \ {\rm and}  \
\log(\phi_i)=\gamma_1,
\label{model2_AIS_DC}
$ where  $\mu_i$ and $\phi_i$ are the mean and precision of the response variable BFP for the $i$-th athlete, $i=1,\ldots,37$. We fitted the model using the MLE and SMLE for the full data and excluding combinations of the most prominent outliers:  \{16\}, \{30\},  and \{16, 30\}. The optimal value selected for the tuning constant $q$ is $0.82$ for the full data, $0.88$ and $0.84$ for the data without observations 16 and 30, respectively, and 1 when both outliers are jointly deleted. These values reveal that the proposed selection method of the tuning constant correctly captures the fact that observations 16 and 30 are, individually and jointly, highly influential in the model fit.
Figure \ref{scatter_AISdata} displays the scatter plot of LBM versus BFP along with the MLE and SMLE fitted lines for the full data and reduced data. We notice that the outlying observations lead to considerable changes in the fitted regression curves under the MLE fit. On the other hand, the SMLE fitted regression curves are almost indistinguishable.

Table \ref{estimates_AIS_1} shows the estimates, standard errors, $z$-statistics (estimate divided by standard error) and asymptotic $p$-values of the Wald-type test of nullity of coefficients (i.e. based on the null asymptotic distribution of the square of the $z$-statistic, which is $\chi^2_1$) for the full data and the reduced data. We also report $p$-values evaluated from parametric bootstrap with $500$ bootstrap replicates. In this application, the asymptotic and the bootstrap $p$-values coincide up to the fourth decimal point.

We notice that the robust estimates present slight changes in the presence of the outlying observations 16 and 30 (individually or jointly). However, maximum likelihood estimates change drastically when outlying observations are excluded. For instance, the MLE for the coefficient of LBM is $-0.027$ for the full data and $-0.038$ for the data without observations 16 and 30, a relative change of $41\%$. The corresponding robust estimates are $-0.037$ and $-0.038$.  Also, the MLE of the mean and precision intercepts suffer major changes when the two atypical observations are excluded. The estimates obtained through the robust method remain almost unchangeable when the outlying observations are deleted. Overall, the robust estimates and the corresponding standard errors for the full data and for the reduced data are close to those obtained by the MLE without outliers.

\begin{table}
\centering
\caption{Estimates, standard errors,  $z$-stat and $p$-values (bootstrap $p$-values between parenteses) for beta regression with constant precision -- AIS data.}\vspace{0.2cm}
\label{estimates_AIS_1}
\scalefont{0.7}
\def\arraystretch{0.65}
\begin{tabular}{>{\raggedleft\arraybackslash}m{2.1cm}>{\raggedleft\arraybackslash}m{1.2cm}>{\raggedleft\arraybackslash}m{1.2cm}>{\raggedleft\arraybackslash}m{1.2cm}>{\raggedleft\arraybackslash}m{1.7cm}>{\raggedleft\arraybackslash}m{0.05cm}>{\raggedleft\arraybackslash}m{1.2cm}>{\raggedleft\arraybackslash}m{1.2cm}>{\raggedleft\arraybackslash}m{1.2cm}>{\raggedleft\arraybackslash}m{1.7cm}}\hline
 & \multicolumn{4}{c}{MLE for the full dataset}                              && \multicolumn{4}{c}{SMLE for the full dataset} \\  \cline{2-5}\cline{7-10}
                 & Estimate &  Std. error& $z$-stat  &$p$-value              && Estimate   & Std. error&$z$-stat   &$p$-value   \\  \cline{2-5}\cline{7-10}
{\it mean submodel}    &&&&&&&&\\
Intercept        &  $0.098$  &  $0.253$  &  $0.387$  &     -                 &&  $0.782$  &  $0.175$  &  $4.460$  &   - \\
LBM              & $-0.027$  &  $0.004$  & $-7.029$  &  $0.000$ ($0.000$)    && $-0.037$  &  $0.003$  & $-13.627$ &  $0.000$ ($0.000$) \\
{\it precision submodel}    &&&&&&&&\\
Intercept        &  $4.571$  &  $0.232$  &  $19.686$ & -                     &&  $5.366$  &  $0.242$  &  $22.163$  &  -  \\ \cline{2-5}\cline{7-10}

& \multicolumn{4}{c}{MLE without observation 16}                              && \multicolumn{4}{c}{SMLE without observation 16} \\  \cline{2-5}\cline{7-10}
                 & Estimate  & Std. error& $z$-stat  &$p$-value               && Estimate&Std. error&$z$-stat&$p$-value    \\ \cline{2-5}\cline{7-10}
{\it mean submodel}&&&&&&&&\\
Intercept        &  $0.392$  &  $0.256$  &  $1.529$  &       -                &&   $0.791$  &  $0.191$  &  $4.136$  &   -       \\
LBM              & $-0.032$  &  $0.004$  & $-8.076$  &  $0.000$ ($0.000$)     &&  $-0.037$  &  $0.003$  &  $-12.694$&   $0.000$ ($0.000$)  \\
{\it precision submodel}&&&&&&&&\\
Intercept        &  $4.742$  &  $0.235$  &  $20.140$ &  -                     &&   $5.366$  &  $0.240$  &  $22.371$ &   -     \\ \cline{2-5}\cline{7-10}

& \multicolumn{4}{c}{MLE without observation 30}                              && \multicolumn{4}{c}{SMLE without observation 30} \\  \cline{2-5}\cline{7-10}
                 & Estimate   &Std. error& $z$-stat &   $p$-value             && Estimate  & Std. error&  $z$-stat &$p$-value    \\  \cline{2-5}\cline{7-10}
{\it mean submodel}&&&&&&&&\\
Intercept        &  $0.382$   & $0.219$  &  $1.744$  &  -                     &&  $0.777$  &  $0.182$  &  $4.264$  &   -      \\
LBM              & $-0.031$   & $0.003$  & $-9.341$  &  $0.000$ ($0.000$)     && $-0.037$  &  $0.003$  & $-13.200$ &   $0.000$ ($0.000$)      \\
{\it precision submodel}&&&&&&&&\\
Intercept        &  $4.953$   & $0.235$  &  $21.032$ &  -                     &&  $5.370$  &  $0.243$  & $22.069$  &   -      \\ \cline{2-5}\cline{7-10}

& \multicolumn{4}{c}{MLE without observations 16 and 30}                      &&\multicolumn{4}{c}{SMLE without observations 16 and 30} \\  \cline{2-5}\cline{7-10}
                 & Estimate  &Std. error &$z$-stat  &$p$-value                && Estimate  &Std. error &$z$-stat   &$p$-value    \\ \cline{2-5}\cline{7-10}
{\it mean submodel}    &&&&&&&&\\
Intercept        &  $0.838$  &  $0.188$  &  $4.467$  &   -                    &&  $0.838$  &  $0.188$  &  $4.467$  &   -\\
LBM              & $-0.038$  &  $0.003$  & $-13.285$ &   $0.000$ ($0.000$)    && $-0.038$  &  $0.003$  & $-13.285$ &   $0.000$ ($0.000$)      \\
{\it precision submodel}    &&&&&&&&\\
Intercept        &  $5.507$  &  $0.239$  &  $23.047$ &   -                    &&  $5.507$  &  $0.239$  &  $23.047$ &   -\\ \hline
\end{tabular}
\end{table}

It is common practice in beta regression models to employ normal probability plots of residuals with simulated envelope to assess the  goodness of fit \citep{espinheira2008beta, ospina2012general, pereira2019quantile}. Following \cite{espinheira2008beta}, we consider the `standardized weighted residual 2' given by
\begin{eqnarray*}
	r_i = \frac{y^{\star}_i-{\widecheck{\mu}}^{\star}_{i}}{\sqrt{\widecheck{v}_{i}(1-\widecheck{h}_{ii})}},
\end{eqnarray*}
where  $\widecheck{\mu}^{\star}_{i}$, $\widecheck{v}_{i}$, and  $\widecheck{h}_{ii}$ are estimates of $\mu^{\star}_i$, $v_i$, and $h_{ii}$, respectively; $h_{ii}$ is the $i$-th diagonal element of $H$, where $H=W^{1/2}X(X^\top WX)^{-1}X^\top W^{1/2}$, $W = \mbox{diag}\{w_1,\ldots, w_n\}$, $w_i = \phi_iv_i[g'_{\mu}(\mu_i)]^{-2}$, and $v_i$ is given in the Appendix. Figure \ref{ALL_ENVELOPES1_AIS} shows normal probability plots of the residuals with simulated envelopes for the MLE and SMLE fits and a plot of the estimated weights employed by the SMLE. The residual plot for the MLE reveals some lack of fit, but the outlying observations, 16 and 30, fall only slightly outside of the envelope. Hence, it masks the influence of these observations. Differently, the residual plot for the SMLE fit suggests better fit for the majority of the data and strongly highlights observations 16 and 30 as outliers. In fact, robust estimators are expected to produce good fit for most of the data but not necessarily for the outliers. Figure \ref{ALL_ENVELOPES1_AIS}(c) shows that observations with bigger residuals tend to receive smaller weights in the robust estimation. As expected observations 16 and 30 received the smallest weights.

\begin{figure}
\centering
\subfigure[][]{\includegraphics[width=5.0cm,height=4.5cm]{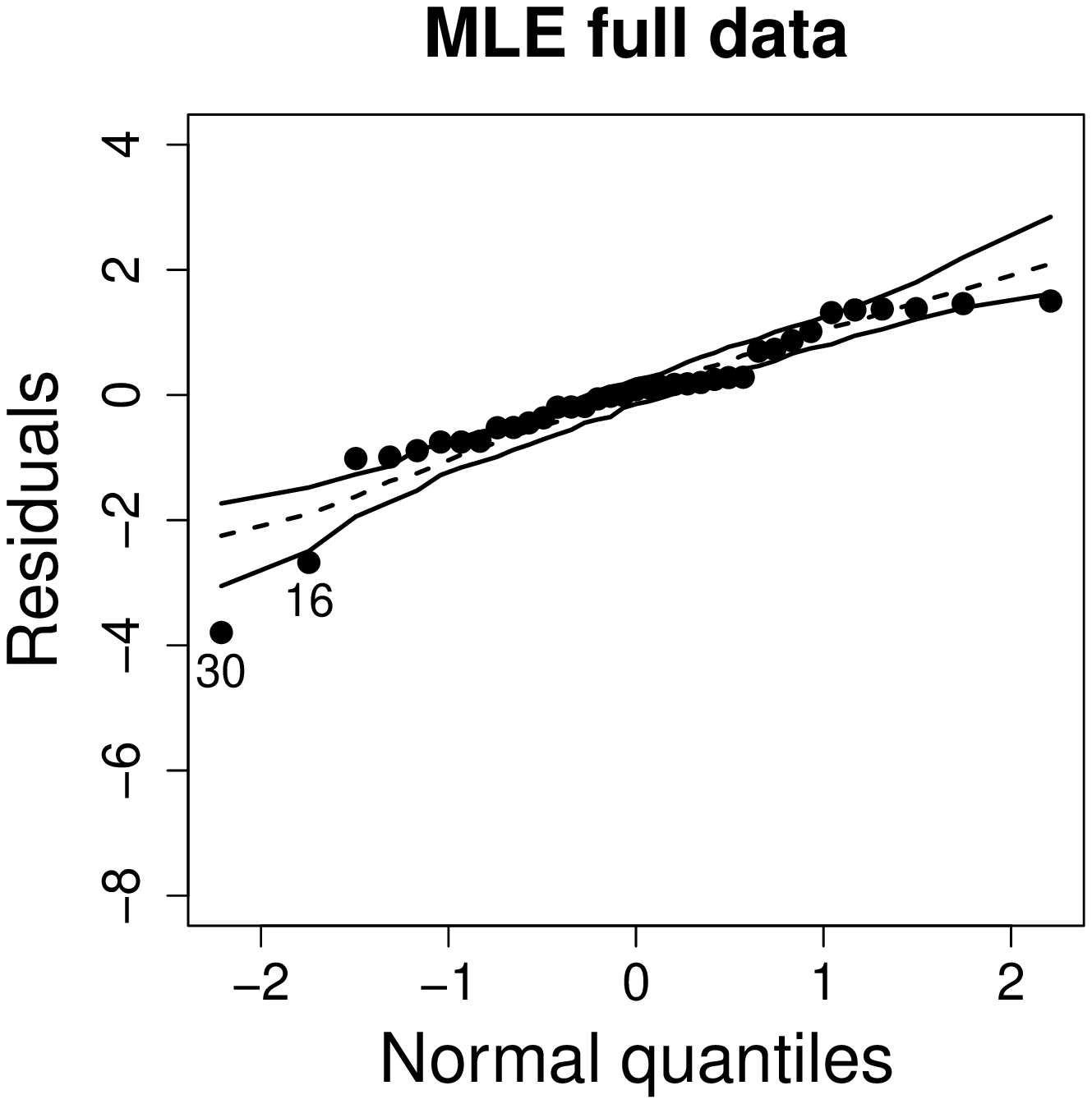}}
\qquad
\subfigure[][]{\includegraphics[width=5.0cm,height=4.5cm]{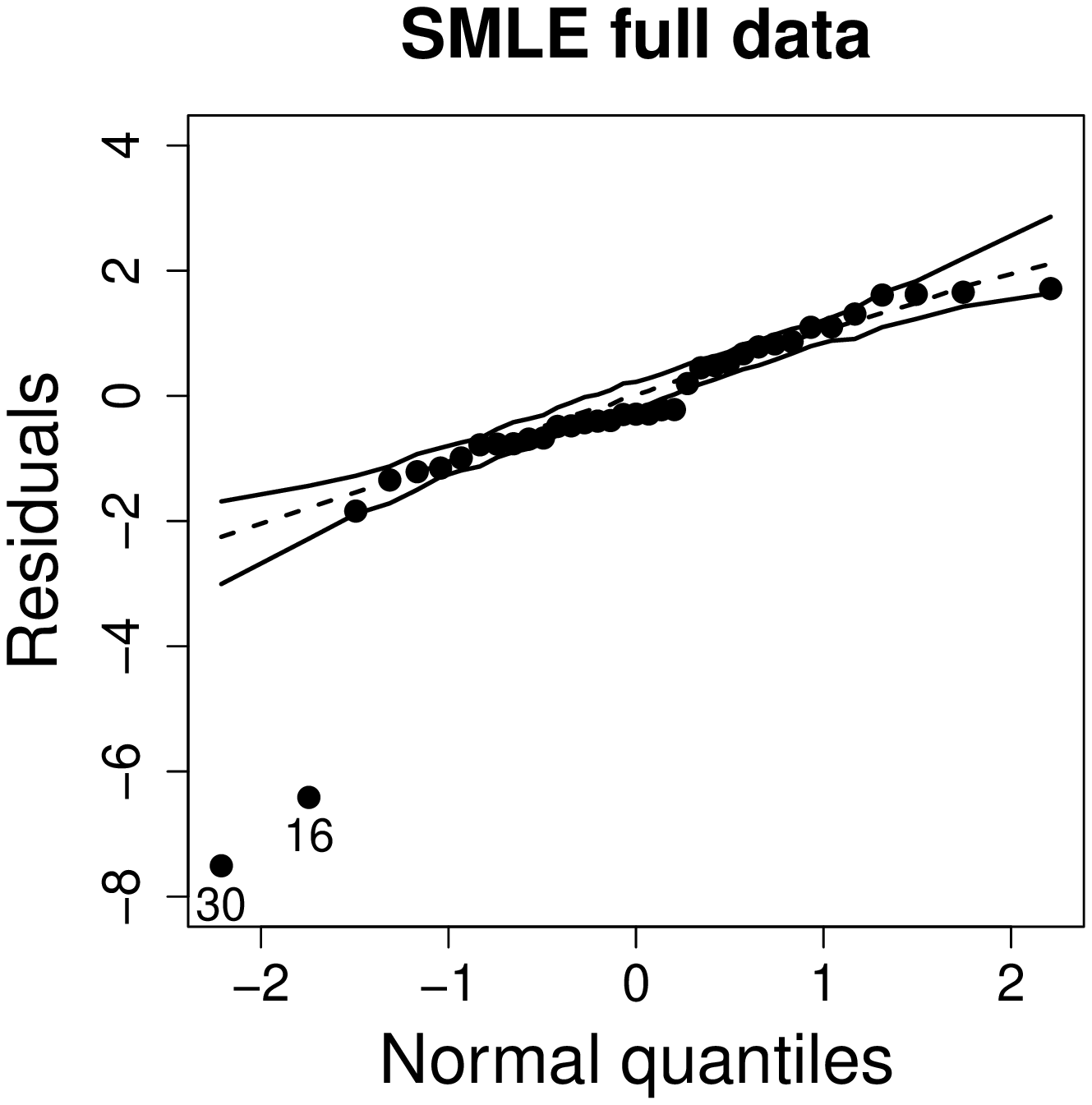}}
\qquad
\subfigure[][]{\includegraphics[width=5.0cm,height=4.5cm]{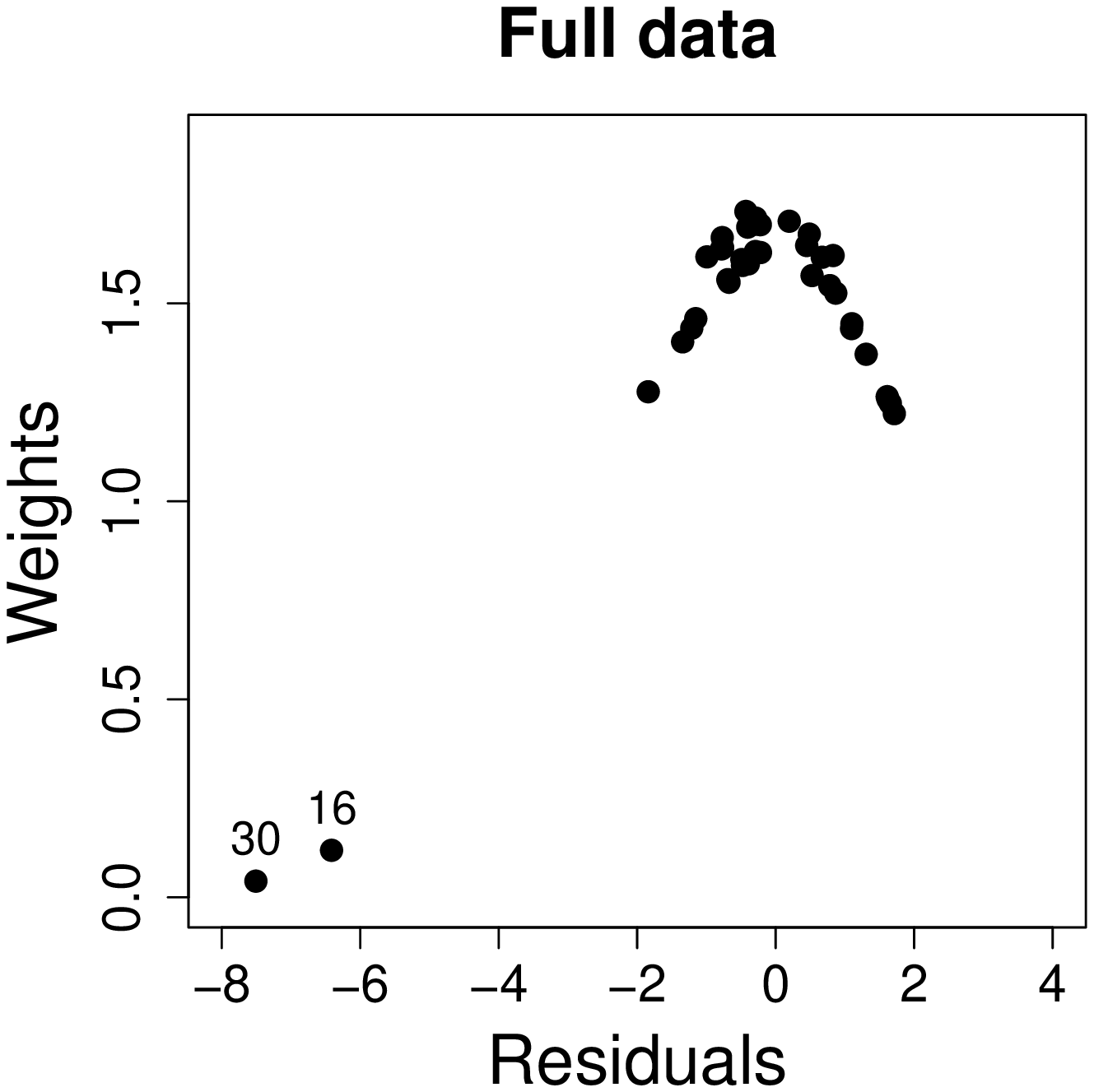}}
\caption{Normal probability plots of the residuals for MLE and SMLE and plot of estimated weights for SMLE -- AIS data.}
\label{ALL_ENVELOPES1_AIS}
\end{figure}

\subsection{Tuna dataset}

We consider a subset of the the data available in the Supplementary Material of \cite{monllor2017shift}  (https://doi.org/10.1371/journal.pone.0178196).
Here, the response variable is the tropical tuna percentage (TTP) in longliner catches and the covariate is the sea surface temperature (SST). The data correspond to 77 observations of longliner catches in different points of the southern Indian ocean in the year 2000. One of the observations on TTP equals 1, which means that 100\% of the catch is of tropical tuna, well above the second largest observed value, which is $0.35$. We have no information of whether this is a correct observation or not, and hence a robust model fit that is not much influenced by this observation is required. An additional difficulty is that beta regressions are not suited for data with observations at the boundary of the unit interval. A practical strategy is to subtract or add a small value to such observations in order to locate them inside the open interval $(0,1)$. Here, we replaced $1$ by $0.999$. Since the influence function of the MLE grows unbounded when an observation approaches zero or one, it may be highly influenced by observations near the boundaries. Recall that the SMLE has bounded influence function and, hence, it may be a suitable alternative to the MLE when observations at the boundary are present and slightly moved inside the open unit interval.

\begin{figure}[h]
\centering
\subfigure{\includegraphics[width=5cm,height=5cm]{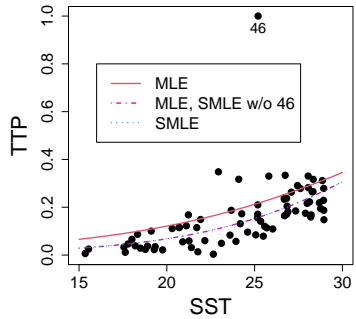}}
\caption{Scatter plots of TTP versus SST along with the fitted lines based on the MLE and SMLE for the full data and the data without an outlier; constant precision model -- Tuna data.}
\label{scatter_TUNA_DC_2}
\end{figure}

First, consider a constant precision beta regression model with $\log({\mu_i}/{(1-\mu_i)}) = \beta_1+\beta_2\mbox{SST}_i$ and $\log(\phi_i)=\gamma_1$, where  $\mu_i$ and $\phi_i$ are the mean and precision of TTP for the $i$-th coordinate, $i=1,\ldots,77$. Figure \ref{scatter_TUNA_DC_2} displays the scatter plot of TTP versus SST along with the MLE and SMLE fitted lines for the full data and the data without observation 46, that corresponds to the largest TTP value. A downward displacement of the MLE curve when observation 46 is excluded is apparent from the plot. The SMLE remains virtually unchanged with the exclusion of this data point.

\begin{figure}[!htb]
	\centering
	\subfigure{\includegraphics[width=5.0cm,height=4.5cm]{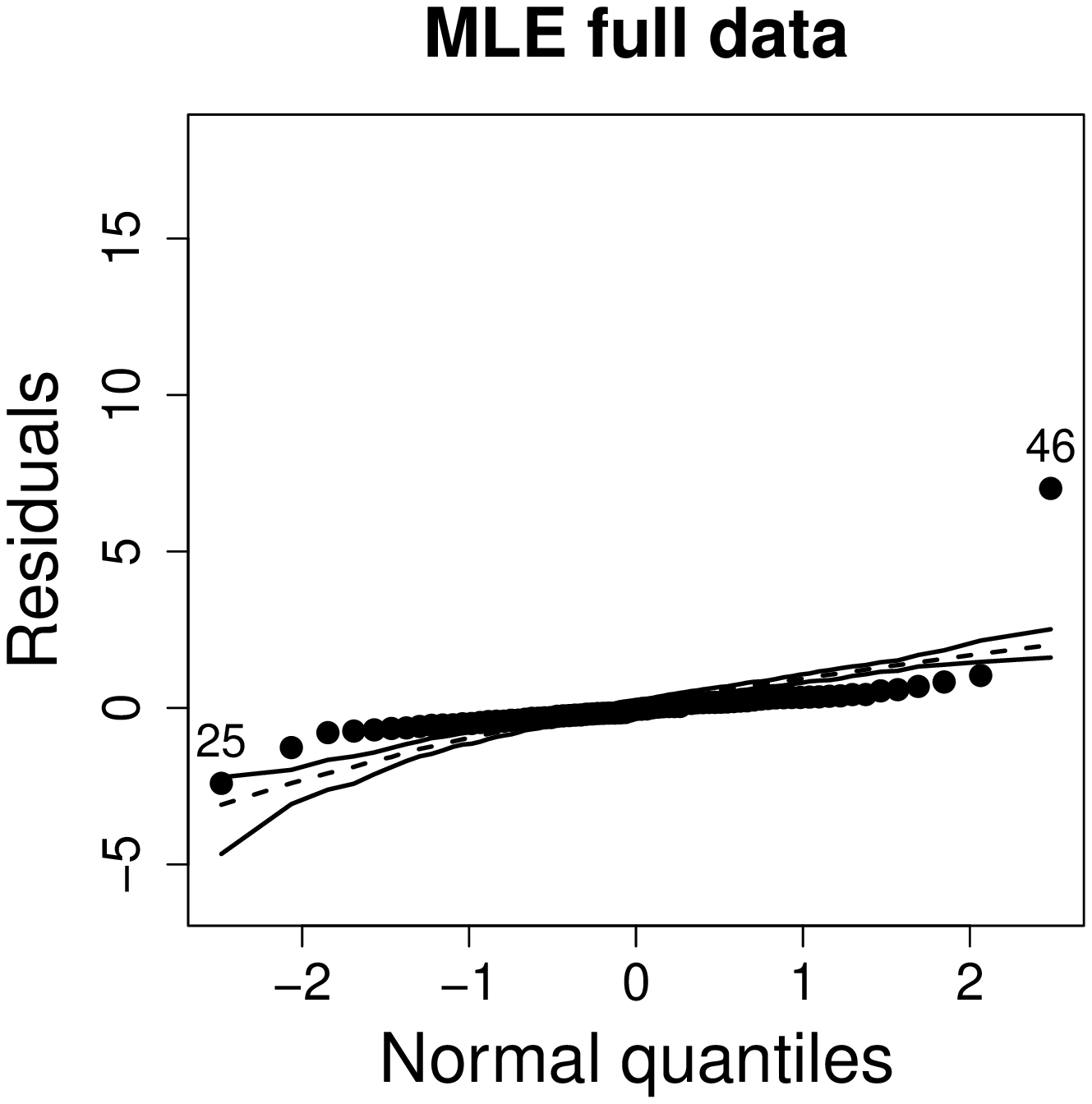}}
	\qquad
	\subfigure{\includegraphics[width=5.0cm,height=4.5cm]{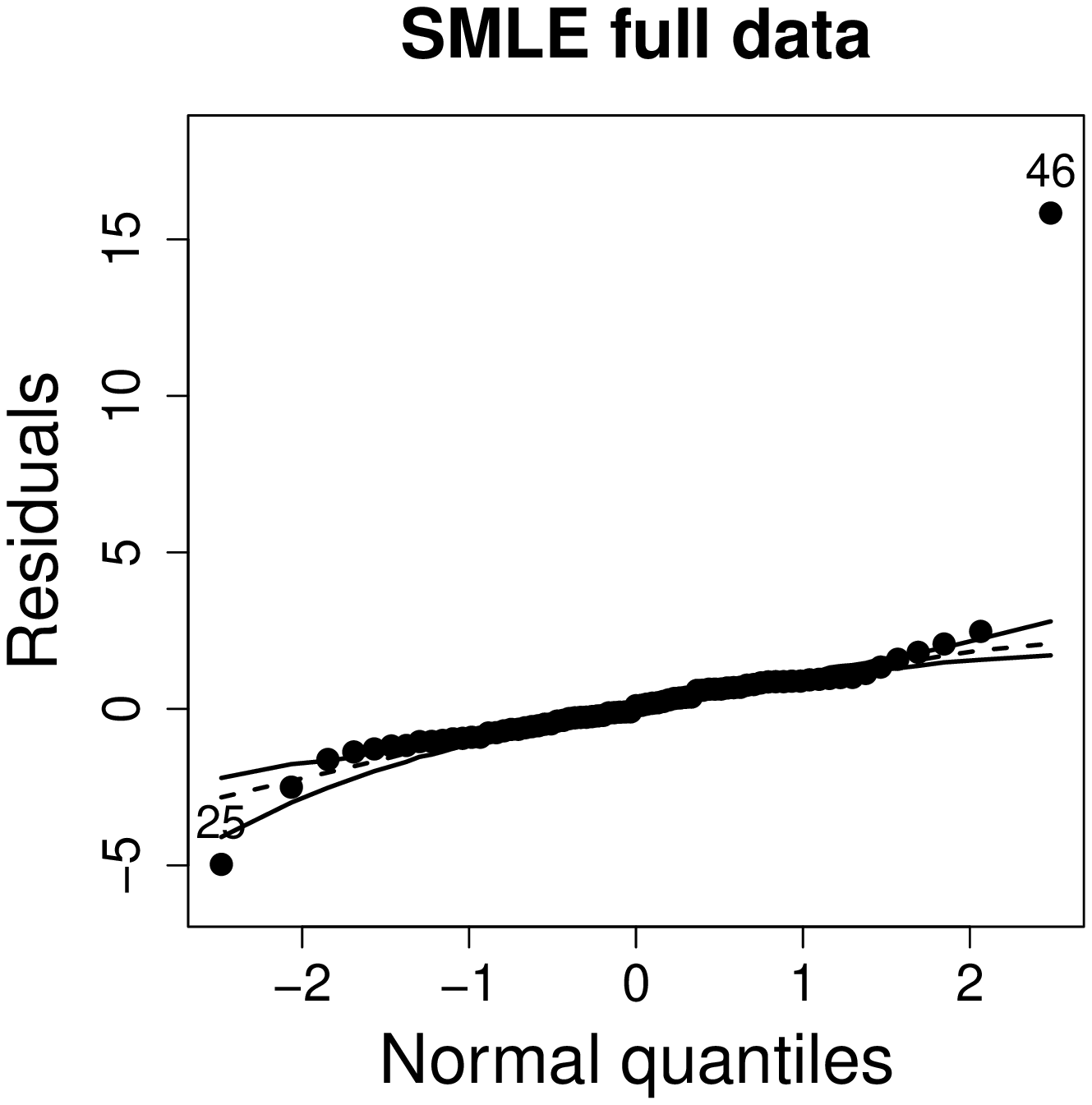}}
	\qquad
	\subfigure{\includegraphics[width=5.0cm,height=4.5cm]{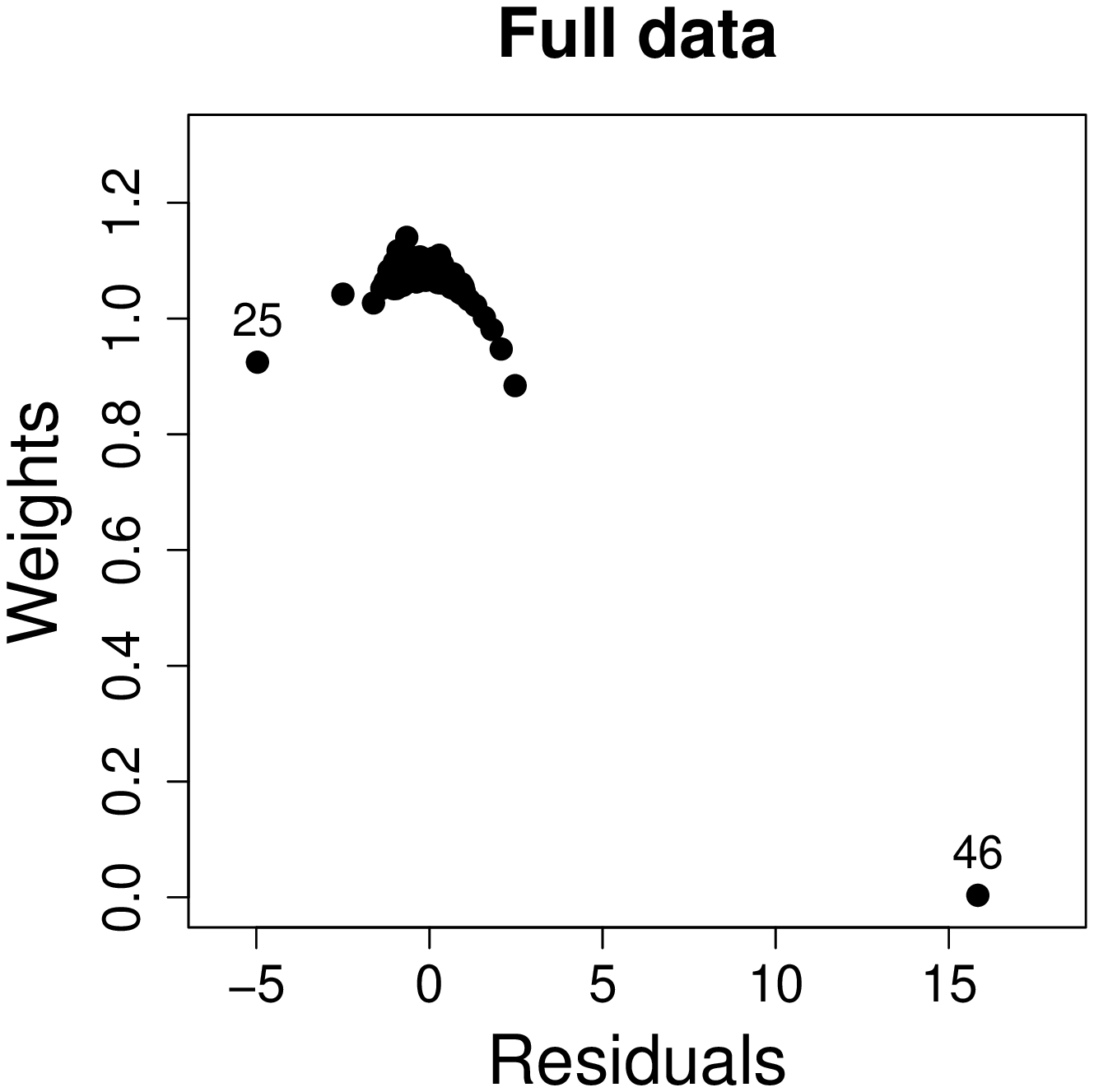}}
    \qquad
	\subfigure{\includegraphics[width=5.0cm,height=4.5cm]{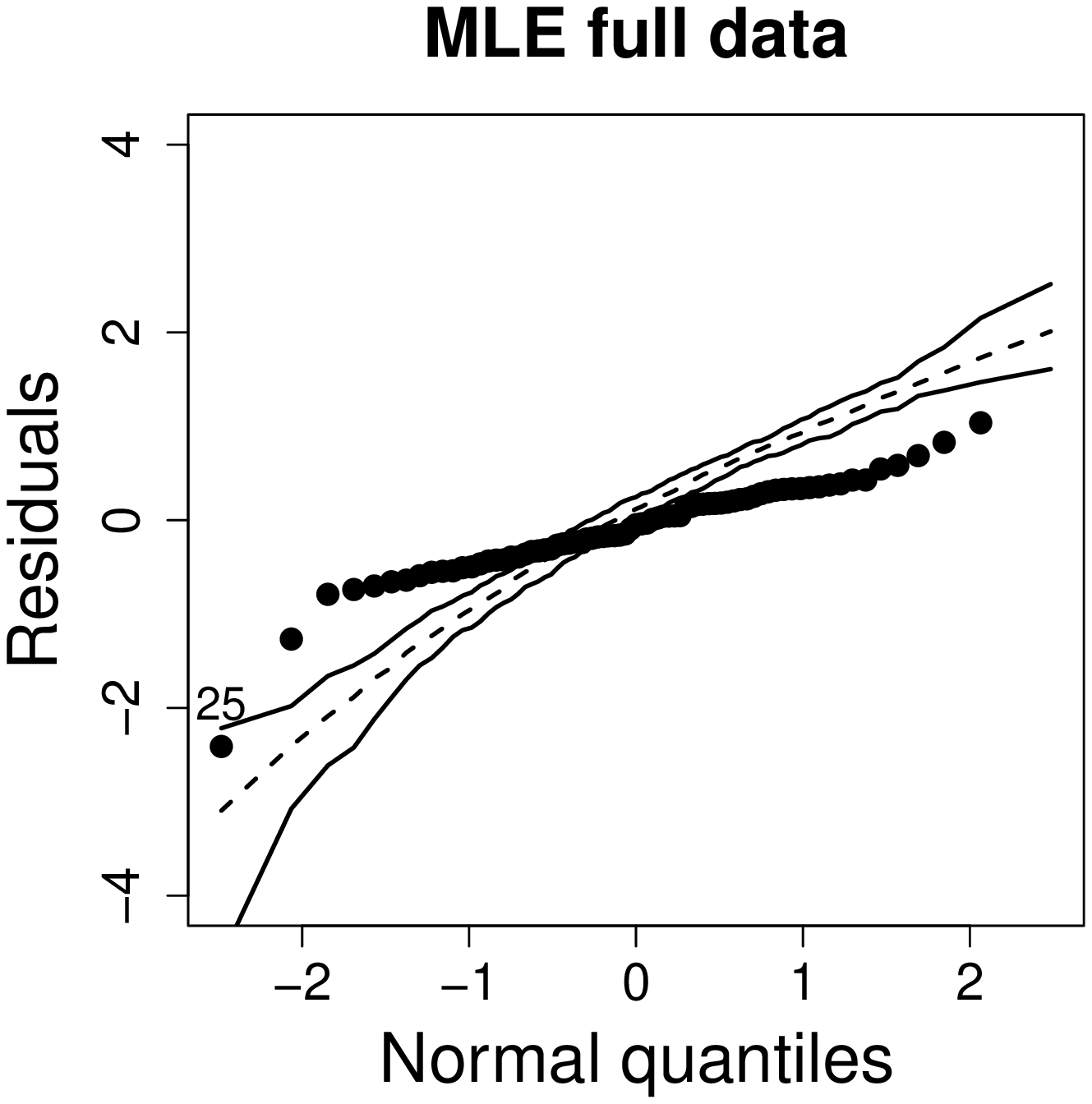}}
	\qquad
	\subfigure{\includegraphics[width=5.0cm,height=4.5cm]{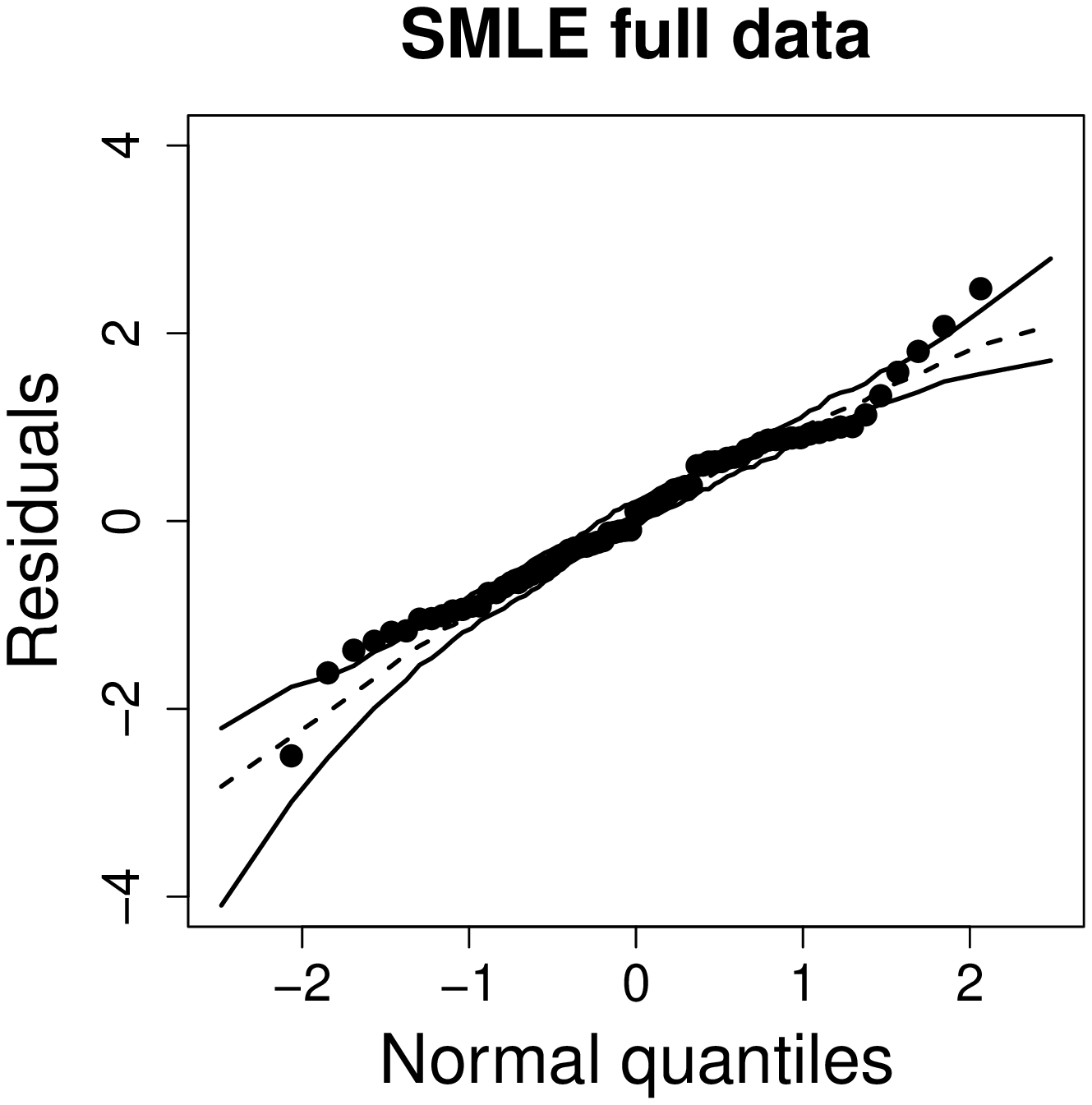}}
	\qquad
	\subfigure{\includegraphics[width=5.0cm,height=4.5cm]{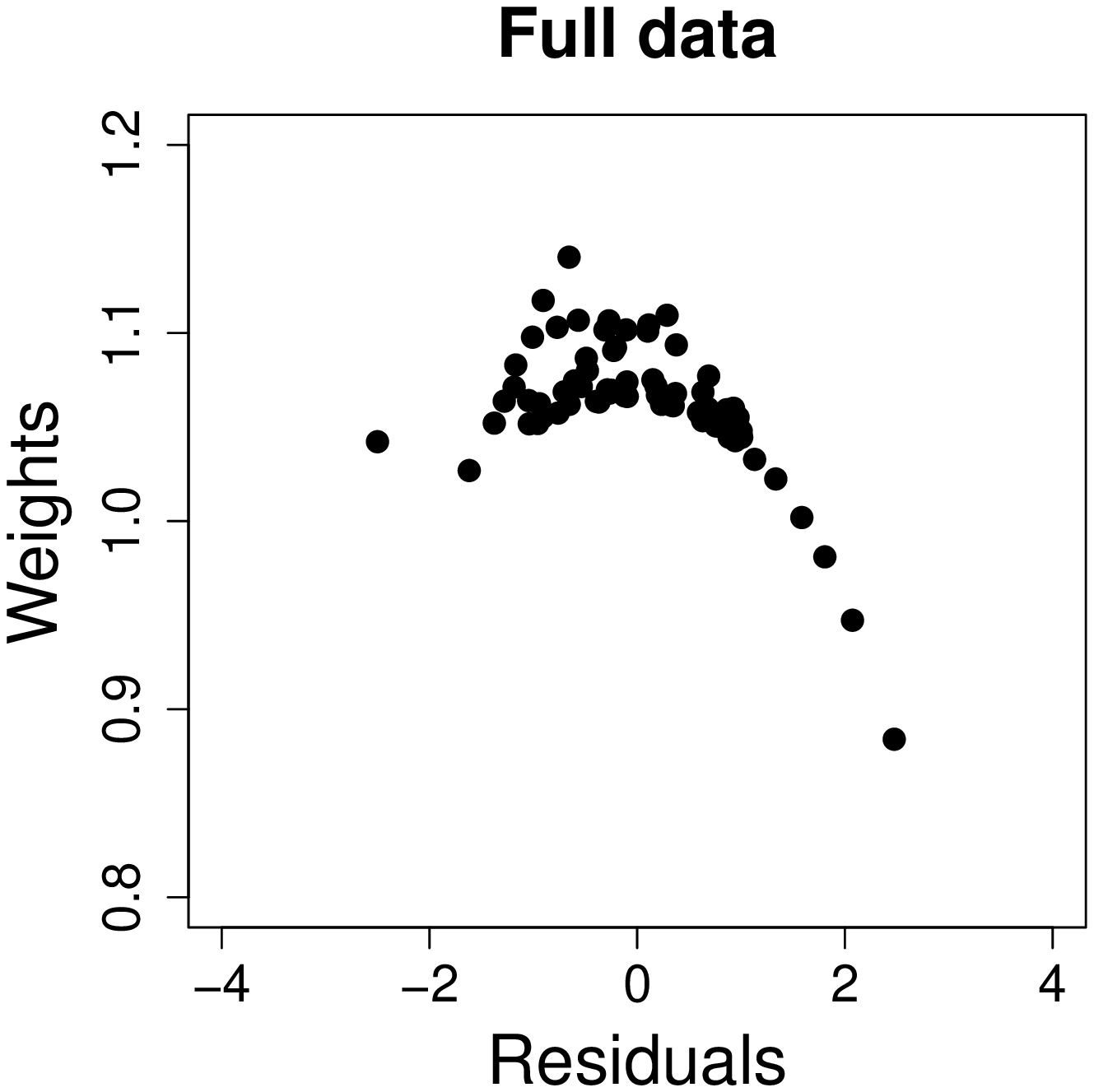}}
 	\caption{Normal probability plots of the residuals for MLE and SMLE and plot of estimated weights for SMLE; constant precision model -- Tuna data. The plots in the second row are zoomed versions of those in the first row.}
	\label{ALL_ENVELOPES_TUNA_DC}
\end{figure}

Figure \ref{ALL_ENVELOPES_TUNA_DC} shows normal probability plots of the residuals and estimated weights for the fitted models employing MLE and SMLE.
The panels in the left column give strong indication of lack of fit of the maximum likelihood estimation and highlights observation 46 as outlier. The
panels in the central column show reasonable fit of the $L_q$-likelihood estimation for all the observations except for case 46, that receives a weight close to zero. Results and plots not included here show that observation 25, which has the second largest absolute residual and falls slightly outside the envelope in the SMLE fit, does not have relevant impact in the MLE fit and, hence, its weight is close to the weights of the majority of the observations.

\begin{figure}[h]
\centering
\subfigure[][]{\includegraphics[width=5cm,height=5cm]{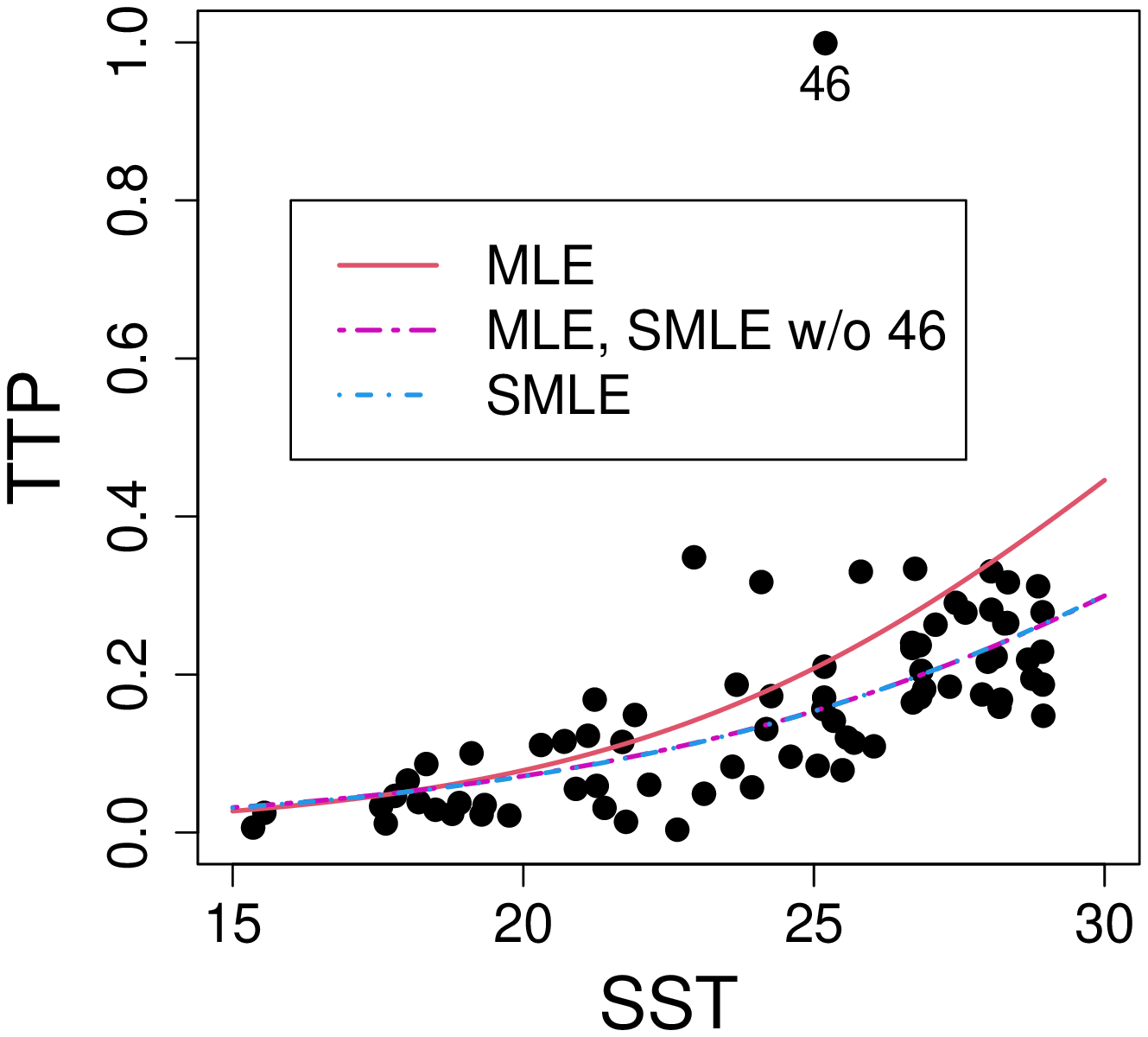}}
\qquad
\subfigure[]{\includegraphics[width=5cm,height=5cm]{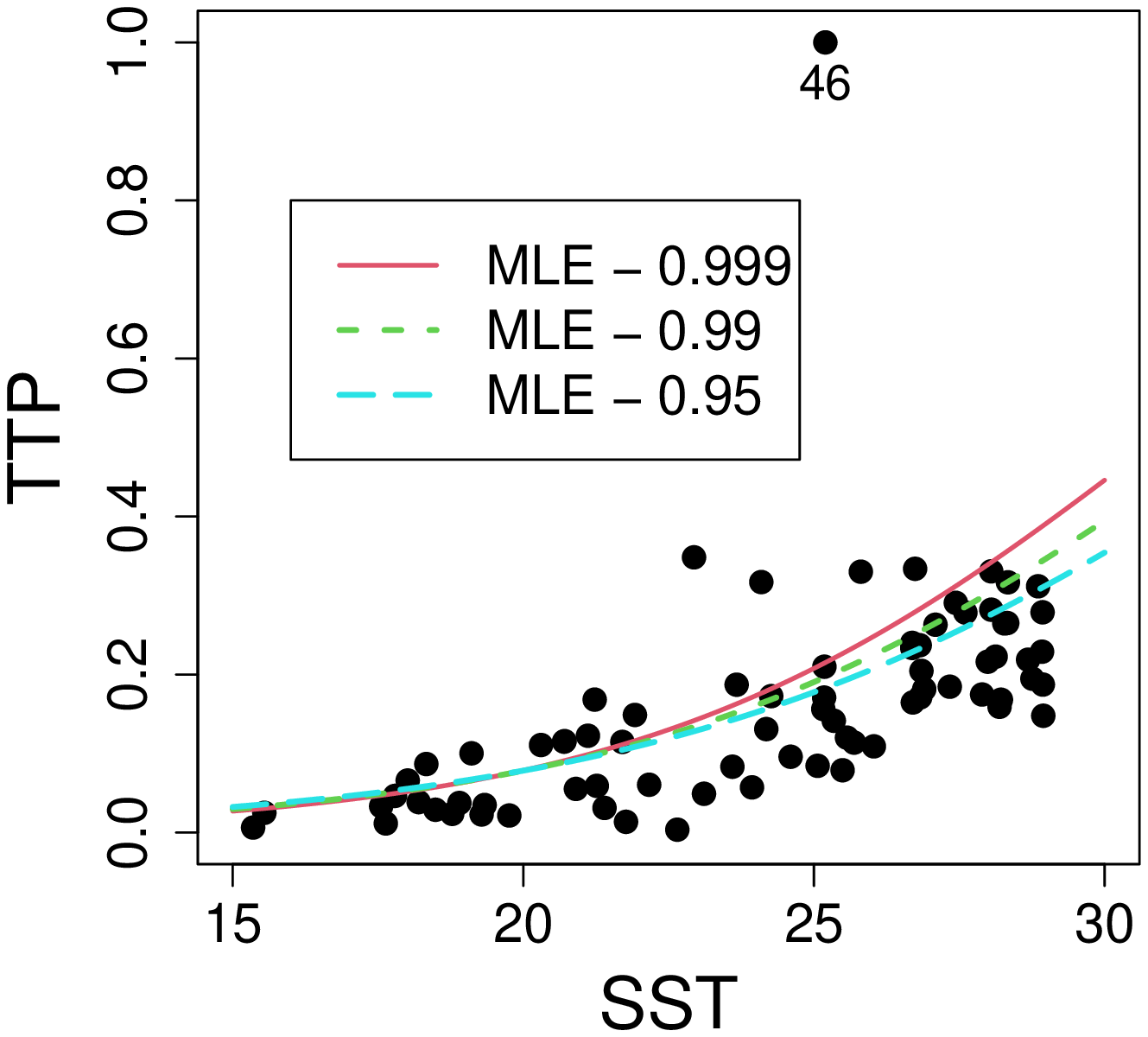}}
\qquad
\subfigure[][]{\includegraphics[width=5cm,height=5cm]{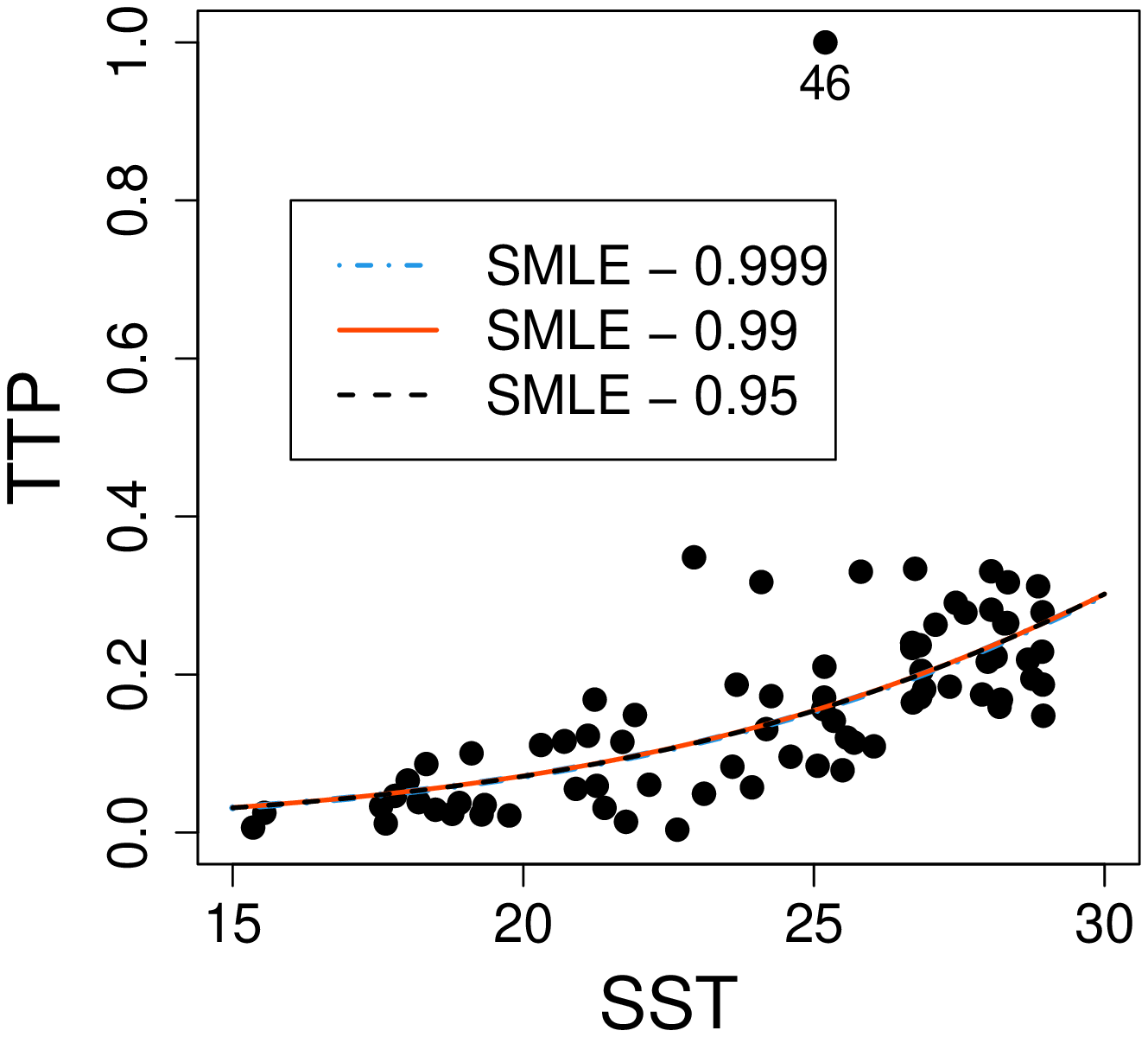}}
\caption{Scatter plots of TTP versus SST along with the fitted lines based on the MLE and SMLE for the full data and the data without an outlier; varying precision model -- Tuna data.}
\label{scatter_TUNA_DV_2}
\end{figure}

It is of interest to investigate whether the influence of observation 46 is weakened when the precision is modeled as a function of the covariate. A varying precision model might also produce better MLE fit. The beta regression model considered now assumes that $\log({\mu_i}{(1-\mu_i)}) =\beta_1+\beta_2\mbox{SST}_i,$ and $\log(\phi_i)=\gamma_1 + \gamma_2\mbox{SST}_i.$ The optimum value of the tuning constant of the SMLE is $0.96$ for the full data and 1 for the reduced data. Hence, when observation 46 is deleted, the SMLE and the MLE coincide. Figure \ref{scatter_TUNA_DV_2}(a) displays the scatter plot of TTP versus SST along with the fitted lines based on the SMLE and MLE for the full data and the data without observation 46. As in the constant precision modeling, this observation has disproportionate influence in the MLE fit and virtually does not affect the robust estimation.

Table \ref{estimates_TUNA_DV} shows the estimates, standard errors, $z$-statistics (estimate divided by standard error) and $p$-values of the Wald-type test of nullity of coefficients for the full data and the reduced data. Note the substantial change caused by observation 46 in the estimate of the intercept of the precision submodel when the MLE is employed; it moves the estimate from $7.022$ to $2.171$, a relative change of $69\%$. For the SMLE the estimates are, respectively, $2.452$ and $2.171$; the later coincides with the MLE. The maximum likelihood estimated coefficient of SST in the precision submodel is $-0.213$ (bootstrap $p$-value=$0.000$) for the full data and $0.047$ (bootstrap $p$-value=$0.338$) for the data without observation 46. Note the sign change in the estimate. Also, it is clear that a single observation causes a relevant change in an inferential conclusion: for the full data the coefficient of SST in the precision submodel is highly significant, while the MLE points to a constant precision model when the outlier is deleted. The corresponding SMLE estimates are $0.036$ (bootstrap $p$-value=$0.414$) for the full data and $0.047$ (bootstrap $p$-value=$0.304$) for the data without observation 46. In both cases, there is no evidence to reject the constant precision model.

We now investigate the sensitivity of the model fits to the choice of the replacement value for the response of observation 46. Recall that we replaced 1 by $0.999$. Figure \ref{scatter_TUNA_DV_2}(b)-(c) show the MLE and SMLE fitted curves for three choices of value for TTP$_{46}$, namely $0.999$, $0.99$, and $0.95$. While the SMLE fitted curves remain unchanged, the MLE fit is markedly influenced by the chosen value. Moreover, the MLE of the SST coefficient in the precision submodel moves from $-0.213$ (bootstrap $p$-value=$0.000$) when TTP$_{46}=0.999$ to $-0.094$ (bootstrap $p$-value=$0.030$) when TTP$_{46}=0.95$, weakening the significance of SST in the precision submodel.

\begin{table}[!htb]
	\centering
	\caption{Estimates, standard errors,  $z$-stat and $p$-values (bootstrap $p$-values between parenteses) for beta regression with varying precision -- Tuna data.}\vspace{0.2cm}
	\label{estimates_TUNA_DV}
	\scalefont{0.7}
\def\arraystretch{0.65} \begin{tabular}{>{\raggedleft\arraybackslash}m{2.1cm}>{\raggedleft\arraybackslash}m{1.2cm}>{\raggedleft\arraybackslash}m{1.2cm}>{\raggedleft\arraybackslash}m{1.2cm}>{\raggedleft\arraybackslash}m{1.7cm}>{\raggedleft\arraybackslash}m{0.05cm}>{\raggedleft\arraybackslash}m{1.2cm}>{\raggedleft\arraybackslash}m{1.2cm}>{\raggedleft\arraybackslash}m{1.2cm}>			{\raggedleft\arraybackslash}m{1.7cm}}\hline
		& \multicolumn{4}{c}{MLE for the full dataset}                                   && \multicolumn{4}{c}{SMLE for the full dataset} \\  \cline{2-5}\cline{7-10}
		                 & Estimate&Std. error&$z$-stat &$p$-value                       && Estimate  &Std. error &$z$-stat  &$p$-value    \\ \cline{2-5}\cline{7-10}
		{\it mean submodel}&&&&&&&&\\
		Intercept        & $-6.945$  &  $0.687$  & $-10.114$  &  -                       && $-6.022$  &  $0.521$  & $-11.557$ &  -       \\
		SST              &  $0.224$  &  $0.029$  &  $7.815$   &  $0.000$ ($0.000$)       &&  $0.173$  &  $0.020$  &  $8.637$  &  $0.000$ ($0.000$)\\
	    {\it precision submodel}&&&&&&&&\\	
		Intercept        &  $7.022$  &  $1.046$  &  $6.711$  &  -                        &&  $2.452$  &  $1.051$  &  $2.333$ &  -     \\
		SST              & $-0.213$  &  $0.042$  & $-5.066$  &  $0.000$ ($0.000$)        &&  $0.036$  &  $0.043$  &  $0.837$ &   $0.402$  ($0.414$) \\\cline{2-5}\cline{7-10}
				
& \multicolumn{4}{c}{MLE without observation 46}                                         && \multicolumn{4}{c}{SMLE without observation 46} \\   \cline{2-5}\cline{7-10}
	                     & Estimate&Std. error&$z$-stat&$p$-value                        && Estimate  &Std. error &  $z$-stat&  $p$-value    \\  \cline{2-5}\cline{7-10}
		{\it mean submodel}&&&&&&&&\\
		Intercept        & $-5.990$  &  $0.530$  & $-11.307$ &  -                        && $-5.990$  &  $0.530$  & $-11.307$ &  - \\
		SST              &  $0.171$  &  $0.020$  &  $8.464$  &  $0.000$ ($0.000$)        &&  $0.171$  &  $0.020$  &  $8.464$  &  $0.000$ ($0.000$) \\
        {\it precision submodel}&&&&&&&&\\
		Intercept        &  $2.171$  &  $1.047$  &  $2.073$  &  -                        &&  $2.171$  &  $1.047$  &  $2.073$  &  - \\
		SST              &  $0.047$  &  $0.043$  &  $1.099$  &  $0.272$  ($0.338$)       &&  $0.047$  &  $0.043$  &  $1.099$  &  $0.272$  ($0.304$) \\ \hline		
	\end{tabular}
\end{table}

\begin{figure}[!htb]
	\centering
	\subfigure{\includegraphics[width=5.0cm,height=4.5cm]{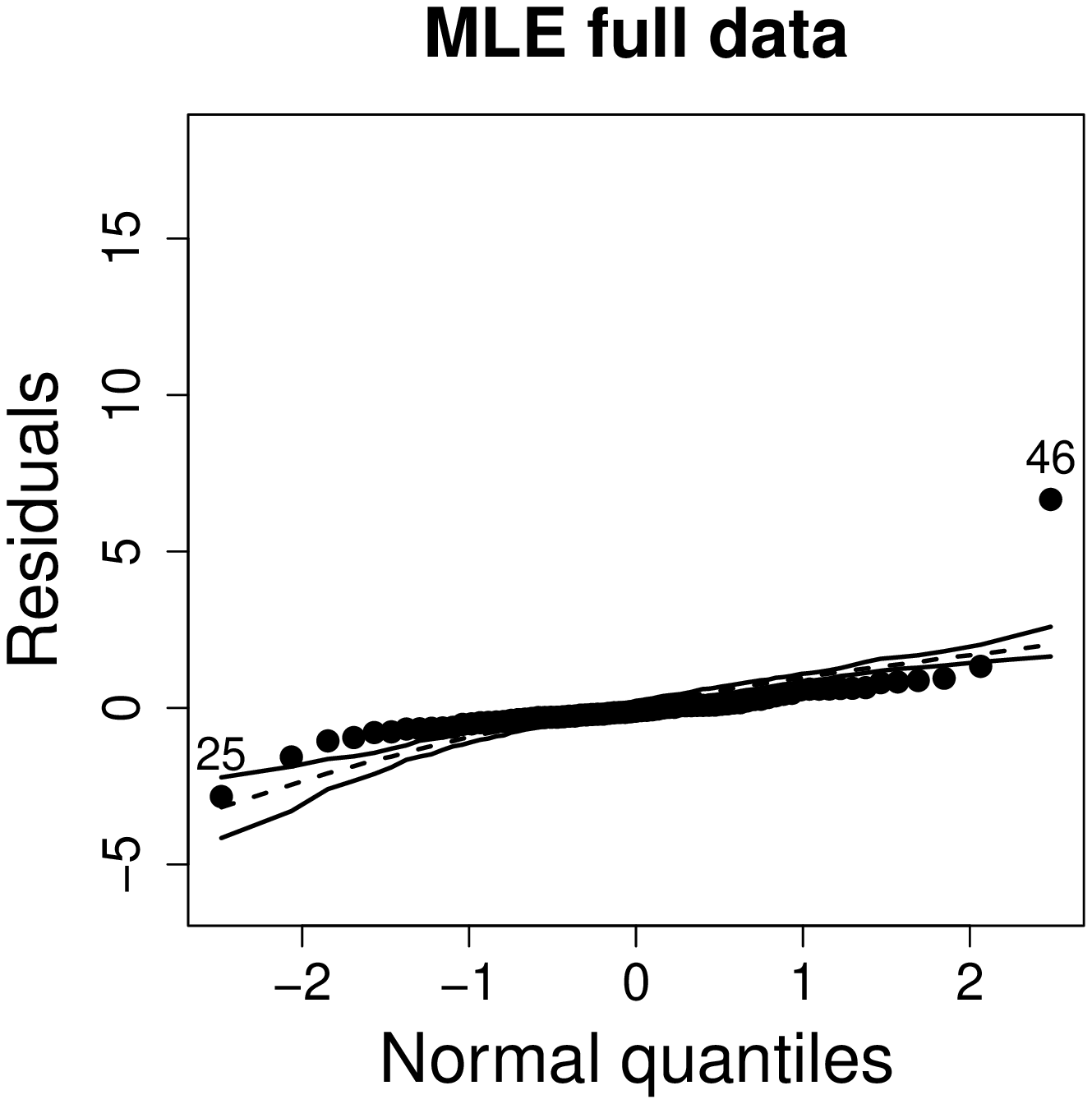}}
	\qquad
	\subfigure{\includegraphics[width=5.0cm,height=4.5cm]{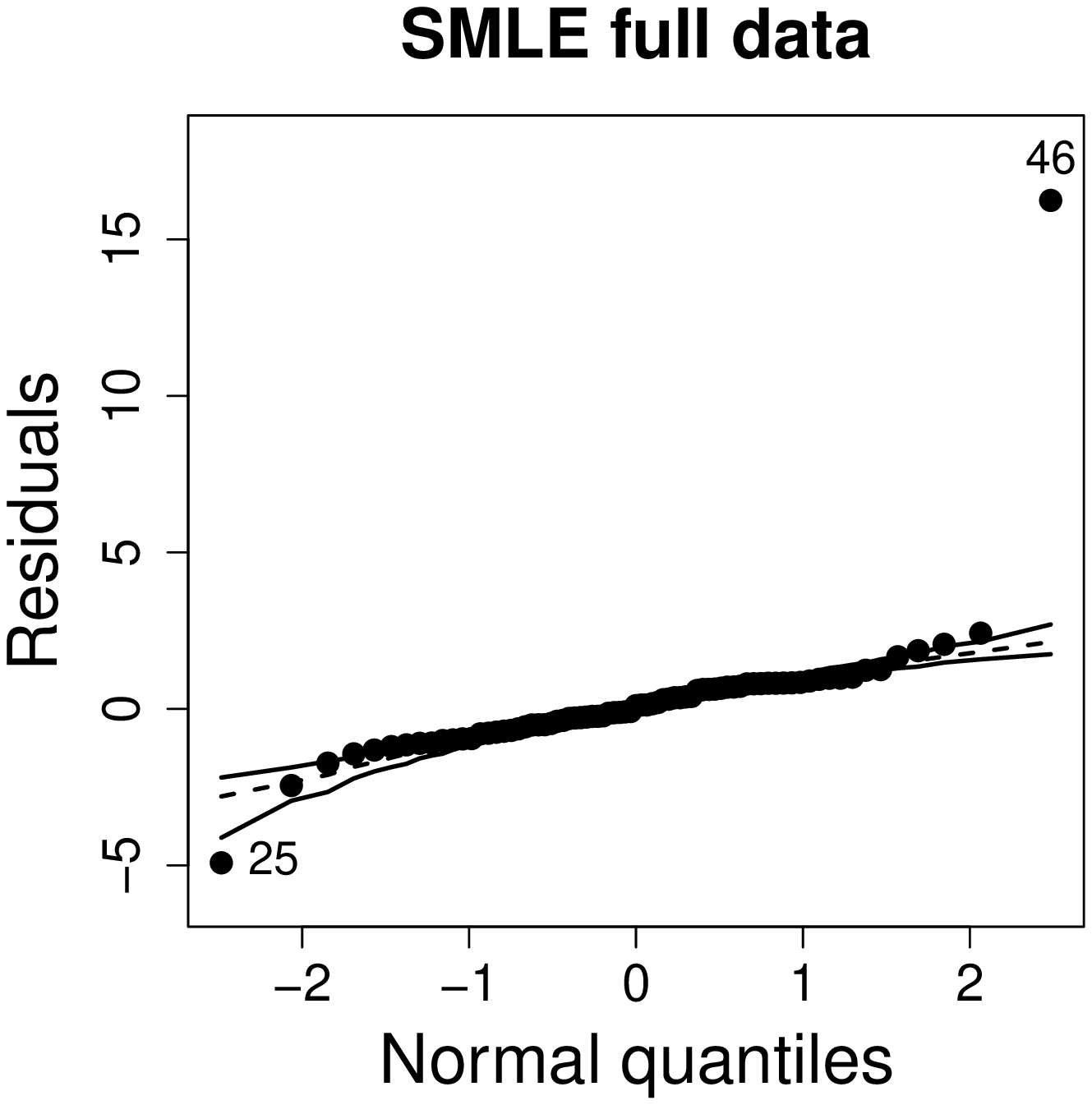}}
	\qquad
	\subfigure{\includegraphics[width=5.0cm,height=4.5cm]{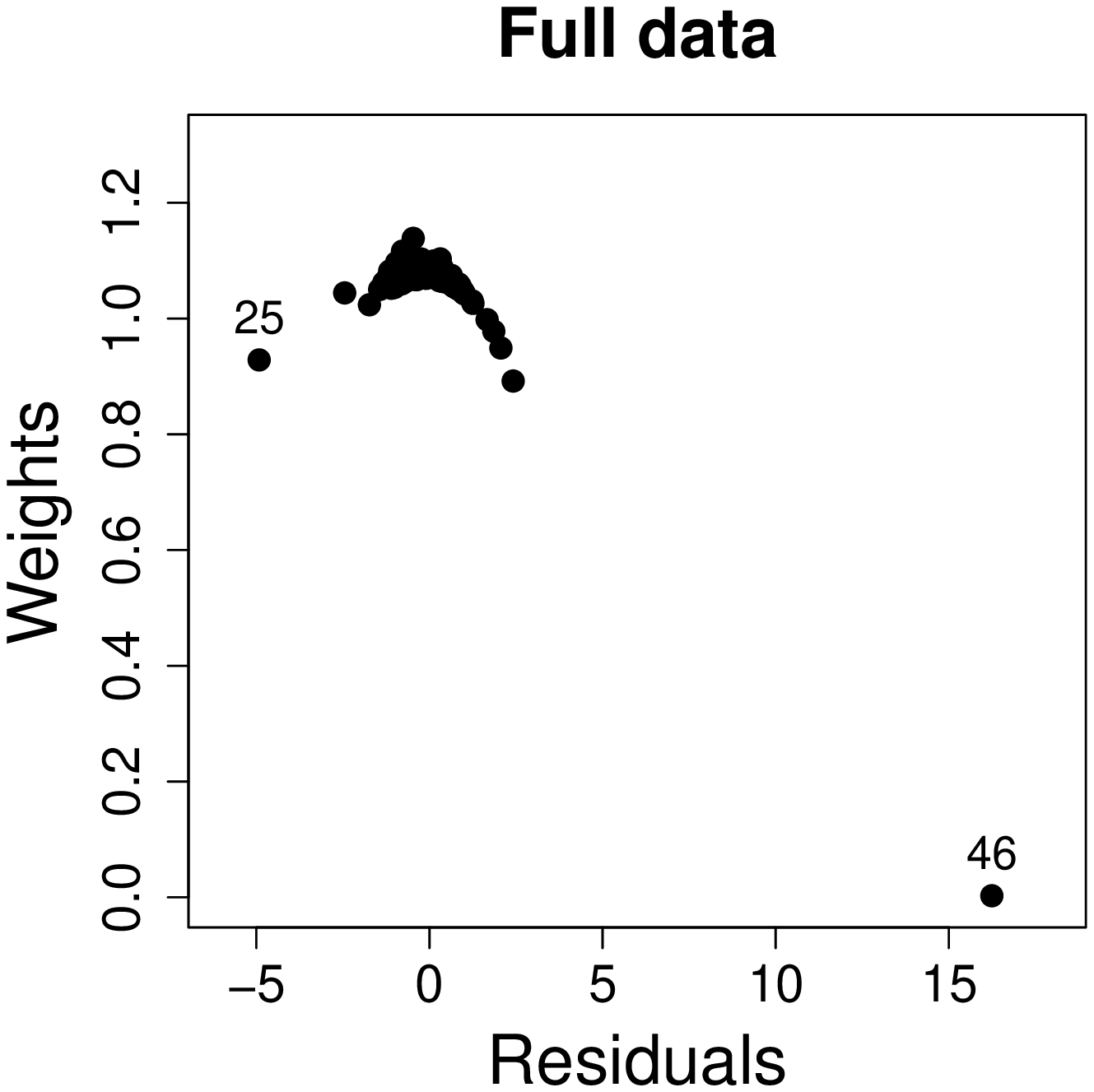}}
    \qquad
	\subfigure{\includegraphics[width=5.0cm,height=4.5cm]{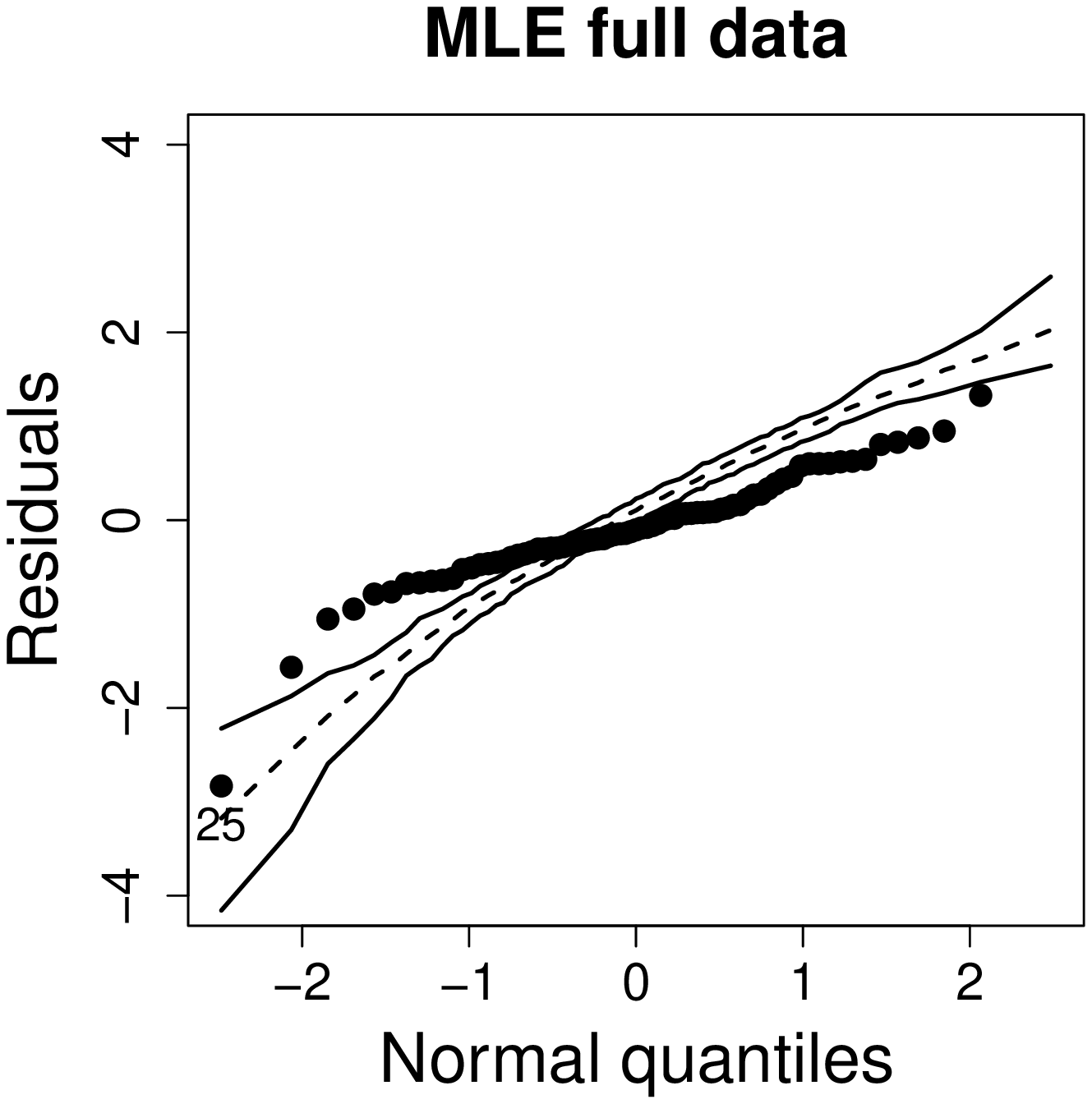}}
	\qquad
	\subfigure{\includegraphics[width=5.0cm,height=4.5cm]{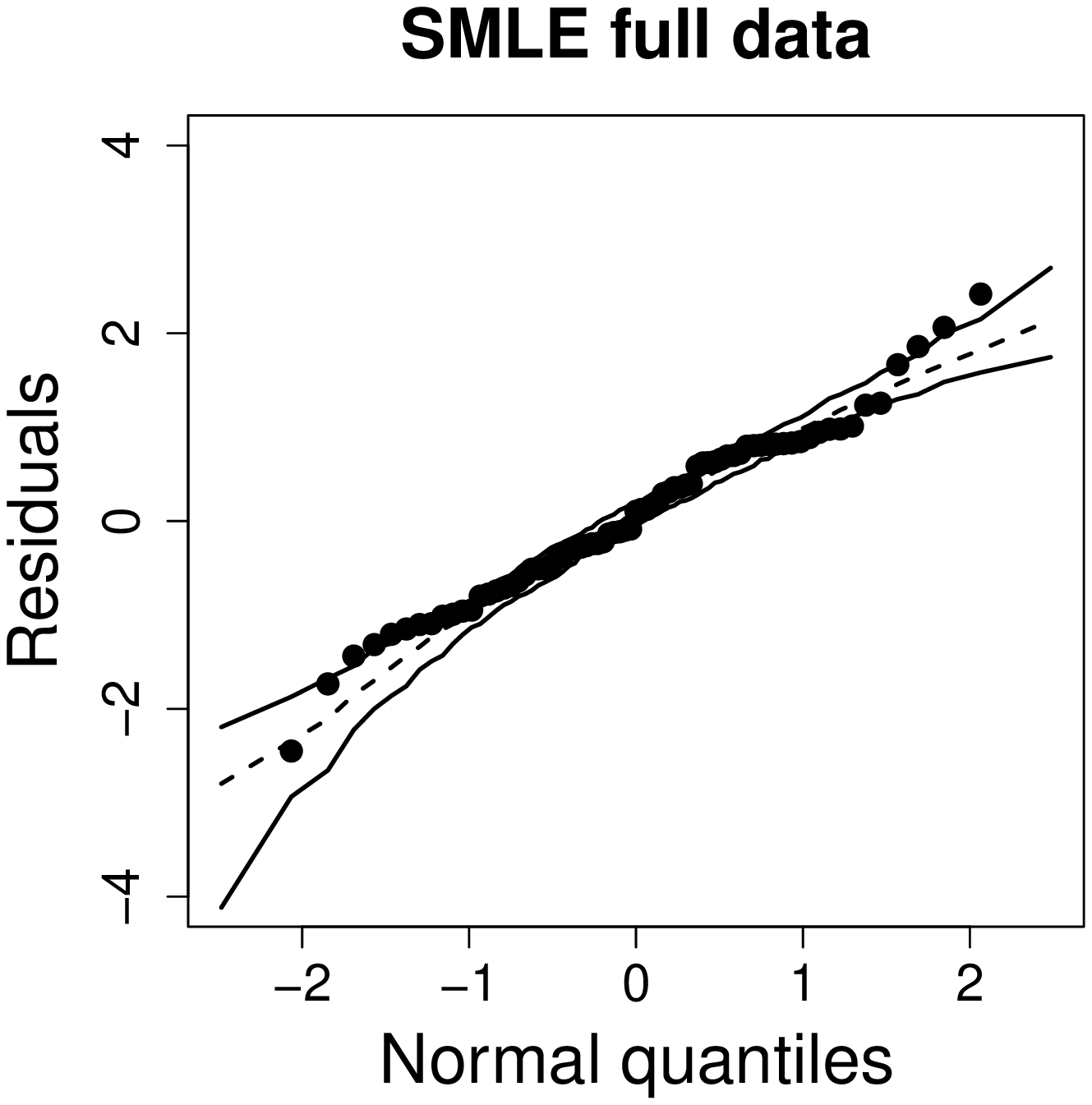}}
	\qquad
	\subfigure{\includegraphics[width=5.0cm,height=4.5cm]{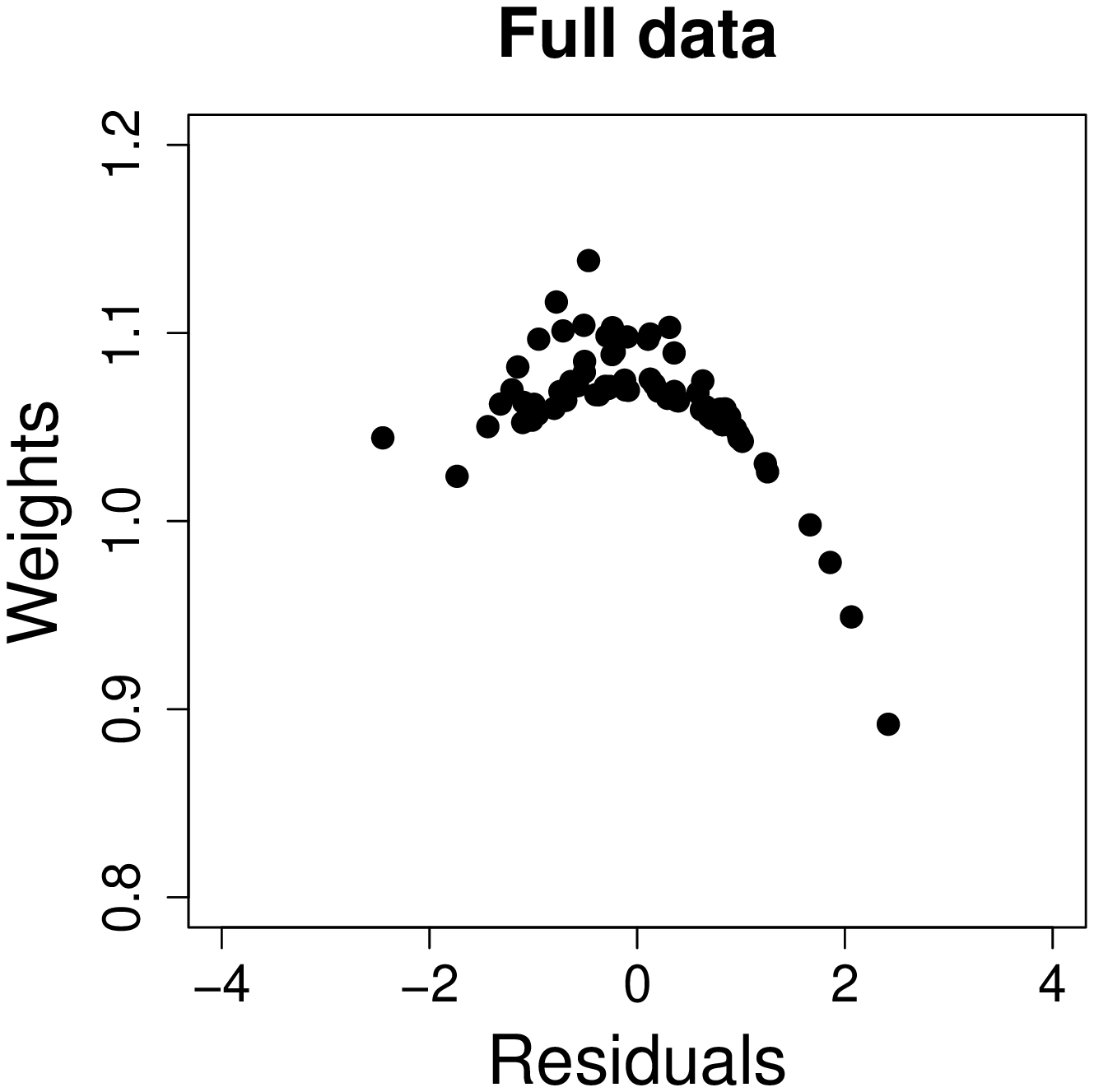}}
    \qquad
	\caption{Normal probability plots of the residuals for MLE and SMLE and plot of estimated weights for SMLE; varying precision model -- Tuna data. The plots in the second row are zoomed versions of those in the first row.}
\label{ALL_ENVELOPES_TUNA_DV}
\end{figure}

Figure \ref{ALL_ENVELOPES_TUNA_DV} show plots of residuals and estimated weights for the MLE and SMLE fits of the varying precision beta regression   model. The plots are similar to the corresponding plots of the constant precision model fits (Figure \ref{ALL_ENVELOPES_TUNA_DC}). We conclude that allowing the precision to be modeled as a function of SST does not improve the MLE fit and that observation 46 remains highly influential. A constant precision beta regression model coupled with the maximum reparameterized $L_q$-likelihood estimation we propose lead to a robust and suitable fit to the data.

\subsection{The Firm cost data}\label{FirmCost}

\begin{figure}[h]
\centering
\subfigure[][]{\includegraphics[width=5cm,height=5cm]{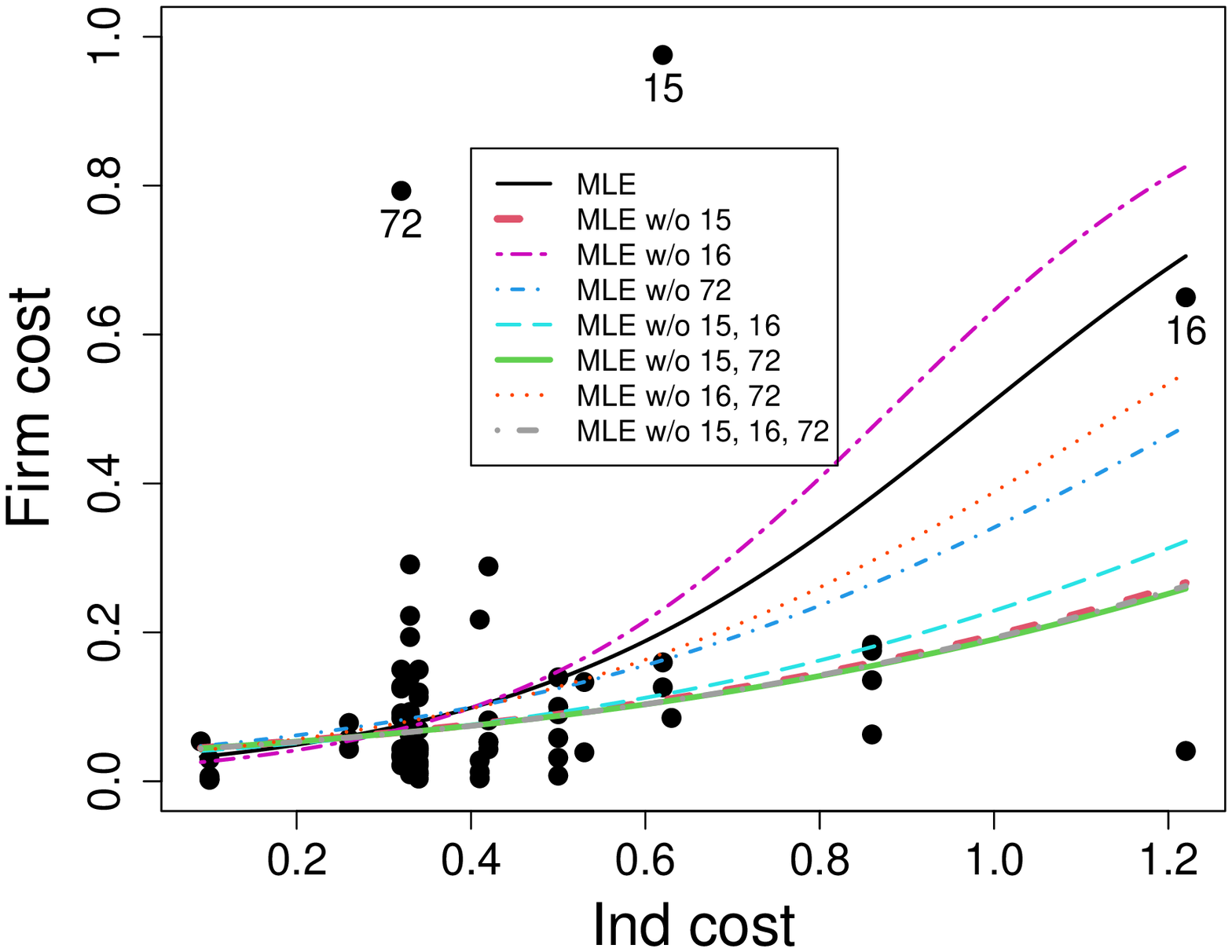}}
\qquad
\subfigure[][]{\includegraphics[width=5cm,height=5cm]{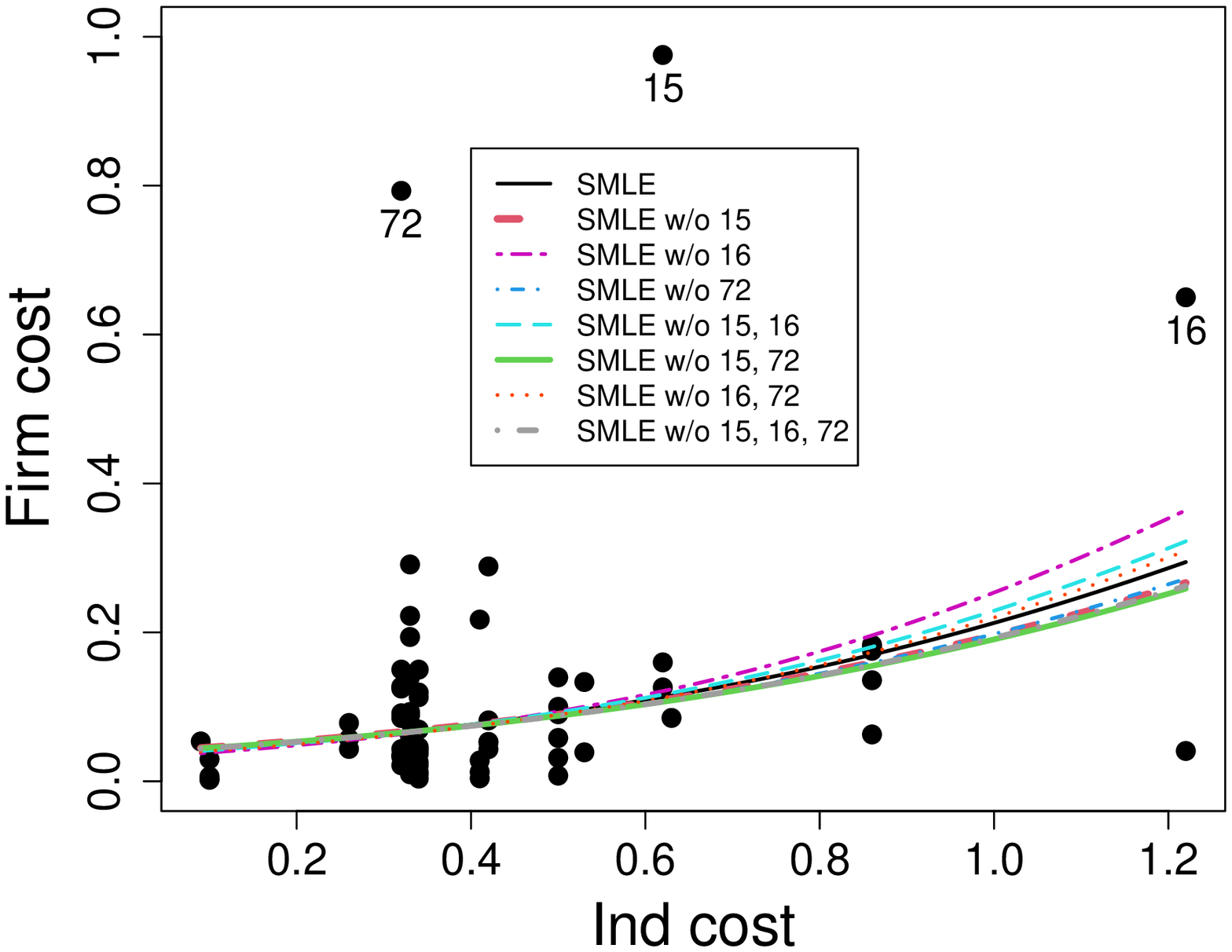}}
\\
\subfigure[][]{\includegraphics[width=5cm,height=5cm]{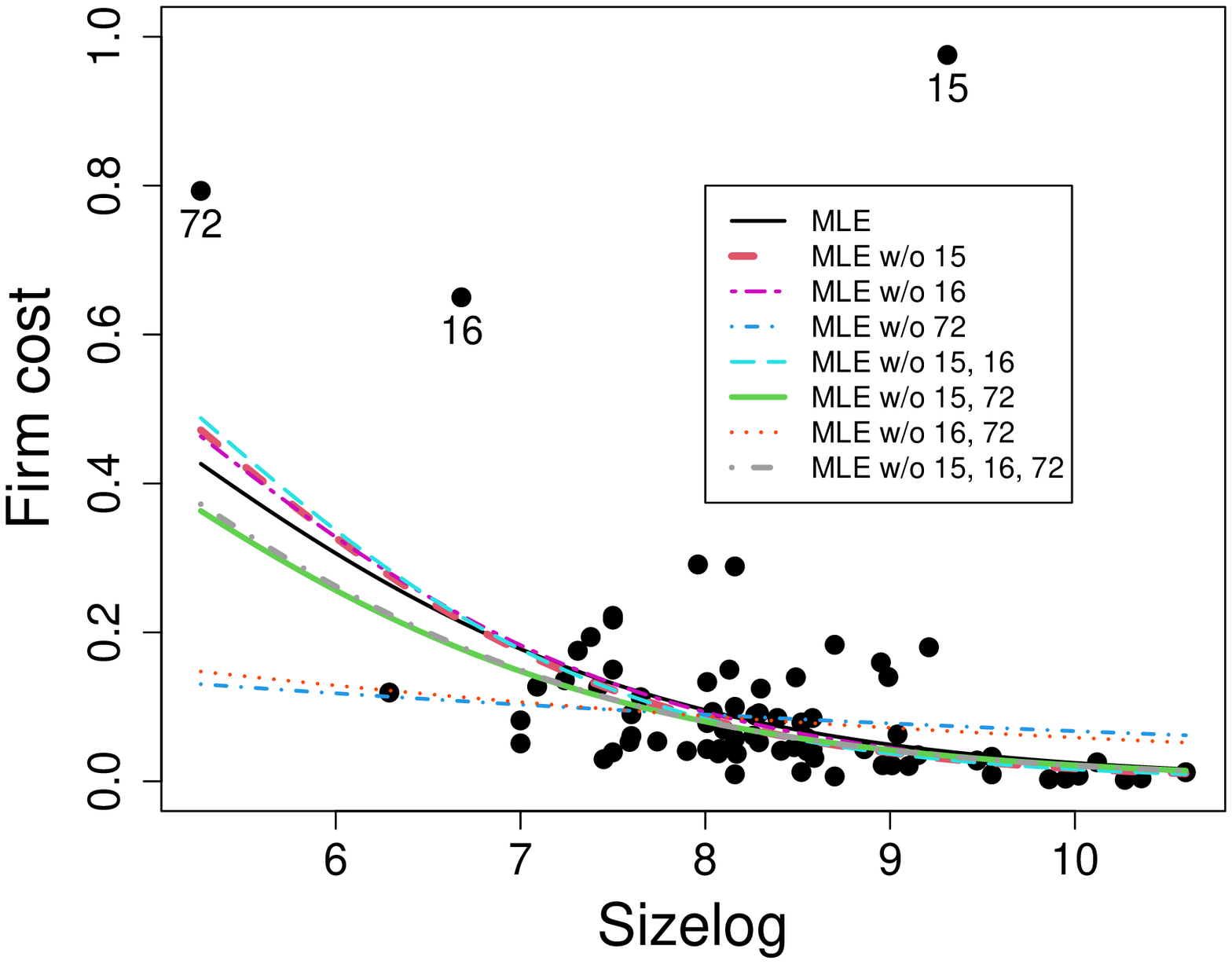}}
\qquad
\subfigure[][]{\includegraphics[width=5cm,height=5cm]{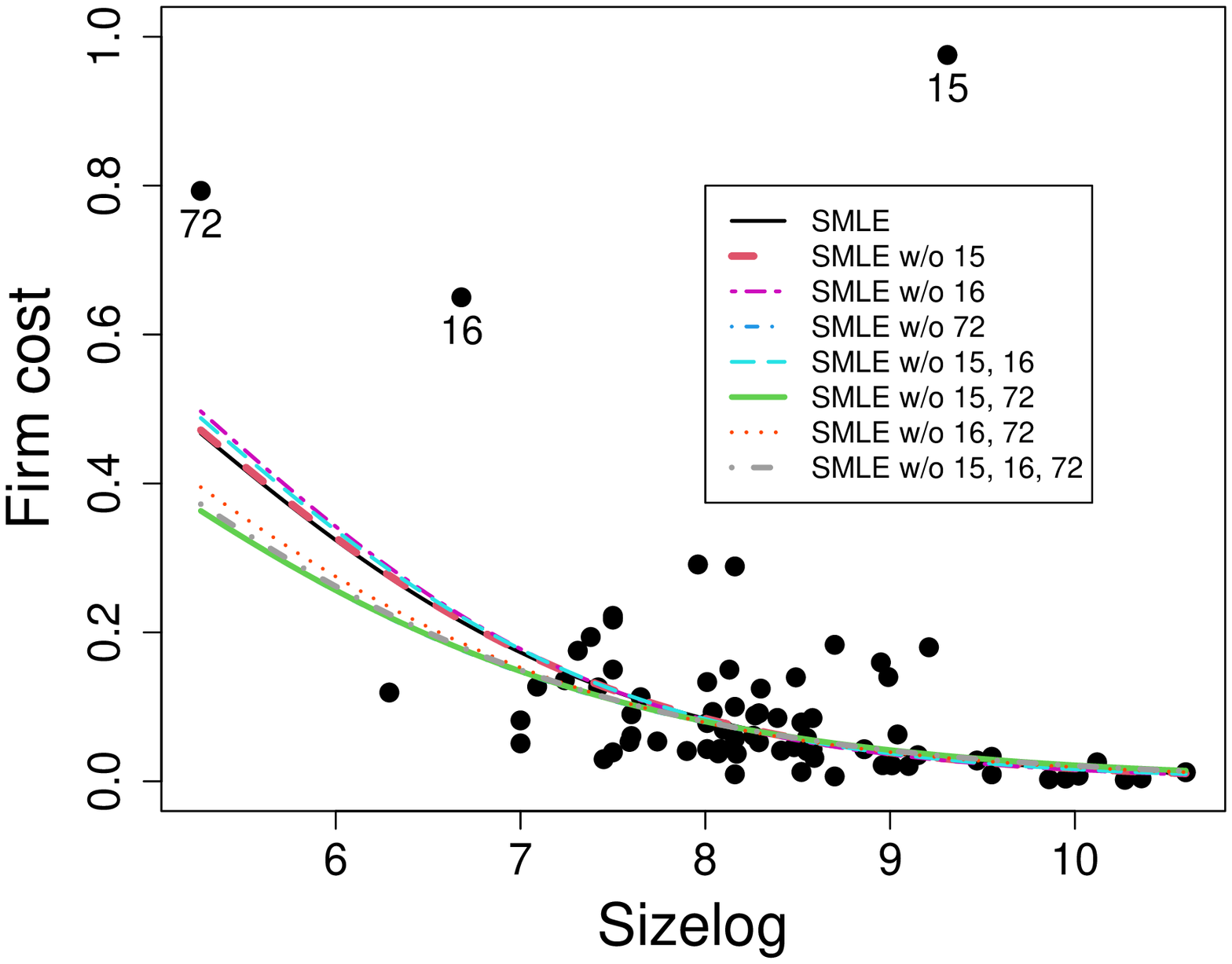}}
\caption{Scatter plots for Firm cost dataset along with the fitted lines based on the MLE (a)-(c) and SMLE (b)-(d) for the full data and the data without outliers.}
\label{scatter_Firmcost}
\end{figure}

We now turn to the application introduced in Section \ref{Int}.\footnote{Dataset taken from the personal page of Dr Edward W. Frees: \\ \mbox{https://instruction.bus.wisc.edu/jfrees/jfreesbooks/Regression\%20Modeling/BookWebDec2010/CSVData/RiskSurvey.csv}.} Consider a beta regression model with
	$\log(\mu_i/(1-\mu_i)) =\beta_1+\beta_2\mbox{Ind cost}_i +\beta_3\mbox{Size log}_i$ and
	$\log(\phi_i)=\gamma_1+\gamma_2\mbox{Ind cost}_i+\gamma_3\mbox{Size log}_i,$
where  $\mu_i$ and $\phi_i$ are the mean and precision of the response variable Firm cost for the $i$-th firm, $i=1,\ldots,73$. We fitted the model using the MLE and the SMLE for the full data and the data without combinations of the most eye-catching outliers, namely observations 15, 16, and 72. The optimal value selected for the tuning constant $q$ is $0.96$ for the full data and $0.94$ for the data without observations 16 or/and 72. When observation 15 is excluded, either individually or with one or both other outliers, the selected $q$ turns to be 1, i.e., in these cases, the SMLE and the MLE coincide. Figure \ref{scatter_Firmcost} displays the scatter plots of Firm cost versus the covariates along with the fitted lines based on the MLE and the SMLE for the full
data and the reduced data. As anticipated in Section \ref{Int} the MLE fit is highly sensitive to the outliers. Differently, the outliers have much smaller impact on the SMLE fitted curves.

Table \ref{estimates_DV_firmCost_2covariates} shows the estimates, standard errors, $z$-statistics, and $p$-values of the Wald-type test of nullity of coefficients for the full data and the data without the subsets of observations \{15\} and \{15, 16, 72\}. The SMLE are much less influenced by the presence of  outliers than the MLE. In fact, some MLE change considerably when observation 15 is excluded. For instance, the MLE for the coefficient of Size log in the precision submodel is $0.244$ (bootstrap $p$-value=$0.200$) for the full data and $0.652$ (bootstrap $p$-value=$0.002$) for the data without observation 15. Note the considerable relative change in the estimated coefficient of Size log (RC = 167\%) and the change of the inferential conclusion; Size log is highly significant for the data without observation 15 and non-significant for the full data at the usual nominal levels. A similar behavior occurs when the subset \{15, 16, 72\} is excluded. On the other hand, the corresponding robust estimate is $0.619$ (bootstrap $p$-value=$0.014$) for the full data and it is equal to the MLE when observation 15 is excluded. Note that the robust inference leads to the conclusion that Size log is strongly significant in the precision model regardless of whether outliers are excluded or not. Additionally, the maximum likelihood estimated coefficient of Ind cost in the mean submodel decreases to half when observation 15 is excluded, while, in the precision submodel, it jumps from $-5.718$ to $-1.596$. Overall, we notice that the robust estimates and the corresponding standard errors for the full data and for the reduced data are close to those obtained by the maximum likelihood method applied to the data without outliers. This suggests that the inference based on the robust estimation is more reliable than the maximum likelihood estimation.

\begin{table*}
	\centering
	\caption{Estimates, standard errors,  $z$-stat and $p$-values (bootstrap $p$-values between parenteses) for beta regression with varying precision -- Firm cost data.}
	\label{estimates_DV_firmCost_2covariates}
	\begin{threeparttable}
	\scalefont{0.7}
\def\arraystretch{0.65}	
 \begin{tabular}{>{\raggedleft\arraybackslash}m{2.1cm}>{\raggedleft\arraybackslash}m{1.2cm}>{\raggedleft\arraybackslash}m{1.2cm}>{\raggedleft\arraybackslash}m{1.2cm}>{\raggedleft\arraybackslash}m{1.7cm}>{\raggedleft\arraybackslash}m{0.05cm}>{\raggedleft\arraybackslash}m{1.2cm}>{\raggedleft\arraybackslash}m{1.2cm}>{\raggedleft\arraybackslash}m{1.2cm}>{\raggedleft\arraybackslash}m{1.7cm}}\hline
		& \multicolumn{4}{c}{MLE for the full dataset}                                      && \multicolumn{4}{c}{SMLE for the full dataset} \\  \cline{2-5}\cline{7-10}
		                      &Estimate   &Std. error &$z$-stat   &$p$-value               && Estimate  & Std. error&  $z$-stat &$p$-value    \\ \cline{2-5}\cline{7-10}
		{\it mean submodel}   &&&&&&&&\\
		Intercept             &  $2.188$  &  $0.917$  &  $2.386$  &  -                      &&  $3.557$  &  $0.846$  &  $4.207$  & -\\
		Ind cost              &  $3.763$  &  $0.551$  &  $6.830$  &  $0.000$  ($0.000$)     &&  $1.978$  &  $0.447$  &  $4.424$  &  $0.000$  ($0.004$)\\
		Size log              & $-0.714$  &  $0.109$  & $-6.580$  &  $0.000$  ($0.000$)     && $-0.828$  &  $0.098$  & $-8.427$  &  $0.000$  ($0.000$) \\
		{\it precision submodel}&&&&&&&&\\
		Intercept             &  $2.789$  &  $1.476$  &  $1.890$  &  -                      && $-1.145$  &  $1.522$  & $-0.753$  &      -\\
        Ind cost              & $-5.718$  &  $0.620$  & $-9.216$  &  $0.000$  ($0.000$)     && $-1.934$  &  $0.734$  & $-2.635$  &  $0.008$  ($0.058$)\\
		Size log              &  $0.244$  &  $0.170$  &  $1.435$  &  $0.151$  ($0.200$)     &&  $0.619$  &  $0.174$  &  $3.562$  &  $0.000$  ($0.014$) \\\cline{2-5}\cline{7-10}

& \multicolumn{4}{c}{MLE and SMLE without observation 15}                               && \multicolumn{4}{c}{MLE and SMLE without observations 15, 16 and 72} \\  \cline{2-5}\cline{7-10}
		                      & Estimate  &Std. error &  $z$-stat &$p$-value\tnote{a}       && Estimate   &Std. error&$z$-stat&$p$-value    \\  \cline{2-5}\cline{7-10}
		{\it mean submodel}&&&&&&&&\\
		Intercept             &  $3.654$  &  $0.846$  &  $4.321$  &  -                      &&   $2.563$  &  $0.917$  &  $2.795$  &  -\\
		Ind cost              &  $1.814$  &  $0.432$  &  $4.196$  &  $0.000$  ($0.008$)     &&   $1.814$  &  $0.476$  &  $3.809$  & $0.000$  ($0.008$)\\
		Size log              & $-0.832$  &  $0.098$  & $-8.498$  &  $0.000$  ($0.000$)     &&  $-0.703$  &  $0.108$  & $-6.508$  & $0.000$  ($0.000$)\\
		{\it precision submodel}&&&&&&&&\\
		Intercept             & $-1.522$  &  $1.530$  & $-0.995$  &       -                 &&   $0.462$  &  $1.732$  &  $0.267$  &       -\\
        Ind cost              & $-1.596$  &  $0.747$  & $-2.136$  &  $0.033$  ($0.118$) &&  $-1.704$  &  $0.868$  & $-1.963$  & $0.050$  ($0.140$) \\
		Size log              &  $0.652$  &  $0.175$  &  $3.728$  &  $0.000$  ($0.002$)     &&   $0.421$  &  $0.200$  &  $2.107$  &  $0.035$ ($0.068$)  \\\hline
\end{tabular}
 \begin{tablenotes}
	\item[a] The bootstrap $p$-values for the coefficients of Ind cost are those computed under the SMLE. The respective bootstrap $p$-values under the MLE are $0.006$ and $0.116$. The remaining bootstrap $p$-values under the SMLE and the MLE coincide.
\end{tablenotes}
\end{threeparttable}
\end{table*}

\begin{figure}[!htb]
\centering
\subfigure[][]{\includegraphics[width=5.0cm,height=4.5cm]{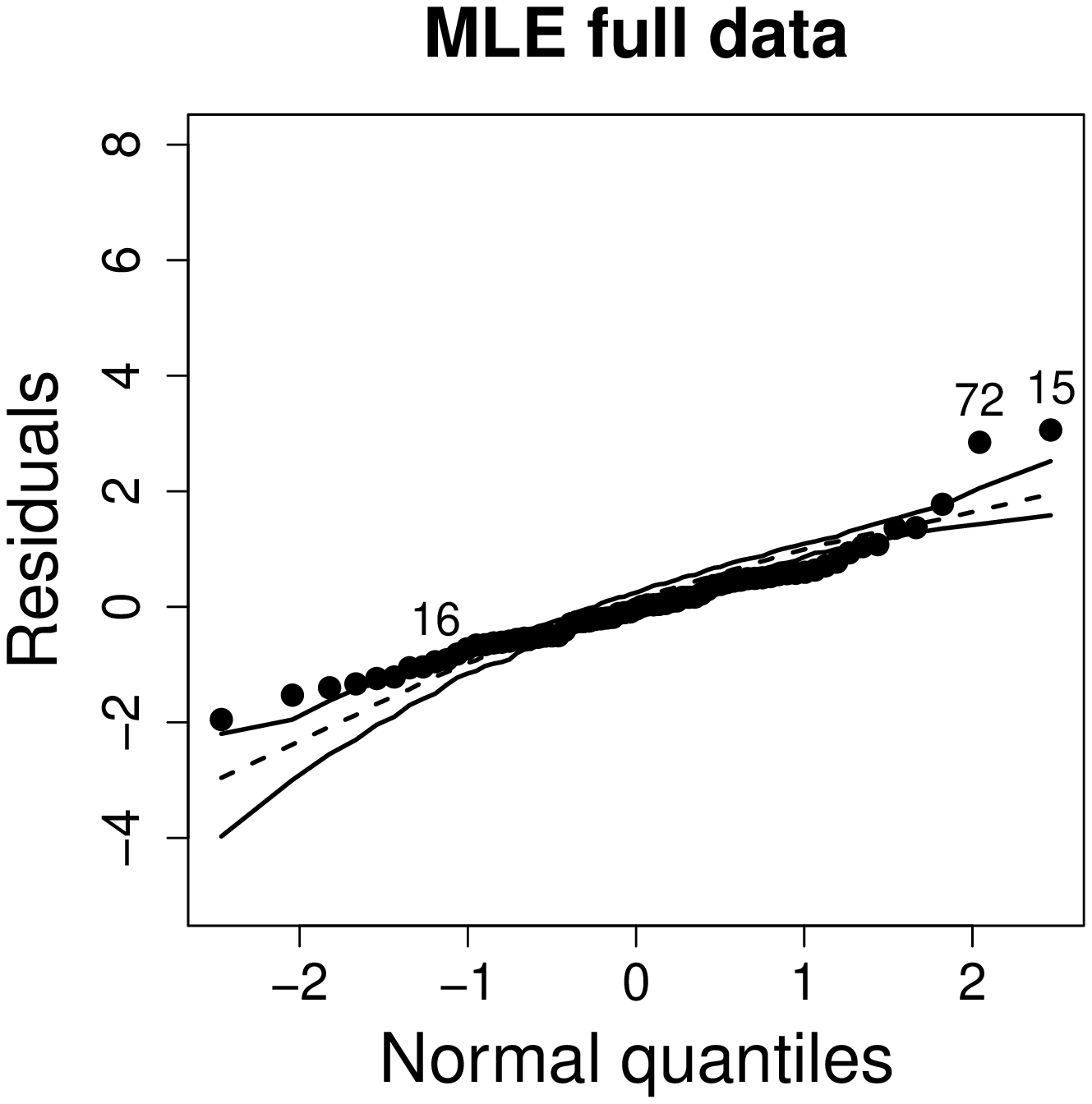}}
\qquad
\subfigure[][]{\includegraphics[width=5.0cm,height=4.5cm]{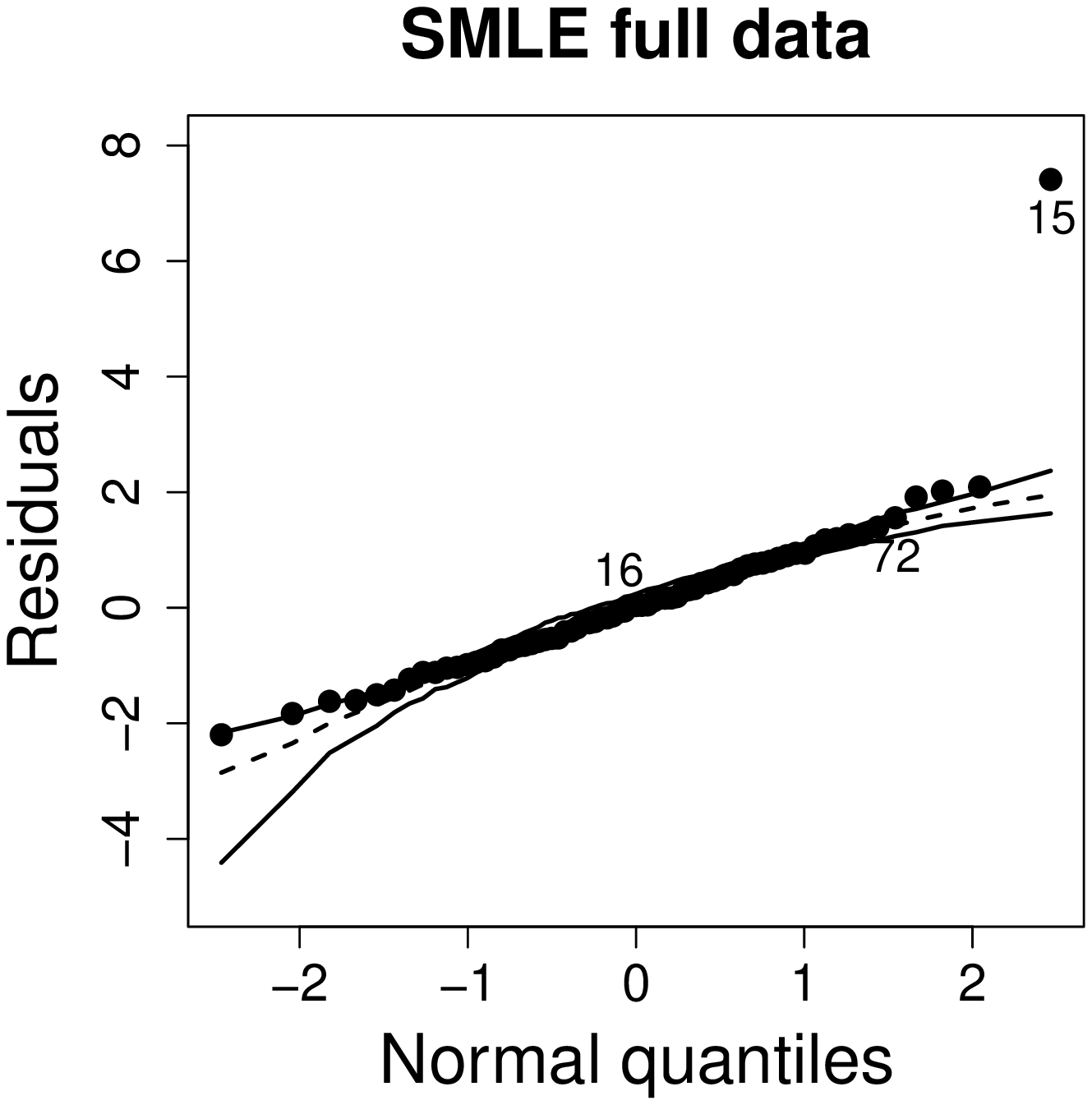}}
\qquad
\subfigure[][]{\includegraphics[width=5.0cm,height=4.5cm]{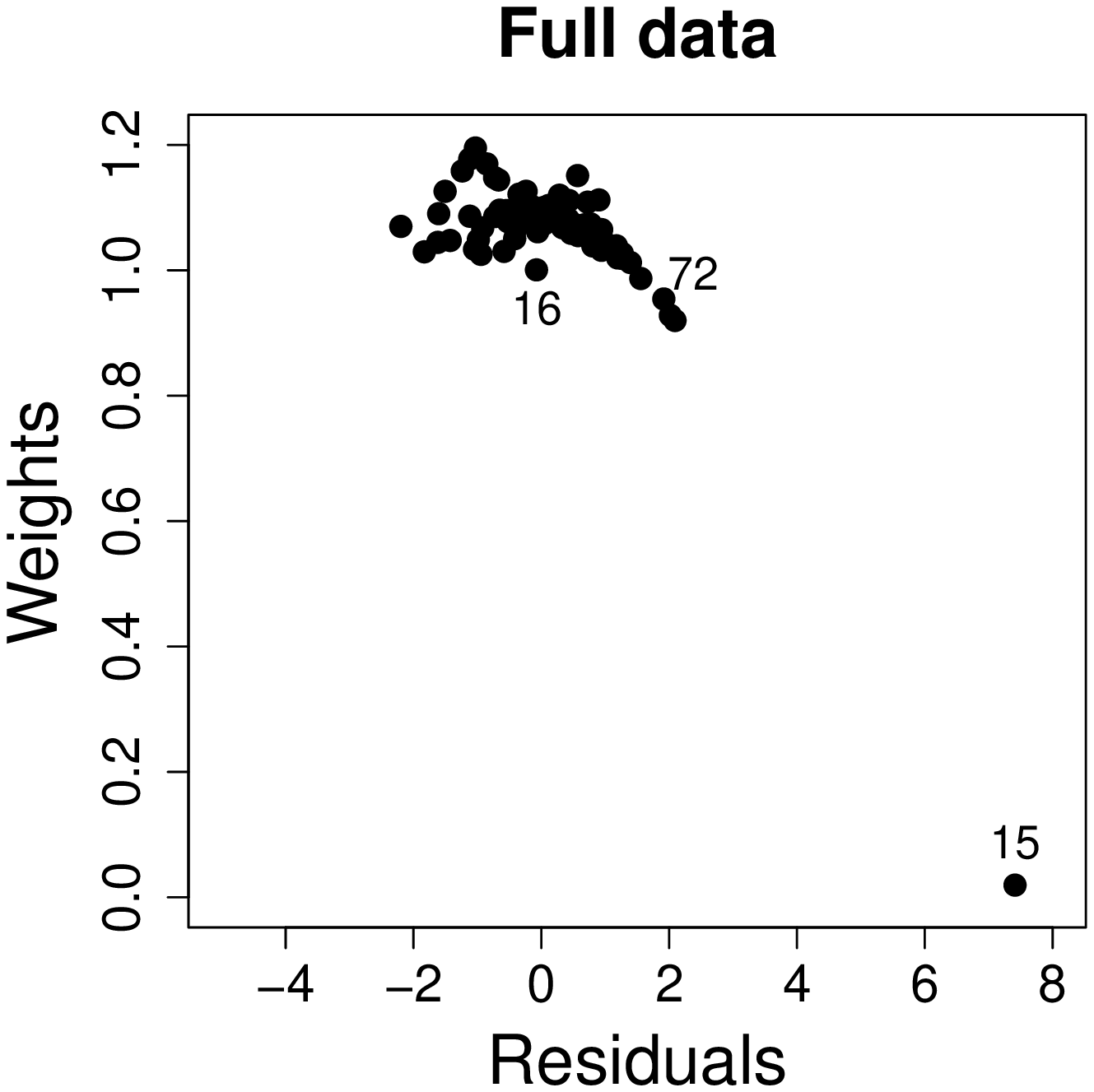}}
\qquad
\subfigure[][]{\includegraphics[width=5.0cm,height=4.5cm]{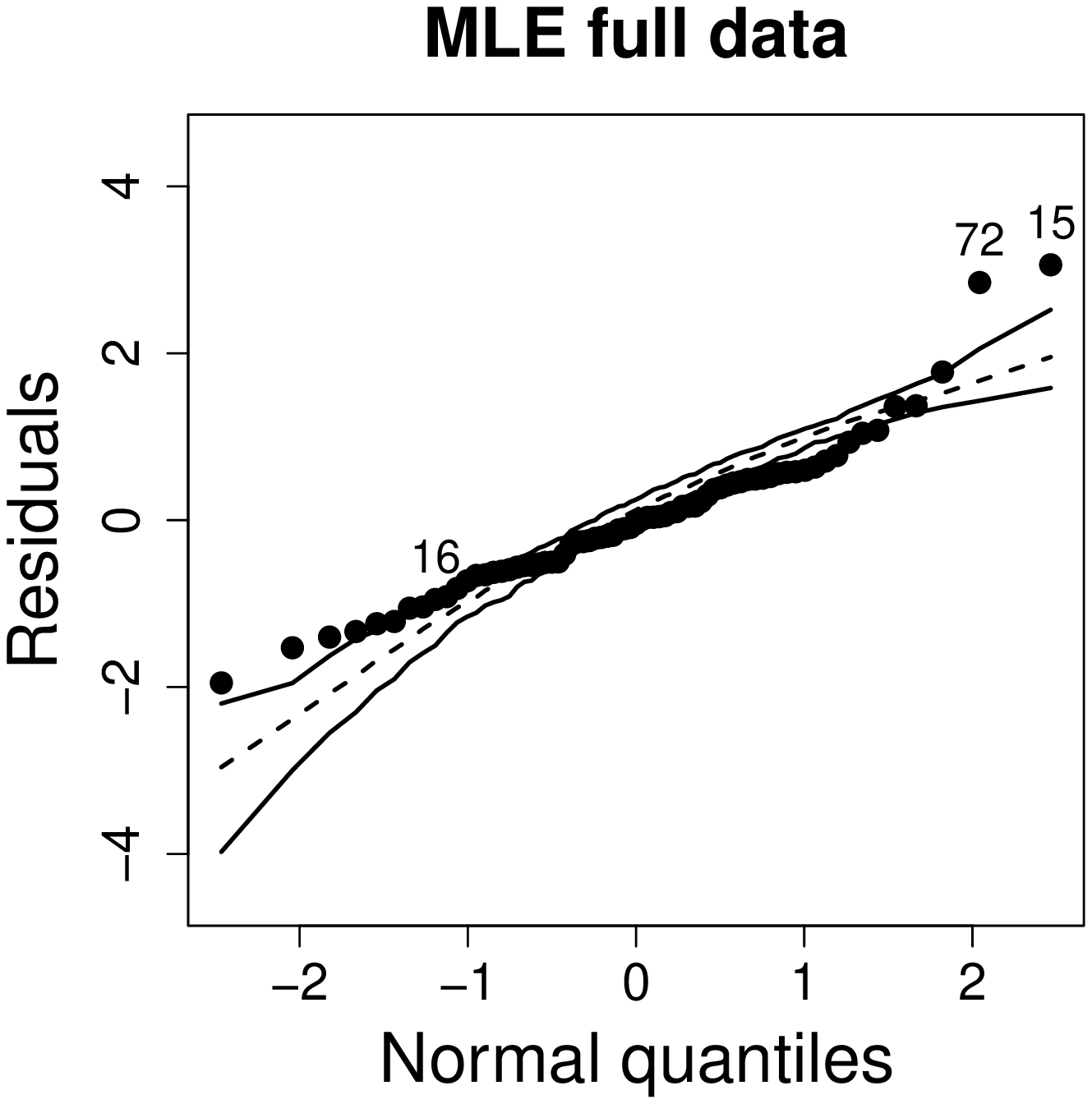}}
\qquad
\subfigure[][]{\includegraphics[width=5.0cm,height=4.5cm]{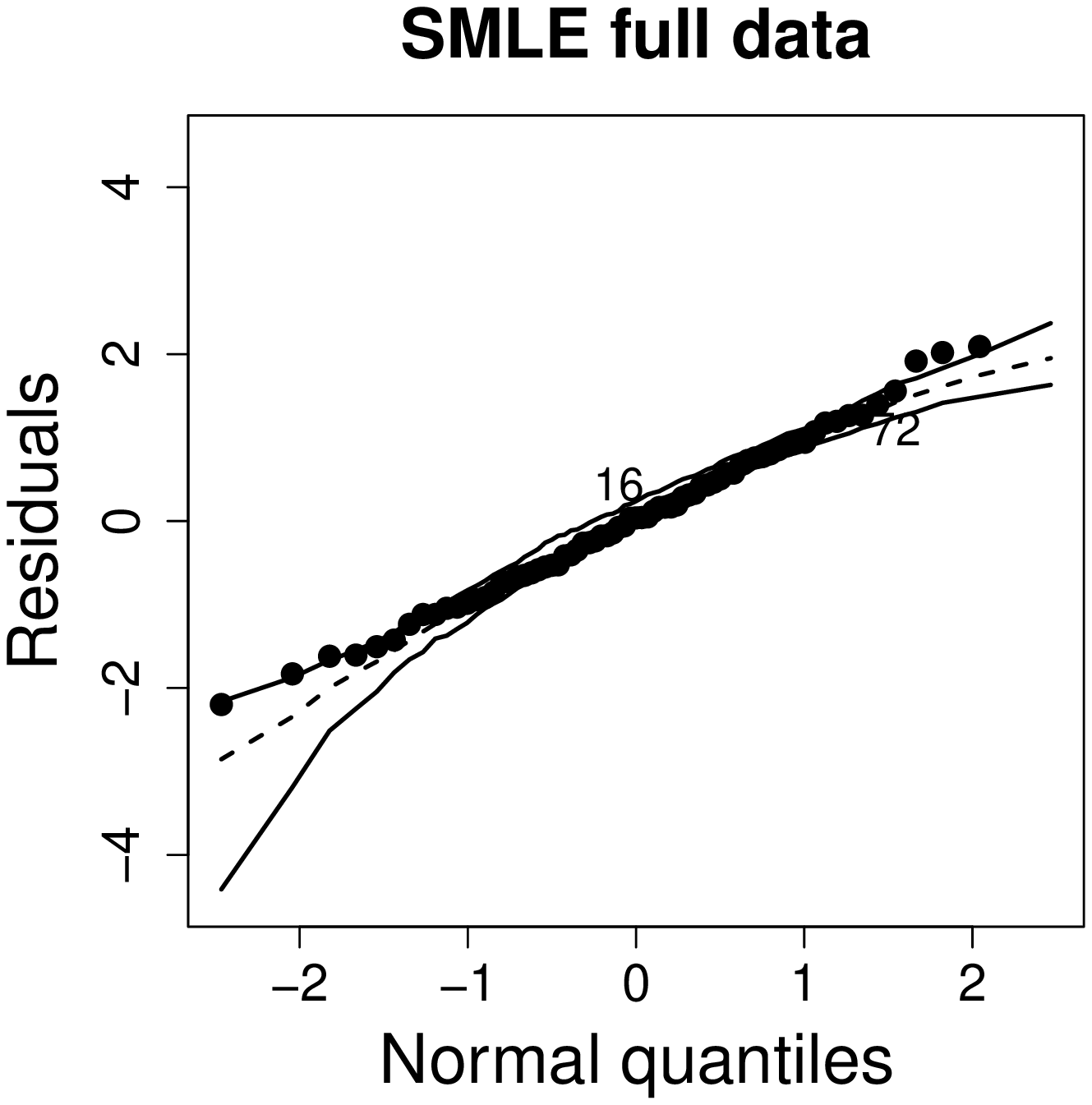}}
\qquad
\subfigure[][]{\includegraphics[width=5.0cm,height=4.5cm]{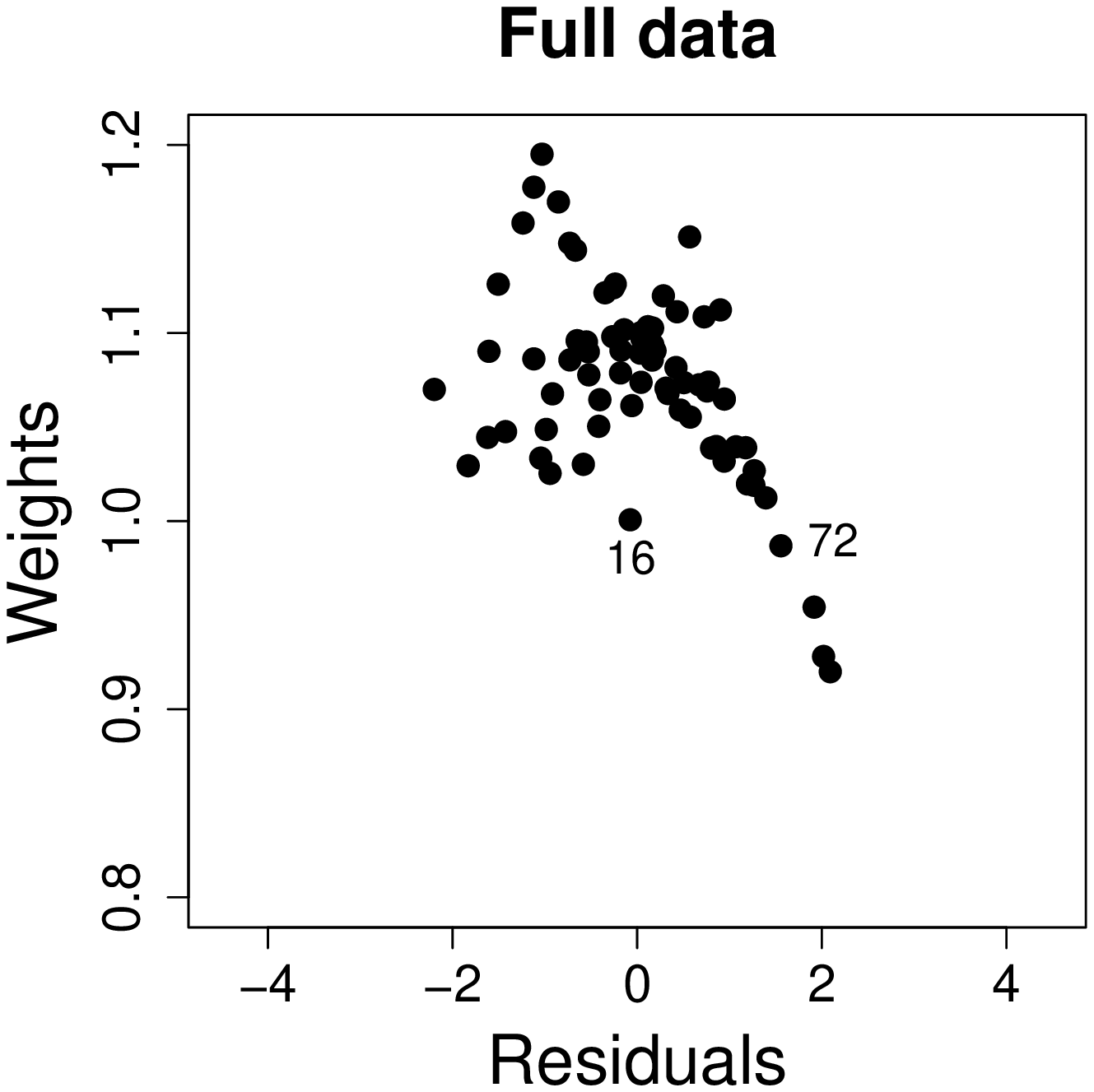}}
\caption{Normal probability plots of the residuals for MLE and SMLE and plot of estimated weights for SMLE; varying precision model -- Firm cost data. The plots in the second row are zoomed versions of those in the first row.}
\label{ALL_ENVELOPES1}
\end{figure}

Figure \ref{ALL_ENVELOPES1} shows normal probability plots of the residuals and estimated weights for the fitted models employing the MLE and the SMLE. Plots (a) and (d) suggests lack of the MLE fit and slightly displays observations 15 and 72 as outliers. Plots (b) and (e), that correspond to the SMLE fit for the full data, show that the majority of the data, including observations 16 and 72, is accommodated within the envelope. Observation 15 is clearly highlighted as an outlier. Indeed, only observation 15 receives a small weight in the robust fit; see plot (c) and (f). This behavior is expected because observations 16 and 72 do not have substantial influence in the MLE fit when observation 15 is not present in the data; see plots (a) and (c) in Figure \ref{scatter_Firmcost}.

\section{Concluding remarks}\label{concl}

Maximum likelihood estimation in beta regression is heavily influenced by outliers. We have proposed a robust inference method based on a reparameterized L$_q$-likelihood. The reparameterization is achieved through simple modifications in the mean and precision link functions. We have demonstrated robustness properties of the proposed estimator for bounded beta densities. The robust method depends on a tuning constant that balances robustness and efficiency. The selection of such a constant is a crucial issue for practical applications. We have proposed an algorithm to select the tuning constant based on the data. The algorithm is designed to choose $q=1$, that corresponds to maximum likelihood estimation, when there is no contamination in the data, and to choose a suitable value of $0<q<1$ when a robust fit is required. We have improved on the minimum power divergence estimator \citep{ghosh2019robust} by implementing our tuning constant selection scheme. Monte Carlo simulations and real data applications have illustrated the advantages of the robust methodology over the usual likelihood-based inference. As a by-product of the proposed approach, diagnostic tools such as normal probability plots of residuals based on robust fits have demonstrated effectiveness in highlighting outliers that would be masked under a non-robust procedure.

 We emphasize that the codes and datasets used in this paper are available at the GitHub repository
 \longurl{https://github.com/terezinharibeiro/RobustBetaRegression}. Practitioners and researchers can replicate the simulations and applications and employ the proposed robust beta regression approach in their own data analyses.

The lack of robustness of the maximum likelihood estimation under data contamination renders Wald-type tests extremely oversized, hence unreliable. The proposed estimator leads to usable Wald-type tests. Likelihood ratio, score and gradient tests based on the reparameterized L$_q$-likelihood and the corresponding weighted modified score function in beta regression will be the subject of our future research. 

A limitation of the present work is that robustness is achieved only for bounded beta densities, which fortunately are the most common in applied research. The second author has been working in an alternative robust method that preserves robustness even for beta densities that grow unboundedly at one or both of the boundaries of the support set. The work is in progress and the results will be reported elsewhere.

A usefull extension of the present work concerns inflated beta regression models \citep{ospina2012general}, that are employed when the response variable includes  a non-negligible amount of observations at zero and/or one. This topic is beyond the scope of the present paper and will be dealt with in the near future.

\vspace{1cm}
\section*{Acknowledgments}

This study was financed in part by the Coordenação de Aperfeiçoamento de Pessoal de Nível Superior - Brazil (CAPES) - Finance Code 001 and by the Conselho
Nacional de Desenvolvimento Científico e Tecnológico - Brazil (CNPq). The second author gratefully acknowledges funding provided by CNPq (Grant No. 305963-2018-0).

\appendix
\section*{Appendix}

We define the following diagonal matrices with their respective diagonal elements:
$B_1 = \diag\{b_{1,1},\ldots, b_{n,1} \}$ and $B_2 = \diag\{b_{1,2},\ldots, b_{n,2} \}$, with 
\begin{eqnarray*}
b_{i,1} = \frac{B(\mu_i\phi_i, (1-\mu_i)\phi_i)^q}{B(\mu_{i,q}\phi_{i,q}, (1-\mu_{i,q})\phi_{i,q})}, \quad 
b_{i,2} = \frac{B(\mu_{i,2-q}\phi_{i,2-q}, (1-\mu_{i,2-q})\phi_{i,2-q})}{B(\mu_i\phi_i, (1-\mu_i)\phi_i)^{2(1-q)}B(\mu_{i,q}\phi_{i,q}, 
(1-\mu_{i,q})\phi_{i,q})};
\end{eqnarray*}
$T_\mu = \diag\{t_{1,\mu},\ldots, t_{n,\mu}\}$ and $T_\phi = \diag\{t_{1,\phi},\ldots, t_{n,\phi}\}$, with 
$$t_{i,\mu} = [g_{\mu}'(\mu_{i,q})]^{-1}, \quad 
t_{i,\phi} = [g_{\phi}'(\phi_{i,q})]^{-1};$$ 
$\Phi_q = \diag\{\phi_{1,q},\ldots, \phi_{n,q} \}$; 
$V = \diag\{v_{1},\ldots, v_{n} \}$ and $V_{2-q} = \diag\{v_{1,2-q},\ldots, v_{n,2-q} \}$, with 
\begin{eqnarray*}
 v_{i} = \mbox{Var}(y^{\star}_i) = \psi'(\mu_{i}\phi_{i}) + \psi'((1-\mu_{i})\phi_{i}), \quad 
v_{i, 2-q} = \psi'(\mu_{i, 2-q}\phi_{i, 2-q}) + \psi'((1-\mu_{i, 2-q})\phi_{i, 2-q});
\label{defV}
\end{eqnarray*}
$C^{*} = \diag\{c^{*}_{1},\ldots, c^{*}_{n} \}$ and $C^{*}_{2-q} = \diag\{c^{*}_{1,2-q},\ldots, c^{*}_{n,2-q} \}$, with
\begin{eqnarray*}
c^{*}_{i} &=& \phi_{i,q}[\mu_{i,q}\psi'(\mu_{i}\phi_{i}) -(1-\mu_{i,q}) \psi'((1-\mu_{i})\phi_{i})];\\
c^{*}_{i, 2-q} &=& \phi_{i,q}[\mu_{i,q}\psi'(\mu_{i, 2-q}\phi_{i, 2-q}) -(1-\mu_{i,q})\psi'((1-\mu_{i, 2-q})\phi_{i, 2-q})];
\end{eqnarray*}
$D^{*} = \diag\{d^{*}_{1},\ldots, d^{*}_{n} \}$ and $D^{*}_{2-q} = \diag\{d^{*}_{1,2-q},\ldots, d^{*}_{n,2-q} \}$, with
\begin{eqnarray*}
d^{*}_{i} &=& \mu_{i,q}^2\psi'(\mu_{i}\phi_{i}) +(1-\mu_{i,q})^2 \psi'((1-\mu_{i})\phi_{i}) - \psi'(\phi_{i}),\\
d^{*}_{i,  2-q} &=& \mu_{i,q}^2\psi'(\mu_{i,2-q}\phi_{i,2-q}) +(1-\mu_{i,q})^2 \psi'((1-\mu_{i,2-q})\phi_{i,2-q}) - \psi'(\phi_{i,2-q});
\end{eqnarray*}
\begin{eqnarray*}
M_1 &=& \diag\{(\mu^{\star}_{1,2-q}-\mu^{\star}_{1}),\ldots, (\mu^{\star}_{n,2-q}-\mu^{\star}_{n})\},\\
M_2 &=& \diag\{\mu^{d}_{1,2-q},\ldots, \mu^{d}_{n,2-q} \}, \\
M_3 &=& \diag\{\mu^{d}_{1,2-q}\phi_{1,q}(\mu^{\star}_{1,2-q}-\mu^{\star}_{1}),\ldots, \mu^{d}_{n,2-q}\phi_{n,q}(\mu^{\star}_{n,2-q}-\mu^{\star}_{n})\},
\end{eqnarray*}
where
\begin{eqnarray*}
\mu^{\star}_{i,2-q} &=& \psi(\mu_{i,2-q}\phi_{i,2-q})-\psi((1-\mu_{i,2-q})\phi_{i,2-q}),\\
\mu^{\dagger}_{i,2-q} &=& \psi((1-\mu_{i,2-q})\phi_{i,2-q})-\psi(\phi_{i,2-q}),\\
\mu^{d}_{i,2-q} &=& \mu_{i,q}(\mu^{\star}_{i,2-q}-\mu^{\star}_{i})+\mu^{\dagger}_{i,2-q}-\mu^{\dagger}_{i}.
\end{eqnarray*}
All the above matrices are well-defined provided that $0<\mu_{i,2-q}<1$ and $\phi_{i,2-q}>0$. For the beta regression model (\ref{fdpbeta})-(\ref{ligphi}), $J_q(\mat{\theta})$ and $K_q(\mat{\theta})$ 
we get after some algebra
\begin{eqnarray*}
J_q(\mat{\theta}) = -q^{-1}\left(
\begin{array}{cccc}
X^\top B_1T_{\mu}^2\Phi_q^2 V X & X^\top B_1 T_{\mu}T_{\phi}C^{*} Z\\
Z^\top B_1 T_{\mu}T_{\phi}C^{*}X & Z^\top B_1T_{\phi}^2D^{*}Z \\
\end{array}
\right),
\label{J_q}
\end{eqnarray*}
\begin{eqnarray*}
K_q(\mat{\theta}) = q^{-2} \left(
\begin{array}{cccc}
X^\top B_2T_{\mu}^2\Phi_q^2 (V_{2-q} + M^2_1)X & X^\top B_2 T_{\mu}T_{\phi}(C^{*}_{2-q}+ M_3)Z\\
Z^\top B_2T_{\mu}T_{\phi}(C^{*}_{2-q}+ M_3)X & Z^\top B_2T_{\phi}^2(D^{*}_{2-q}+M_2^2)Z \\
\end{array}
\right),
\label{K_q}
\end{eqnarray*}
where $X=(\mat{X}_1,\ldots,\mat{X}_{p_1})$, $Z=(\mat{Z}_1,\ldots,\mat{Z}_{p_2})$; see the Supplementary Material for details.

For instance, setting $q=1$
\begin{eqnarray*}
	K_1(\mat{\theta}) = \left(
	\begin{array}{cccc}
		X^\top T_{\mu}^2\Phi^2VX & X^\top  T_{\mu}T_{\phi}CZ\\
		Z^\top T_{\mu}T_{\phi}CX & Z^\top T_{\phi}^2DZ \\
	\end{array}
	\right),
	\label{K_1}
\end{eqnarray*}
where $\Phi = \diag\{\phi_{1},\ldots, \phi_{n} \}$; $C = \diag\{c_{1},\ldots, c_{n} \}$ and $D = \diag\{d_{1},\ldots, d_{n} \}$ with
\begin{eqnarray*}
	c_{i} &=& \phi_{i}[\mu_{i}\psi'(\mu_{i}\phi_{i}) -(1-\mu_{i}) \psi'((1-\mu_{i})\phi_{i})];\\
	d_{i} &=& \mu_{i}^2\psi'(\mu_{i}\phi_{i}) +(1-\mu_{i})^2 \psi'((1-\mu_{i})\phi_{i}) - \psi'(\phi_{i}).
\end{eqnarray*}

\bibliographystyle{model2-names}

\bibliography{mybibfile}

\begin{thebibliography}{27}
\expandafter\ifx\csname natexlab\endcsname\relax\def\natexlab#1{#1}\fi
\providecommand{\url}[1]{\texttt{#1}}
\providecommand{\href}[2]{#2}
\providecommand{\path}[1]{#1}
\providecommand{\DOIprefix}{doi:}
\providecommand{\ArXivprefix}{arXiv:}
\providecommand{\URLprefix}{URL: }
\providecommand{\Pubmedprefix}{pmid:}
\providecommand{\doi}[1]{\href{http://dx.doi.org/#1}{\path{#1}}}
\providecommand{\Pubmed}[1]{\href{pmid:#1}{\path{#1}}}
\providecommand{\bibinfo}[2]{#2}
\ifx\xfnm\relax \def\xfnm[#1]{\unskip,\space#1}\fi
\bibitem[{Basu et~al.(1998)Basu, Harris, Hjort and Jones}]{basu1998}
\bibinfo{author}{Basu, A.}, \bibinfo{author}{Harris, I.R.},
  \bibinfo{author}{Hjort, N.L.}, \bibinfo{author}{Jones, M.},
  \bibinfo{year}{1998}.
\newblock \bibinfo{title}{Robust and efficient estimation by minimising a
  density power divergence}.
\newblock \bibinfo{journal}{Biometrika} \bibinfo{volume}{85},
  \bibinfo{pages}{549--559}.
\bibitem[{Bayes et~al.(2012)Bayes, Baz{\'a}n and Garc{\'\i}a}]{bayes2012new}
\bibinfo{author}{Bayes, C.L.}, \bibinfo{author}{Baz{\'a}n, J.L.},
  \bibinfo{author}{Garc{\'\i}a, C.}, \bibinfo{year}{2012}.
\newblock \bibinfo{title}{A new robust regression model for proportions}.
\newblock \bibinfo{journal}{Bayesian Analysis} \bibinfo{volume}{7},
  \bibinfo{pages}{841--866}.
\bibitem[{Box and Cox(1964)}]{box1964analysis}
\bibinfo{author}{Box, G.E.}, \bibinfo{author}{Cox, D.R.}, \bibinfo{year}{1964}.
\newblock \bibinfo{title}{An analysis of transformations}.
\newblock \bibinfo{journal}{Journal of the Royal Statistical Society B}
  \bibinfo{volume}{26}, \bibinfo{pages}{211--252}.
\bibitem[{Di~Brisco et~al.(2020)Di~Brisco, Migliorati and
  Ongaro}]{dibrisco2019}
\bibinfo{author}{Di~Brisco, A.M.}, \bibinfo{author}{Migliorati, S.},
  \bibinfo{author}{Ongaro, A.}, \bibinfo{year}{2020}.
\newblock \bibinfo{title}{Robustness against outliers: A new variance inflated
  regression model for proportions}.
\newblock \bibinfo{journal}{Statistical Modelling} \bibinfo{volume}{20},
  \bibinfo{pages}{274--309}.
\bibitem[{Espinheira et~al.(2008)Espinheira, Ferrari and
  Cribari-Neto}]{espinheira2008beta}
\bibinfo{author}{Espinheira, P.L.}, \bibinfo{author}{Ferrari, S.L.P.},
  \bibinfo{author}{Cribari-Neto, F.}, \bibinfo{year}{2008}.
\newblock \bibinfo{title}{On beta regression residuals}.
\newblock \bibinfo{journal}{Journal of Applied Statistics}
  \bibinfo{volume}{35}, \bibinfo{pages}{407--419}.
\bibitem[{Espinheira et~al.(2017)Espinheira, Santos and
  Cribari-Neto}]{espinheira2017nonlinear}
\bibinfo{author}{Espinheira, P.L.}, \bibinfo{author}{Santos, E.G.},
  \bibinfo{author}{Cribari-Neto, F.}, \bibinfo{year}{2017}.
\newblock \bibinfo{title}{On nonlinear beta regression residuals}.
\newblock \bibinfo{journal}{Biometrical Journal} \bibinfo{volume}{59},
  \bibinfo{pages}{445--461}.
\bibitem[{Espinheira et~al.(2019)Espinheira, da~Silva, Silva and
  Ospina}]{espinheira2019model}
\bibinfo{author}{Espinheira, P.L.}, \bibinfo{author}{da~Silva, L.C.M.},
  \bibinfo{author}{Silva, A.O.}, \bibinfo{author}{Ospina, R.},
  \bibinfo{year}{2019}.
\newblock \bibinfo{title}{Model selection criteria on beta regression for
  machine learning}.
\newblock \bibinfo{journal}{Machine Learning and Knowledge Extraction}
  \bibinfo{volume}{1}, \bibinfo{pages}{427--449}.
\bibitem[{Ferrari and La~Vecchia(2012)}]{Ferrari2012}
\bibinfo{author}{Ferrari, D.}, \bibinfo{author}{La~Vecchia, D.},
  \bibinfo{year}{2012}.
\newblock \bibinfo{title}{{On robust estimation via pseudo-additive
  information}}.
\newblock \bibinfo{journal}{Biometrika} \bibinfo{volume}{99},
  \bibinfo{pages}{238--244}.
\bibitem[{Ferrari and Yang(2010)}]{ferrari2010maximum}
\bibinfo{author}{Ferrari, D.}, \bibinfo{author}{Yang, Y.},
  \bibinfo{year}{2010}.
\newblock \bibinfo{title}{Maximum $\mbox{L}_q$-likelihood estimation}.
\newblock \bibinfo{journal}{The Annals of Statistics} \bibinfo{volume}{38},
  \bibinfo{pages}{753--783}.
\bibitem[{Ferrari and Cribari-Neto(2004)}]{ferrari2004beta}
\bibinfo{author}{Ferrari, S.L.P.}, \bibinfo{author}{Cribari-Neto, F.},
  \bibinfo{year}{2004}.
\newblock \bibinfo{title}{Beta regression for modelling rates and proportions}.
\newblock \bibinfo{journal}{Journal of Applied Statistics}
  \bibinfo{volume}{31}, \bibinfo{pages}{799--815}.
\bibitem[{Ghosh(2019)}]{ghosh2019robust}
\bibinfo{author}{Ghosh, A.}, \bibinfo{year}{2019}.
\newblock \bibinfo{title}{Robust inference under the beta regression model with
  application to health care studies}.
\newblock \bibinfo{journal}{Statistical Methods in Medical Research}
  \bibinfo{volume}{28}, \bibinfo{pages}{871--888}.
\bibitem[{G{\'o}mez-D{\'e}niz et~al.(2014)G{\'o}mez-D{\'e}niz, Sordo and
  Calder{\'\i}n-Ojeda}]{gomez2014log}
\bibinfo{author}{G{\'o}mez-D{\'e}niz, E.}, \bibinfo{author}{Sordo, M.A.},
  \bibinfo{author}{Calder{\'\i}n-Ojeda, E.}, \bibinfo{year}{2014}.
\newblock \bibinfo{title}{The log--\mbox{L}indley distribution as an
  alternative to the beta regression model with applications in insurance}.
\newblock \bibinfo{journal}{Insurance: Mathematics and Economics}
  \bibinfo{volume}{54}, \bibinfo{pages}{49--57}.
\bibitem[{Hampel et~al.(2011)Hampel, Ronchetti, Rousseeuw and
  Stahel}]{hampel2011robust}
\bibinfo{author}{Hampel, F.R.}, \bibinfo{author}{Ronchetti, E.M.},
  \bibinfo{author}{Rousseeuw, P.J.}, \bibinfo{author}{Stahel, W.A.},
  \bibinfo{year}{2011}.
\newblock \bibinfo{title}{Robust Statistics: The Approach Based on Influence
  Functions}. volume \bibinfo{volume}{196}.
\newblock \bibinfo{publisher}{New York: John Wiley \& Sons}.
\bibitem[{Heritier et~al.(2009)Heritier, Cantoni, Copt and
  Victoria-Feser}]{heritier2009robust}
\bibinfo{author}{Heritier, S.}, \bibinfo{author}{Cantoni, E.},
  \bibinfo{author}{Copt, S.}, \bibinfo{author}{Victoria-Feser, M.P.},
  \bibinfo{year}{2009}.
\newblock \bibinfo{title}{Robust Methods in Biostatistics}. volume
  \bibinfo{volume}{825}.
\newblock \bibinfo{publisher}{New York: John Wiley \& Sons}.
\bibitem[{Heritier and Ronchetti(1994)}]{heritier1994robust}
\bibinfo{author}{Heritier, S.}, \bibinfo{author}{Ronchetti, E.},
  \bibinfo{year}{1994}.
\newblock \bibinfo{title}{Robust bounded-influence tests in general parametric
  models}.
\newblock \bibinfo{journal}{Journal of the American Statistical Association}
  \bibinfo{volume}{89}, \bibinfo{pages}{897--904}.
\bibitem[{Huber(1981)}]{huber1981j}
\bibinfo{author}{Huber, P.}, \bibinfo{year}{1981}.
\newblock \bibinfo{title}{Robust Statistics}.
\newblock \bibinfo{publisher}{New York: John Wiley \& Sons}.
\bibitem[{Huber(1964)}]{huber1964robust}
\bibinfo{author}{Huber, P.J.}, \bibinfo{year}{1964}.
\newblock \bibinfo{title}{Robust estimation of a location parameter}.
\newblock \bibinfo{journal}{The Annals of Mathematical Statistics}
  \bibinfo{volume}{35}, \bibinfo{pages}{73--101}.
\bibitem[{La~Vecchia et~al.(2015)La~Vecchia, Camponovo and
  Ferrari}]{la2015robust}
\bibinfo{author}{La~Vecchia, D.}, \bibinfo{author}{Camponovo, L.},
  \bibinfo{author}{Ferrari, D.}, \bibinfo{year}{2015}.
\newblock \bibinfo{title}{Robust heart rate variability analysis by generalized
  entropy minimization}.
\newblock \bibinfo{journal}{Computational Statistics \& Data Analysis}
  \bibinfo{volume}{82}, \bibinfo{pages}{137--151}.
\bibitem[{Migliorati et~al.(2018)Migliorati, Di~Brisco and
  Ongaro}]{migliorati2018new}
\bibinfo{author}{Migliorati, S.}, \bibinfo{author}{Di~Brisco, A.M.},
  \bibinfo{author}{Ongaro, A.}, \bibinfo{year}{2018}.
\newblock \bibinfo{title}{A new regression model for bounded responses}.
\newblock \bibinfo{journal}{Bayesian Analysis} \bibinfo{volume}{13},
  \bibinfo{pages}{845--872}.
\bibitem[{Monllor-Hurtado et~al.(2017)Monllor-Hurtado, Pennino and
  Sanchez-Lizaso}]{monllor2017shift}
\bibinfo{author}{Monllor-Hurtado, A.}, \bibinfo{author}{Pennino, M.G.},
  \bibinfo{author}{Sanchez-Lizaso, J.L.}, \bibinfo{year}{2017}.
\newblock \bibinfo{title}{Shift in tuna catches due to ocean warming}.
\newblock \bibinfo{journal}{PloS One} \bibinfo{volume}{12},
  \bibinfo{pages}{e0178196}.
\bibitem[{van Niekerk et~al.(2019)van Niekerk, Bekker and
  Arashi}]{niekerk2018beta}
\bibinfo{author}{van Niekerk, J.}, \bibinfo{author}{Bekker, A.},
  \bibinfo{author}{Arashi, M.}, \bibinfo{year}{2019}.
\newblock \bibinfo{title}{Beta regression in the presence of outliers -- a
  wieldy bayesian solution}.
\newblock \bibinfo{journal}{Statistical Methods in Medical Research}
  \bibinfo{volume}{28}, \bibinfo{pages}{3729--3740}.
\bibitem[{Ospina and Ferrari(2012)}]{ospina2012general}
\bibinfo{author}{Ospina, R.}, \bibinfo{author}{Ferrari, S.L.},
  \bibinfo{year}{2012}.
\newblock \bibinfo{title}{A general class of zero-or-one inflated beta
  regression models}.
\newblock \bibinfo{journal}{Computational Statistics \& Data Analysis}
  \bibinfo{volume}{56}, \bibinfo{pages}{1609--1623}.
\bibitem[{Pereira(2019)}]{pereira2019quantile}
\bibinfo{author}{Pereira, G.H.A.}, \bibinfo{year}{2019}.
\newblock \bibinfo{title}{On quantile residuals in beta regression}.
\newblock \bibinfo{journal}{Communications in Statistics - Simulation and
  Computation} \bibinfo{volume}{48}, \bibinfo{pages}{302--316}.
\bibitem[{{R Core Team}(2020)}]{Rreference}
\bibinfo{author}{{R Core Team}}, \bibinfo{year}{2020}.
\newblock \bibinfo{title}{R: A Language and Environment for Statistical
  Computing}.
\newblock \bibinfo{organization}{R Foundation for Statistical Computing}.
  \bibinfo{address}{Vienna, Austria}.
\newblock \URLprefix \url{https://www.R-project.org/}.
\bibitem[{Schmit and Roth(1990)}]{schmit1990cost}
\bibinfo{author}{Schmit, J.T.}, \bibinfo{author}{Roth, K.},
  \bibinfo{year}{1990}.
\newblock \bibinfo{title}{Cost effectiveness of risk management practices}.
\newblock \bibinfo{journal}{Journal of Risk and Insurance}
  \bibinfo{volume}{57}, \bibinfo{pages}{455--470}.
\bibitem[{Smithson and Verkuilen(2006)}]{smithson2006better}
\bibinfo{author}{Smithson, M.}, \bibinfo{author}{Verkuilen, J.},
  \bibinfo{year}{2006}.
\newblock \bibinfo{title}{A better lemon squeezer? \mbox{M}aximum-likelihood
  regression with beta-distributed dependent variables.}
\newblock \bibinfo{journal}{Psychological Methods} \bibinfo{volume}{11},
  \bibinfo{pages}{54--71}.
\bibitem[{Tsallis(1988)}]{tsallis1988possible}
\bibinfo{author}{Tsallis, C.}, \bibinfo{year}{1988}.
\newblock \bibinfo{title}{Possible generalization of
  \mbox{B}oltzmann-\mbox{G}ibbs statistics}.
\newblock \bibinfo{journal}{Journal of Statistical Physics}
  \bibinfo{volume}{52}, \bibinfo{pages}{479--487}.

\end{thebibliography}

\end{document}